\documentclass[11pt,3p,doublespacing]{elsarticle}
\journal{Journal for consideration.}

\usepackage{mathptmx} % Times-compatible math fonts
\usepackage{titlesec}
\usepackage{sectsty}
\usepackage[version=4]{mhchem} % for simple formula, reactions, ions, etc
\usepackage{chemfig} % for useful for organic chemistry diagrams.
\biboptions{sort&compress}

\usepackage[colorlinks]{hyperref}
\AtBeginDocument{%
	\hypersetup{
		plainpages=false,
		bookmarksopen=true,
		bookmarksnumbered=true,
		breaklinks=true,
		linktocpage=true,
		colorlinks=true,
		linkcolor=purple,
		urlcolor=black,
		citecolor=magenta
	}
}

\usepackage{imakeidx}
\makeindex
\usepackage{nomencl}
\makenomenclature
\usepackage[xindy]{glossaries}
\makeglossaries

\usepackage{graphicx,epsfig,epstopdf,psfrag,rotating,pdflscape,float}
\graphicspath{{../}{Figures/}{Figures/Validation/}}
\DeclareGraphicsExtensions{.png,.jpg,.pdf,.eps}

\usepackage{array,multirow,longtable,threeparttable,tablefootnote}
\usepackage{footnote}
\makesavenoteenv{tabular}
\makesavenoteenv{table}

\usepackage{amsmath,amssymb,amsfonts,gensymb,mathtools}
\usepackage{textgreek}
\usepackage{siunitx}

\usepackage[left,displaymath,mathlines]{lineno}
\usepackage[normalem]{ulem} % for \sout
\usepackage[usenames,dvipsnames]{xcolor}
\usepackage{soul} % for \hl, \ul etc.

\usepackage{caption}
\usepackage{subcaption} % modern replacement for subfigure

\usepackage{setspace,enumitem,csquotes}
\usepackage{cancel}
\usepackage{resizegather}
\usepackage[markup=underlined,final]{changes}

\definechangesauthor[name={Aishvarya}, color=blue]{AK}

\setdeletedmarkup{\textcolor{red}{\sout{#1}}}

\setaddedmarkup{\textcolor{blue}{#1}}

\newcommand{\overbar}[1]{\mkern1.5mu\overline{\mkern-1.5mu#1\mkern-1.5mu}\mkern1.5mu}
\newcommand{\fig}{Fig.~}
\newcommand{\figs}{Figs.~}
\newcommand{\tab}{Table~}

\newcommand{\eqn}{Eq.~}
\newcommand{\eqns}{Eqs.~}

\usepackage{xcolor}

\begin{document}
	\begin{frontmatter}
		
	%	\title{Three-dimensional modelling of diffusion-type turbulent combustion in a realistic can-type combustor}
	%	\title{Comprehensive Assessment of RANS-based Turbulence Models for Turbulent Non-Premixed Swirling Combustion in a Realistic Can Combustor}
		\title{Performance Evaluation of RANS-Based Turbulence Models in Predicting Turbulent Non-Premixed Swirling Combustion within a Realistic Can Combustor}

		% (Author, address, abstract etc. go here)
%		\author[labelb,labelc]{Aishvarya Kumar}
		\author[labela]{Aishvarya Kumar} %\emailauthor{aishvarya_pdf@sliet.ac.in}{AK}
		%\emailauthor{aishvarya.kumar@gmail.com}{AK}
		\author[labelb]{Ram Prakash Bharti\corref{cor1}} \emailauthor{rpbharti@iitr.ac.in}{RPB}
		\cortext[cor1]{Corresponding author}
%		\author[labela]{Kamlesh Kumari} %\emailauthor{kamlesh213@sliet.ac.in}{KK}
	
		\address[labela]{Department of Chemical Engineering, Sant Longowal Institute of Engineering and Technology (SLIET), Longowal 148016, Punjab, India}
%		\address[labelb]{Co-Founder, Skyguru, 3/48 Vikas Nagar, Lucknow 226022, Uttar Pradesh, India}		
%		\fntext[labelc]{Independent researcher; Former Doctoral Scholar, Department of Engineering, City St George's, University of London, Northampton Square, London EC1V 0HB, UK}

		\address[labelb]{Complex Fluid Dynamics and Microfluidics (CFDM) Lab, Department of Chemical Engineering, Indian Institute of Technology Roorkee, Roorkee 247667, Uttarakhand, India}
\vspace{-1em}		
		\begin{abstract}
			%\fontsize{11}{16pt}
			\normalsize\onehalfspacing
			This study has presented a comprehensive computational fluid dynamics (CFD) analysis of combustion flow in a realistic can combustor, evaluating the influence of various turbulence models on flow, thermal, and species fields. The non-premixed combustion modeling is performed using a presumed (beta) PDF approach in conjunction with a steady laminar flamelet model employing the San Diego reaction mechanism, and the turbulence is modeled using the RANS approach. The influence of turbulence models (standard $k-\epsilon$, realizable $k-\epsilon$, SST $k-\omega$, LPS-RSM) on the velocity field, such as the mean axial velocity, mean transverse velocity, turbulent kinetic energy (TKE) and shear stress, is analyzed, besides their influence on temperature and species (\ce{C3H8}, \ce{CO2}, and \ce{CO}) concentration. Analysis showed that despite the shortcomings of the isotropic turbulent viscosity formulation of the SST $k-\omega$ model being evident, it predicted the mean axial velocity, mean transverse velocity, turbulent kinetic energy and shear stress more accurately. Additionally, it predicted the flow features expected in a can combustor, such as the central recirculation zone (CRZ) and central vortex core (CVC), more accurately than other models.  Besides, the model predicted a higher temperature in the primary zone, which is supported by a lower prediction of \ce{C3H8}, and elevated TKE, both of which support strong mixing and efficient heat release. Furthermore, the SST $k-\omega$ model predicted the most compact stoichiometric mixture fraction bubble, encompassing CRZ and shear layers, indicating that the majority of the combustion occurs in the primary zone. The corresponding progress variables also indicated high values in the primary zone and shear layers, confirming near completion of the reaction, supported by negligible prediction of \ce{C3H8} and \ce{CO} at the outlet. 
\deleted{To further improve the predictions, the authors recommend employing a turbulence modeling approach which is less dissipative and more capable of predicting adverse pressure gradients, effects of flow curvature and anisotropy like LES.}
	\end{abstract}
	\begin{keyword}
		%\fontsize{11}{18pt}\selectfont
		\normalsize
		%----keywords here, in the form: keyword \sep keyword
		{Gas turbine combustor \sep Turbulence modeling \sep Reynolds-Averaged Navier-Stokes (RANS) \sep Reynolds Stress Model (RSM)\sep Turbulent Diffusion Flame \sep Confined Swirling reacting flows} % \sep Presumed PDF model  
		%----PACS codes here, in the form: \PACS code \sep code
	\end{keyword}
		
	\end{frontmatter}
\clearpage
%\linenumbers
%
\section{Introduction}
\label{sec:intro}
%===============================
The development of gas turbine engines can be traced back to the early 20th century with the pioneering work of Sir Frank Whittle, who patented the concept in 1930 \cite{Leyes1999}. His design employed a compressor to pressurize incoming air, which was mixed with fuel and ignited in a combustor to produce high-velocity gases that powered a turbine. Initially deployed in military aircraft during the second world war  \cite{Saravanamuttoo2017}, gas turbines were later adopted for civil aviation and have since expanded into power generation and marine propulsion. Continuous advancements in engineering, aerodynamics, and materials technology have significantly improved efficiency, power density, reliability, and emissions control, cementing gas turbines as vital components of modern transport and energy systems \cite{Saravanamuttoo2017,Mattingly2005}. Despite these advances, the fundamental operating principle remains unchanged. At the core of every gas turbine lies the combustor, which mixes and burns fuel with compressed air to generate the high-temperature high-pressure (HTHP) gases that drive the turbines. The combustor must achieve stable and efficient combustion while maintaining exhaust gas temperatures high enough to maximize turbine work extraction, yet not so high as to damage turbine blades. Moreover, it must minimize pollutant emissions, particularly \ce{NO_x}, \ce{CO}, and unburned hydrocarbons  (UHCs), while withstanding extreme operating conditions of temperature, pressure, velocity, and corrosive byproducts \cite{Lefebvre2010,Turns2012}.
\newline
The design of modern combustors is further complicated by combustion instabilities, including flame instability, lift-off, blow-off, flashback, and extinction. These phenomena can reduce performance, increase emissions, and even compromise structural integrity. Their occurrence is influenced by turbulence, fuel-air mixing, acoustic resonance, and swirler design \cite{Lieuwen2005,Huang2009}. Consequently, advanced combustor design relies heavily on detailed analysis of combustor aerodynamics and reacting flows. Aerodynamic optimization governs flow distribution, fuel-air mixing, and flame anchoring \cite{Poinsot2022}, while combustion analysis provides insight into kinetics, flame structure, and flame-holder interactions. Together, these approaches guide the development of low-emission, high-efficiency, and durable combustors.
\newline
Building upon this understanding, numerous experimental and numerical studies have been conducted to investigate combustor flow dynamics and combustion processes, aiming to improve flame stability, minimize emissions, and enhance overall gas turbine performance \cite{Turns2012,Lefebvre2010,Lieuwen2005}. Therefore, understanding both non-reacting and reacting flows, i.e., combustor aerodynamics and combustion flow analysis, is crucial for effective combustor design. Combustor aerodynamics is essential for optimizing the flow field, promoting uniform fuel-air mixing, and evaluating swirler geometry, all of which help minimize turbulence, improve flame anchoring, and prevent flame extinction. On the other hand, combustion flow analysis provides a deeper understanding of combustion kinetics, flame structure, and flame-holder interactions, thereby facilitating the development of stable, efficient, and low-emission combustors.
%
%===============================
\section{Literature Review}
\label{sec:lit-review}
%===============================
%
Early investigations into gas turbine combustion centered on simplified axisymmetric geometries to dissect fundamental processes of flow, flame stabilization, and emissions. \citet{friswell1972emissions} studied transparent water and combustion rigs, employing polystyrene tracers to map flow patterns and residence time distributions (RTDs) in flame tube primary zones. They identified distinct recirculation and vortex structures, noting that mean residence time strongly influenced \ce{NO_x} formation, while primary air-fuel ratio (AFR) exerted minimal effect on emissions for fixed geometries, attributed to localized constant AFRs or diffusion-dominated burning. \citet{katsuk1976emissions} corroborated these findings, confirming that elevated inlet temperatures accelerated combustion (reducing \ce{CO}) but minimally impacted \ce{NO_x}, underscoring complexities from local temperature and equivalence ratio variations.  \citet{katsuk1976emissions2} presented a two-dimensional axisymmetric analytical model of can-type gas turbine combustors simultaneously accounting finite rates of turbulent diffusion and chemical reactions. \citet{friswell1979influence} later revealed that smoke diminished with lower hydrogen content but became inert to fuel composition shifts above $\sim 10$ bar, an insight pivotal for high-altitude engine design. \citet{noyce1981measurements} emphasized primary port air as the dominant controller of exit pollutant levels, while \citet{green1983isothermal} demonstrated how jet arrangement reshaped vortex size and wall circulation via water using laser doppler anemometry (LDA) experiments. \citet{jones1989velocity} confirmed the recirculation zone as central to flame anchoring, with uniform species distributions prevailing at the exit. Collectively, these studies established geometry, fuel composition, and airflow distribution as cornerstone drivers of flow and emissions in simplified combustors.
\newline
Building on axisymmetric foundations, experiments transitioned to realistic geometries mirroring engine combustors, concentrating on Rolls-Royce Spey and Tay systems. \citet{jones1983temperature} probed the Spey combustor using thermocouples and gas probes, exposing that equilibrium prevailed solely in fuel-lean zones. Crucially, they revealed that \ce{CO} oxidation and fuel breakdown were kinetically controlled, with persistent unburned hydrocarbons (UHCs) dominating the primary zone, highlighting mixing and residence time limitations over reaction rates. \citet{bicen1986velocity} mapped the Spey’s flow field via LDA, disclosing solid-body swirl near the axis and free-vortex behavior near walls. Combustion contracted the primary vortex, amplifying turbulence and axial velocity; reduced AFR intensified velocity fluctuations by $\sim 75$\% at the exit. %\added{Added citation of Heitor PhD thesis.Please check, Sir.}\citet{heitor1985experiments, heitor1986velocity} 
\citet{heitor1986velocity} expanded these insights, noting combustion strengthened yet narrowed the vortex, with downstream momentum gradually suppressing swirl near the exit. Secondary air diluted reactants, quenching \ce{CO} oxidation, a phenomenon echoed in later studies. Mixture fraction fields in the primary zone remained driven by flow dynamics rather than combustion processes, reinforcing jet-induced mixing as paramount. Subsequent studies \citep{tse1988flow,bicen1990combustion} characterized the Rolls-Royce Tay combustor, distinguished by fewer/larger dilution holes (6 vs. 12 in Spey) and staggered ports. Their key revelations included $\sim 45$\% of primary air reversing upstream due to swirler-jet interactions and exceptional  ($\sim 98$\%) combustion efficiency, despite incomplete CO burnout downstream of dilution holes. Recent refinements amplify these insights: \citet{shah2016thermal} optimized swirler vane angles (\ang{45}) for Spey-like combustors, flattening temperature profiles and minimizing emissions. \citet{sadatakhavi2020experimental} confirmed stable stoichiometric combustion near the flame zone using liquid kerosene. \citet{liu2022numerical} linked mixing quality to \ce{NO_x}/\ce{CO} trade-offs in biogas-fueled combustors, noting vane angle enhancements but diminishing returns with blade count. 
\newline
Contemporary breakthroughs inject fresh vitality: \citet{zhang2023hybrid} pioneered a hybrid RANS/LES framework for can combustors, refining \ce{NO_x} predictions by resolving small-scale turbulence; \citet{patel2023ml} deployed machine learning (ML) to optimize swirler vane angles, slashing computational costs by 40$\%$ while sustaining predictive fidelity; \citet{chen2024advanced} harnessed ultrafast laser diagnostics, i.e., planar laser-induced fluorescence (PLIF) of hydroxyl radical (\ce{OH}),  PLIF-\ce{OH^*}, to map real-time flame-fluctuation interactions, uncovering novel stabilization mechanisms. These advances underscore the intricate behaviour of swirl, dilution design, AFR, and fuel in can combustor performance.
\newline
Numerical simulations evolved from rudimentary turbulence-chemistry models to sophisticated flamelet-based frameworks. \citet{sampath1987numerical} pioneered the eddy break-up (EBU) model with $k-\epsilon$ turbulence, achieving broad agreement for velocity and temperature profiles. \citet{Biswas_1997} extended this to Magnussen’s eddy dissipation concept (EDC), balancing fuel-oxygen-product dissipation imbalances and validating against benchmark data. \citet{Chakraborty_2000} employed Direct Numerical Simulation (DNS) to evaluate combustion models for high-speed hydrogen-air flows, offering nuanced insights into model performance. Later, flamelet-based approaches ascended: \citet{di2004large} applied a steady laminar flamelet model (SLFM) with presumed PDF to the Tay combustor via large eddy simulation (LES), capturing temperatures but underpredicting \ce{CO}/\ce{CO2}/\ce{O2}. \citet{meloni2013pollutant} replicated this for an industrial FRAME 6B combustor, achieving robust species agreement albeit with overestimated UHCs. Limitations of equilibrium models surfaced: \citet{krieger2015numerical} paired Reynolds stress model (RSM) turbulence with an equilibrium PDF for the Spey combustor, matching velocities but faltering in species predictions due to neglected finite-rate kinetics, strain effects, and extinction. Similar constraints plagued \citep{sharma2018new} in kerosene spray simulations. Cutting-edge innovations redefine simulation paradigms: \citet{li2023ai} integrated artificial neural network (ANN) with flamelet models, slashing CPU time by 65$\%$ while boosting \ce{NO_x} accuracy by 22$\%$; \citet{wang2024transported} developed a transported PDF method with adaptive mesh refinement, capturing flame-strain interactions in Tay combustors with unprecedented resolution; \citet{garcia2023review} presented a review on hybrid turbulence models (RANS / LES / DNS), spotlighting emergent frameworks like the scale-adaptive simulation (SAS); \citet{Benim_2016} synergized URANS / LES with flamelet modeling in OpenFOAM, successfully predicting syngas flashback tendencies.
Recent work \citep{kumar2024assessment} evaluated RANS-based turbulence models for simulating thermally confined swirling flow in a can-annular combustor configuration, comparing standard two-equation models ($k-\epsilon$, $k-\omega$, SST $k-\omega$) and seven-equation Reynolds stress models (linear pressure-strain: LPS-RSM) against experimental data at primary and dilution planes. Two-equation models generally struggled to predict confined swirling flow accurately, though the SST $k-\omega$ model performed best among standard options. Seven-equation model, LPS-RSM, showed partial promise but exhibited discrepancies due to explicit consideration of anisotropy (and no assumption of isotropic turbulence), for confined flows, inadequate capture of vortex intricacies and turbulence-kinetic-energy interactions, overestimation of velocity and shear stress due to linear pressure-strain assumptions, and insufficient treatment of diffusion or third-order convective term approximations in MUSCL schemes.  \deleted{The study concluded that while advanced RANS models improve upon standard formulations, higher-fidelity approaches like hybrid turbulence models (RANS-LES/ LES/ DNS) are essential for accurately resolving confined swirling flows in combustors.}
\section{Novelty and Significance}
\label{sec:novelty}
%===============================
%
Despite significant advancements in computational understanding of combustion, the efficacy of RANS-based turbulence models for predicting reacting flow dynamics in realistic gas turbine combustors remains insufficiently explored. Specifically, the ability of these models to accurately predict mean axial and transverse velocities, turbulent kinetic energy, and Reynolds shear stress in critical regions, such as the primary and dilution hole planes, has not been systematically evaluated for CAN-type geometries \added{at reacting conditions, to the best of authors' knowledge}. This lack of clarity impedes their reliable application during the combustor design phase, creating a significant uncertainty in model selection.
\newline
This study directly addresses these gaps by performing simulations of reacting flow within a realistic CAN-type combustor geometry (representing the Rolls-Royce Spey), experimentally characterized by \citep{heitor1985experiments, heitor1986velocity}. The combustion is modeled using a non-premixed probability density function (PDF) approach coupled with the Steady Laminar Flamelet Model (SLFM) and the San Diego Mechanism. This framework is chosen for its efficacy in predicting diffusion flames, which is consistent with literature identifying the Spey combustor as operating under diffusion-controlled combustion conditions \citep{jones1983temperature, heitor1985experiments, bicen1986velocity, heitor1986velocity}.
\newline 
The study systematically assesses the predictive capabilities of RANS-based turbulence models, including standard $k-\epsilon$, realizable $k-\epsilon$, SST $k-\omega$, and the Linear Pressure Strain Reynolds Stress Model (LPS-RSM) for the first time on the given geometry ``Rolls-Royce Spey'' against detailed turbulence statistics, including mean axial and transverse velocity, turbulent kinetic energy, and Reynolds stresses.  This study also presents a critical assessment of models’ performance, identifying their strengths and limitations \added{at reacting conditions} to accurately capture key features of confined swirling flows in a CAN-type combustor, like CRZ (Central Recirculation Zone), CVC (Central Vortex Core), accurate prediction of shear layers, and primary and dilution jet penetration.
%\newline 
Additionally, this study evaluates the influence of turbulence models on the turbulence driving mixing process, by analysing key predicted scalar quantities like temperature, species (including reactants (\ce{C3H8}) and products concentration (\ce{CO2} and \ce{CO})), turbulent thermal diffusivity, mean mixture fraction and progress variables, thereby linking turbulence modelling to combustion-relevant scalar predictions. These findings allow us to establish a direct relation between turbulence model selection and relevant scalar prediction to combustion. This integrated approach allows us to identify the limitations of conventional RANS models in predicting mixing-controlled combustion behavior.
\newline
Thus, the novelty of this study lies in the systematic assessment of RANS-based turbulence models in the reacting gas turbine combustor, thereby comprehensive correlation between turbulence model performance and detailed reacting-flow statistics in a realistic geometry, an analysis not previously attempted for the Rolls-Royce Spey combustor.  Furthermore, the present study bridges the critical gap between computational cost and accuracy. While high-fidelity methods like Large-Eddy Simulation (LES), Detached Eddy Simulation (DES), and Scale-Adaptive Simulation (SAS) are often impractical for industrial design phases due to their high computational cost and long turnaround times, a validated RANS approach offers a necessary practical compromise. Consequently, this work facilitates reliable combustor performance assessment within constrained development timelines.
The presented methodology and insights can be directly applied to the design and development of aeronautical and industrial gas turbines, including microturbines for UAVs and decentralised power systems particularly deployed in remote areas operating with alternative fuels like bio-LPG.
%
%--------------------------------------
\section{Flow Configuration}
\begin{figure}[!b]
	\centering
	\begin{subfigure}[b]{1\textwidth}
		\centering
		\includegraphics[width=\linewidth]{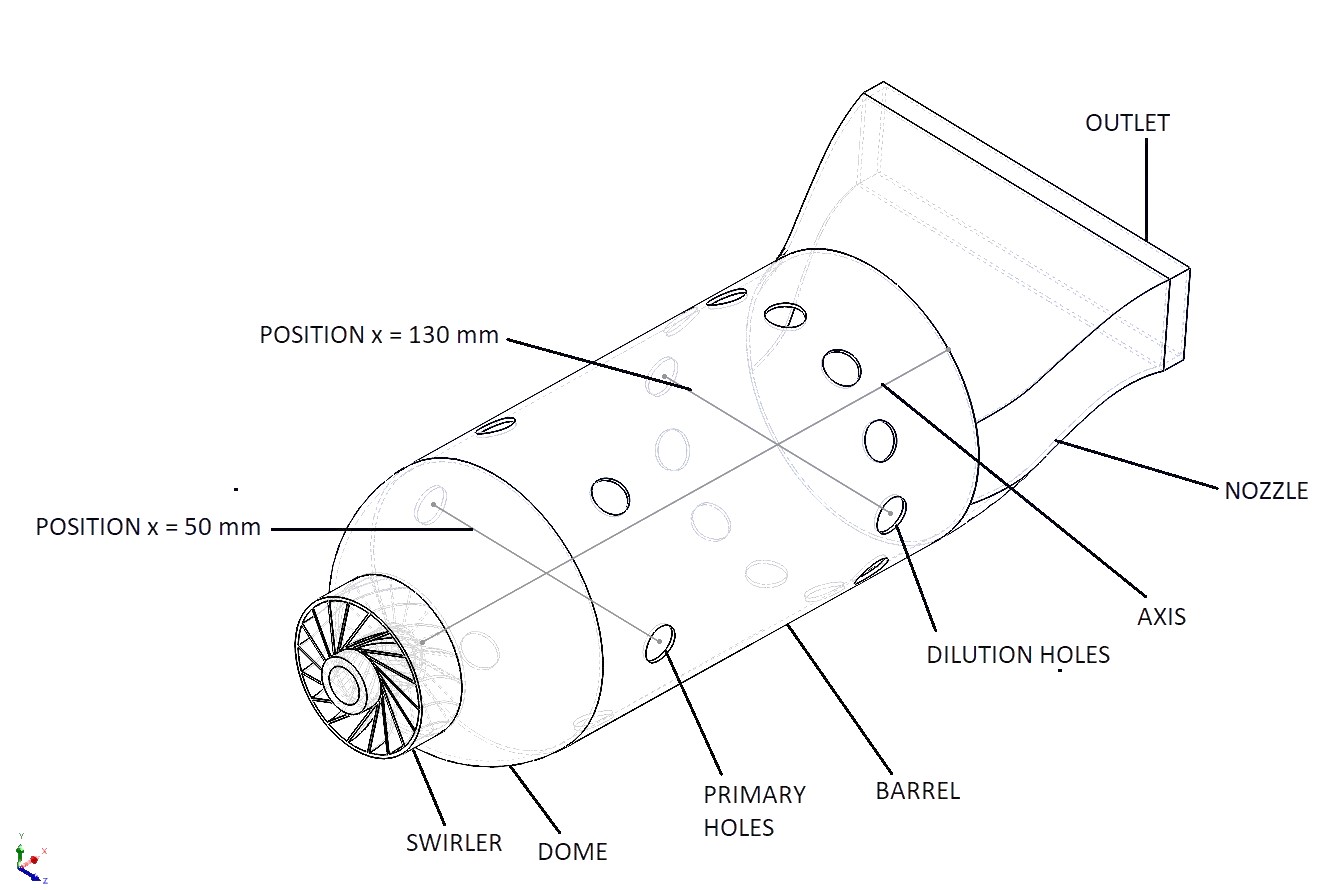}
		\caption{3D view of combustor geometry.}
		\label{fig:1}
	\end{subfigure}
	\hfill
	\begin{subfigure}[b]{0.48\textwidth}
		\centering
		\includegraphics[width=\linewidth]{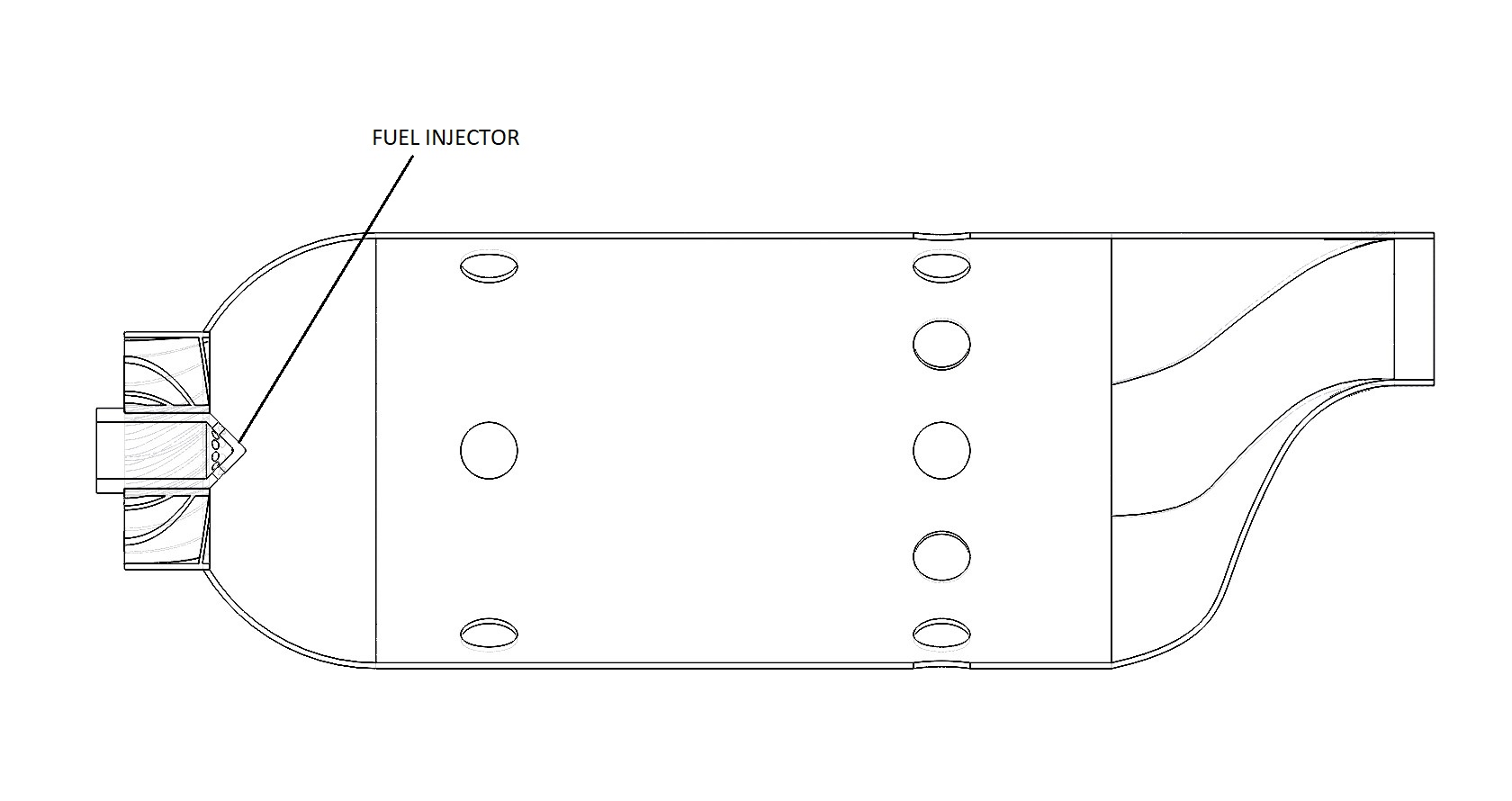}
		\caption{ Sectional side view of the combustor}
		\label{fig:2}
	\end{subfigure}
	\hfill
	\begin{subfigure}[b]{0.48\textwidth}
		\centering
		\includegraphics[width=\linewidth]{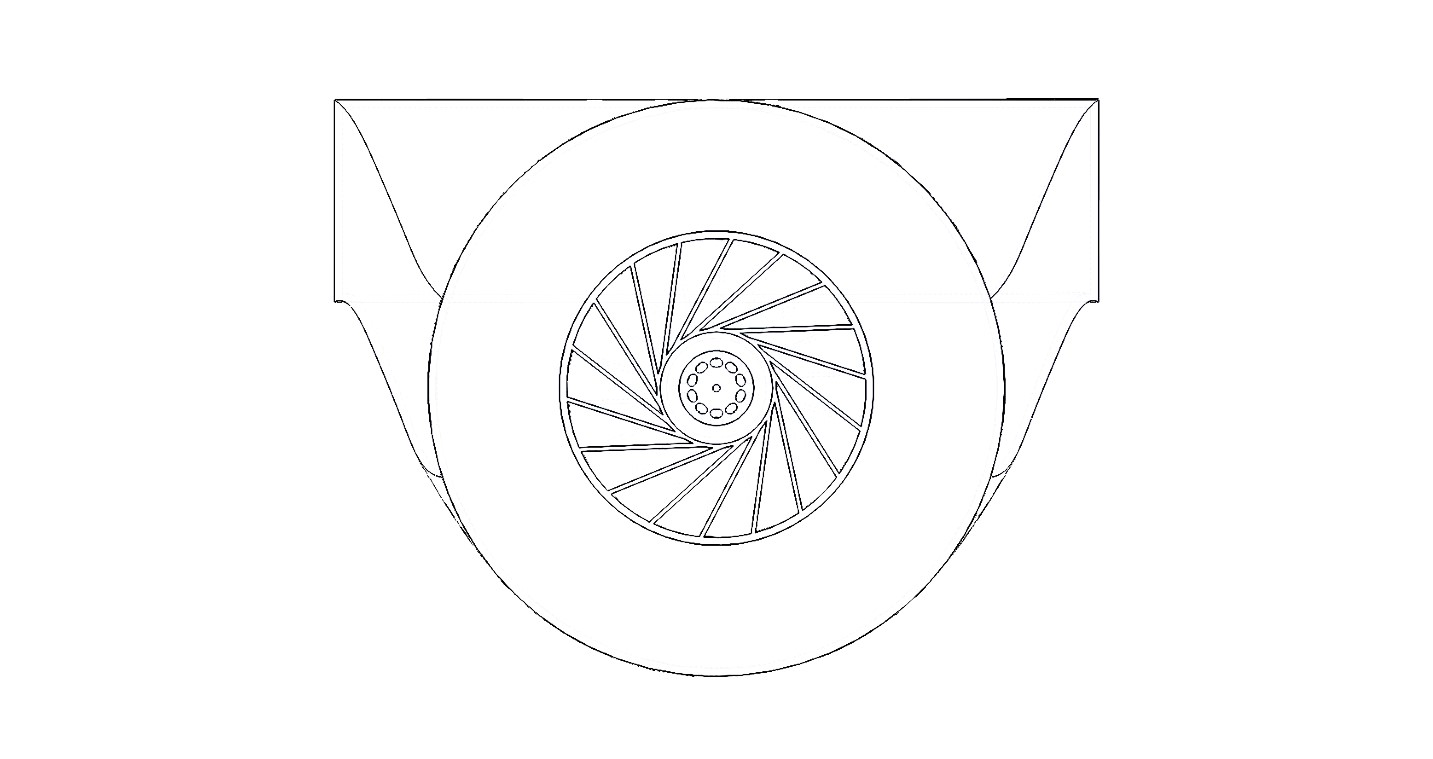}
		\caption{Front view of the combustor.}
		\label{fig:3}
	\end{subfigure}
	\caption{Schematic illustration of a combustor geometry with the primary and dilution hole plane positions for field measurement.}
	\label{fig:1s}
\end{figure}

%--------------------------------------
%
\fig\ref{fig:1s} illustrates the three-dimensional (3D) geometrical model flow configuration of a can-annular combustor chamber representing the Rolls-Royce Spey jet engine combustor (\fig\ref{fig:1}), with the sectional side view (\fig\ref{fig:2}) and the front view (\fig\ref{fig:3}) of the combustor, as used in the earlier experimental studies \citep{heitor1985experiments, heitor1986velocity}. Since the geometrical and experimental details for reference case have been thoroughly elaborated in literature \citep{heitor1985experiments, heitor1986velocity}, only the salient features are included here to avoid repetition.
The combustor, made from\deleted{ Transply}, a porous ceramic translucent material developed by Rolls Royce \added{PLC}, ideal for high-temperature applications, consisted of a hemispherical head integrated with a circular barrel  (diameter of combustor, $D_c=75$ mm). The total liner length (swirler excluded) of combustor is 210 mm. 
The barrel featured two sets of holes: six primary holes (PH, each of 10 mm diameter) followed by 12 dilution holes (DH, each of 20 mm diameter), separated by the distance of 80 mm. A circular-to-rectangular transition nozzle was attached to the downstream end of the barrel. 
\begin{table}[!t]
	\centering
	\renewcommand{\arraystretch}{1.5}
	\caption{Experimental conditions}
	\label{tab:1}
	\resizebox{0.8\textwidth}{!}{
		\begin{tabular}{|l|c|l|c|}
			\hline
			Total fuel inlet through injector nozzle (g/s) & 1.63 & Working Pressure ($P_0$, atm) & 1 \\ \hline
			Total air inlet to combustor (g/s) & 85 & Air inlet temperature (K) & 318 \\ \hline
			Air-fuel ratio (AFR) & \multicolumn{3}{c|}{52.1} \\ \hline
			\textit{Air inlet section / boundary} & \textit{Flow distribution} (\%) & \multicolumn{2}{c|}{\textit{Flow rate} (g/s)} \\ \hline
			Swirler & 24 & \multicolumn{2}{c|}{20.4} \\ \hline
			Primary holes (PH) & 16 & \multicolumn{2}{c|}{13.6 (= $6\times 2.2667$)} \\ \hline
			Dilution holes (DH) & 33 & \multicolumn{2}{c|}{28.05 (= $12\times 2.3375$)} \\ \hline
			Porous walls of hemispherical dome & 7 & \multicolumn{2}{c|}{5.95} \\ \hline
			Porous walls of cylindrical barrel & 14 & \multicolumn{2}{c|}{11.9} \\ \hline
			Porous walls of discharge nozzle & 6 & \multicolumn{2}{c|}{5.1} \\ \hline
		\end{tabular}
	}
\end{table}
\newline
The combustor head features a central fuel injector with a \ang{90} conical shape which injects the fuel into the chamber (see \fig\ref{fig:2}). The injector has ten uniformly distributed holes (each of 1.7 mm diameter), located circumferentially on a 4.50 mm radius. The fuel injector is surrounded by a curved vane swirler comprising 18 vanes oriented at swirl angle of \ang{45}. The vanes have a maximum thickness of 0.56 mm and produce a swirl number of $S = 0.73$. The combustor was enclosed within a plenum chamber, ensuring a uniform flow distribution through the inlet ports. 
\deleted{To eliminate uncertainties associated with liquid fuel vaporization, a gaseous fuel mixture (over 95\% propane) was employed in the experiments.} The Reynolds number, based on the combustor diameter ($D_c = 2R_c$) and upstream flow condition ($U_b\approx 97$ m/s), was $Re = 7.2 \times 10^4$. Furthermore, the air entered the combustion chamber through various inlet sections/ports (swirler, primary holes, dilution holes, and porous walls of hemispherical head, cylindrical barrel, and discharge nozzle made of Transply material), and the approximate flow distribution through these sections. The experimental conditions \citep{heitor1985experiments, heitor1986velocity} are listed in \tab\ref{tab:1}. 
\newline
The LDV (Laser Doppler Velocimetry) measurements were recorded, as shown in \fig\ref{fig:1}, at the positions of primary holes ($x = 50$ mm) and dilution holes ($x = 130$ mm). 
The flow split was estimated based on prior knowledge of the swirler's performance, hole discharge coefficients, and Transply material porosity, and thus, the flow distribution calculations were approximate\added{d}, with individual flow rates expected to be within $\pm 10$\% of actual values  \citep{heitor1985experiments, heitor1986velocity}.
However, the precise airflow distribution through the porous walls remains uncertain and may necessitate additional empirical research for accurate determination. Therefore, the accuracy of the approximated distribution of the total air flow passing through porous walls (refer \tab\ref{tab:1}) should be treated with caution, as the characteristics of the porous material were not sufficiently clear in the literature  \citep{jones1983temperature, heitor1985experiments, bicen1986velocity, heitor1986velocity, tse1988flow, bicen1990combustion}. 
\begin{table}[t!]
	\caption{Boundary conditions used in this study.}\label{tab:2}	
	\centering
	\renewcommand{\arraystretch}{1.5}
	\resizebox{0.8\textwidth}{!}{
		\begin{tabular}{|l|c|c|c|c|c|}
			\hline
			{Boundary} & {Type} & {Condition} & {Value (kg/s)} &No. of Boundaries & {Total Value (kg/s)}\\ \hline
			Swirler & Inlet & Mass Flow Rate & 20.4  & 1 & 20.4\\ \hline
			Injector Holes & Inlet & Mass Flow Rate &  0.163  & 10&1.63\\ \hline
			Primary Holes & Inlet & Mass Flow Rate &  13.6/6  & 6 &13.6 \\ \hline
			Dilution Holes & Inlet & Mass Flow Rate & 28.05/12  & 12 & 28.05\\ \hline
			Dome & Wall & Adiabatic & --  & --&--\\ \hline
			Barrel & Wall & Adiabatic & --  & --&--\\ \hline
			Nozzle & Wall & Adiabatic & --  & --&--\\ \hline
			Outlet & Outlet & Outflow & --  & --&--\\ \hline
	\end{tabular}}
\end{table}	
%
%\deleted[id=AK]{The fuel flow was monitored using a rotameter, while the air flow was measured using a calibrated orifice meter; both measurements had an uncertainty of approximately $\pm 2$\%. Temperature was measured using 80 $\mu$m bare-wire platinum-13\% rhodium and platinum thermocouples whose output was amplified (250X), digitized, and then processed by a microcomputer.  Concentration measurements of various species were obtained via water-cooled probe sampling,  analyzed using flame ionization, gas chromatography, infrared, and paramagnetic analyzers. Furthermore, the experimental measurements were recorded, as illustrated in \fig\ref{fig:1}, at the locations of the primary holes ($x=20$ mm) and the dilution holes ($x=80$ mm).}
%
\newline
In this study, the boundary conditions used to replicate the experimental setup (\tab\ref{tab:1}) are summarized in \tab\ref{tab:2}. The combustor liner surfaces, including the dome, barrel, and nozzle, are treated as solid walls, consistent with literature \citep{krieger2015numerical,wang2021numerical}. The current assumption is necessitated by the absence of detailed information regarding the properties of the porous media used in constructing the combustor liner, as noted in the original references \citep{heitor1985experiments, heitor1986velocity}. This treatment, however, differs from that adopted in other studies \citep{di2004large,kumar2024assessment}, where alternative boundary assumptions were applied.%
%--------------------------------------
\section{Mathematical Modeling}
%--------------------------------------
%
The mathematical model for the physical problem described above is briefly presented here; the detailed formulation is provided in the Supplementary Information (Appendix \ref{appendixA}).
To simulate turbulent combustion with significant density fluctuations, the Favre-averaging approach \citep{Liou1991} is employed. This density-weighted formulation simplifies non-linear terms in the governing equations, enhancing computational efficiency and numerical stability.

\subsection{Favre-averaged Governing Equations}
In Favre-averaging, instantaneous quantities ($\psi$) are decomposed into a mean ($\tilde{\psi}$) and a fluctuating component ($\psi^{\prime\prime}$), defined as $\tilde{\psi} = \overline{\rho\psi}/\overline{\rho}$. For the mixture, the density is governed by the ideal gas law:
\begin{equation}
	\overbar{\rho} \approx \frac{\bar{p}}{R\,\tilde{T}\,\displaystyle\sum_{k}({\tilde{Y}_k}/{M_k})},
\end{equation}
where $\bar{p}$ is the mean pressure, $\tilde{T}$ is the Favre-averaged temperature, and $\tilde{Y}_k$ is the mass fraction of species $k$.
\newline
The conservation laws for mass, momentum, and thermal energy are expressed as:
\begin{subequations}
	\begin{align}
		&\text{Continuity:} \quad \frac{\partial \overline{\rho}}{\partial t} + \frac{\partial (\overline{\rho} \tilde{u}_j)}{\partial x_j} = 0, \\[5pt]
		&\text{Momentum:} \quad \frac{\partial (\overline{\rho}\,\tilde{u}_i)}{\partial t} + \frac{\partial (\overline{\rho}\,\tilde{u}_i \tilde{u}_j)}{\partial x_j} = -\frac{\partial \overline{p}}{\partial x_i} + \frac{\partial (\overline{\tau}_{ij} - \tilde{R}_{ij})}{\partial x_j} + \overline{\rho} \overline{f_i}, \\[5pt]
		&\text{Energy:} \quad \frac{\partial (\overline{\rho} \tilde{H})}{\partial t} + \nabla \cdot (\overline{\rho} \tilde{\mathbf{u}} \tilde{H}) = \nabla \cdot \left(\frac{\overline{k}_t}{\overline{c}_p} \nabla \tilde{H}\right) + \overline{S}_h.
	\end{align}
\end{subequations}
Here, $\tilde{R}_{ij} = \overline{\rho u_i^{\prime\prime} u_j^{\prime\prime}}$ is the Reynolds stress tensor, and $\overline{\tau}_{ij}$ is the viscous stress tensor. The energy equation is solved for enthalpy $\tilde{H}$ assuming a unity Lewis number, where $\overline{k}_t$ is the turbulent thermal conductivity and $\overline{c}_p$ is the mean specific heat.
\subsection{Mixture Fraction and Turbulence-Chemistry Interaction}
The non-premixed combustion is modeled using the mixture fraction approach. Based on the Simple Chemical Reacting System (SCRS) assumption, a single conserved scalar—the mixture fraction $f$—is introduced. It represents the local fuel-to-air ratio and obeys a transport equation without source terms:
\begin{equation}
	\frac{\partial (\overline{\rho} \tilde{f})}{\partial t} + \frac{\partial (\overline{\rho} \tilde{u_j} \tilde{f} )}{\partial x_j} = \frac{\partial }{\partial x_j} \left( \tilde{\Gamma}_{f} \frac{\partial \tilde{f}}{\partial x_j}\right),
\end{equation}
where $\tilde{\Gamma}_{f} = \mu/\sigma + \mu_t/\sigma_t$.
\newline 
To account for turbulence-chemistry interactions, a presumed $\beta$-probability density function (PDF) is employed. The mean values of scalars (species, temperature, density) are obtained by integrating the instantaneous values over the PDF:
\begin{equation}
	\tilde{\phi}_i = \int_{0}^{1} \phi_i (f, \tilde{H}) p(f) \, df.
\end{equation}
The shape of the $\beta$-PDF is determined by the mean mixture fraction $\tilde{f}$ and its variance $\sigma_f$, where $\sigma_f$ is solved via its own transport equation.
\subsection{Flamelet Model}
The turbulent flame is treated as an ensemble of laminar flamelets (Steady Laminar Flamelet Model, SLFM) embedded in the turbulent flow. This approach is valid for high Damkohler numbers ($Da \gg 1$), where chemical time scales are much shorter than turbulent mixing time scales.
\newline 
Flamelet libraries are generated by solving counterflow diffusion flame equations in mixture fraction space. To capture non-equilibrium effects due to aerodynamic strain, the scalar dissipation rate $\chi = 2\mathcal{D}|\nabla f|^2$ is utilized. The stoichiometric scalar dissipation rate $\chi_{st}$ serves as a parameter characterizing the flame strain.
\newline 
For non-adiabatic systems, enthalpy $\tilde{H}$ is included as an additional parameter. The mean scalars are retrieved from look-up tables generated by convolving the flamelet solutions with the presumed $\beta$-PDF. The mean scalar dissipation rate for RANS simulations is modeled as:
\begin{equation}
	\tilde{\chi}_{\rm st} = \frac{C_{\chi} \epsilon \sigma_{f}}{\tilde{k}},
\end{equation}
with $C_{\chi}=2$.
\subsection{Turbulence Modeling}
To close the RANS equations, several turbulence models are evaluated. The instantaneous velocity is decomposed using Reynolds decomposition, and Reynolds stresses are modeled using the Boussinesq hypothesis (for eddy viscosity models) or solved directly (for RSM).
\subsubsection{Standard $k-\epsilon$ Model}
The transport equations for turbulent kinetic energy $k$ and dissipation rate $\epsilon$ are:
\begin{align}
	\frac{\partial (\overline{\rho} k)}{\partial t} + \frac{\partial (\overline{\rho} \tilde{u}_j k)}{\partial x_j} &= \frac{\partial}{\partial x_j} \left[ \left( \mu + \frac{\mu_t}{\sigma_k} \right) \frac{\partial k}{\partial x_j} \right] + P_k - \tilde{\rho} \epsilon, \\
	\frac{\partial (\overline{\rho} \epsilon)}{\partial t} + \frac{\partial (\overline{\rho} \tilde{u}_j \epsilon)}{\partial x_j} &= \frac{\partial}{\partial x_j} \left[ \left( \mu + \frac{\mu_t}{\sigma_\epsilon} \right) \frac{\partial \epsilon}{\partial x_j} \right] + C_{\epsilon 1} \frac{\epsilon}{k} P_k - C_{\epsilon 2} \frac{\epsilon^2}{k}\overline{\rho},
\end{align}
where $\mu_t = \overline{\rho} C_\mu k^2/\epsilon$ and $P_k = \mu_t (\partial \tilde{u}_i/\partial x_j + \partial \tilde{u}_j/\partial x_i) \partial \tilde{u}_i/\partial x_j$.
\subsubsection{Realizable $k-\epsilon$ Model}
This model ensures the realizability of normal stresses. The $\epsilon$ equation is derived from the mean-square vorticity fluctuation, and $C_\mu$ is variable depending on the mean strain and rotation rates:
\begin{equation}
	C_\mu = \frac{1}{A_0 + A_s (k U^{*}/\epsilon)}, \quad U^{*} = \sqrt{\tilde{S}_{ij}\tilde{S}_{ij} + \tilde{\Omega}_{ij}\tilde{\Omega}_{ij}}.
\end{equation}
\subsubsection{SST $k-\omega$ Model}
The Shear Stress Transport (SST) model blends the $k-\omega$ model near walls with the $k-\epsilon$ model in free streams via blending functions $F_1$ and $F_2$. The turbulent viscosity is limited to account for the transport of the principal shear stress:
\begin{equation}
	\mu_t = \frac{\overline{\rho} k}{\omega} \left( \max \left[ \frac{1}{\alpha^*}, \frac{F_2 \tilde{S}_{ij}}{\alpha_1 \omega} \right] \right)^{-1}.
\end{equation}
\subsubsection{Reynolds Stress Model (RSM)}
The RSM solves transport equations for each component of the Reynolds stress tensor $\tilde{R}_{ij}$ and the dissipation rate $\epsilon$:
\begin{equation}
	\frac{\partial}{\partial t}(\overline{\rho}\tilde{R}_{ij}) + C_{ij} = D_{T,ij} + D_{L,ij} - P_{ij} + \phi_{ij} - \epsilon_{ij},
\end{equation}
where $C_{ij}$ is convection, $D_{T,ij}$ and $D_{L,ij}$ are turbulent and molecular diffusion, $P_{ij}$ is stress production, $\phi_{ij}$ is the pressure-strain correlation (modeled using the Linear Pressure Strain model), and $\epsilon_{ij}$ is dissipation. This approach avoids the isotropic eddy viscosity assumption, providing better accuracy for complex flows.
%
%--------------------------------------
\section{Numerical Approach}
%--------------------------------------
%
In this study, the model governing equations are solved using the ANSYS Fluent (2025R2, student version) solver, which \deleted{supports up to one million (i.e., $10^6$) control volumes (CVs) and} is based on the finite volume method (FVM) \citep{Versteeg2007}. Owing to the quasi-steady nature of the flow \citep{heitor1985experiments, heitor1986velocity}, steady-state simulations have been performed. The pressure-velocity coupling is handled using the COUPLED algorithm \citep{Ghobadian2007}, which provides enhanced stability for strongly interacting flow fields. The pressure field is interpolated using the PRESTO ({PRE}ssure {ST}aggering {O}ption) scheme \citep{Patankar1980}, which is particularly suited for swirling flow \citep{Cellek2018}, while convective terms {for energy, mean mixture fraction and mean mixture variance} are discretized using the second-order upwind (SOU) scheme \citep{Patankar1980}, The turbulent kinetic energy ($k$), turbulent dissipation rate ($\epsilon$), specific dissipation rate ($\omega$) are discretized using first-order upwind (FOU) scheme for stability.
\newline
Combustion is modelled using a non-premixed combustion approach coupled with a presumed beta-PDF formulation in conjunction with the \textit{steady laminar flamelet model} (SLFM). The chemical kinetics are governed by the San Diego Mechanism \citep{williams2010sandiego}, which comprises six elements, 58 species, and 360 reactions. Turbulence is modelled using conventional RANS two-equation models (standard $k$--$\epsilon$, SST $k$--$\omega$) and RSM seven-equation model (LPS-RSM), while near-wall effects are treated using the \textit{enhanced wall treatment} (EWT) method, which combines linear (viscous sub-layers) and logarithmic (turbulent) wall laws through a blending function \citep{Kader1981} to improve accuracy across the near-wall region.
\begin{figure}[!b]
	\centering
	\begin{subfigure}[b]{0.8\linewidth}
		\centering
		\includegraphics[width=\linewidth]{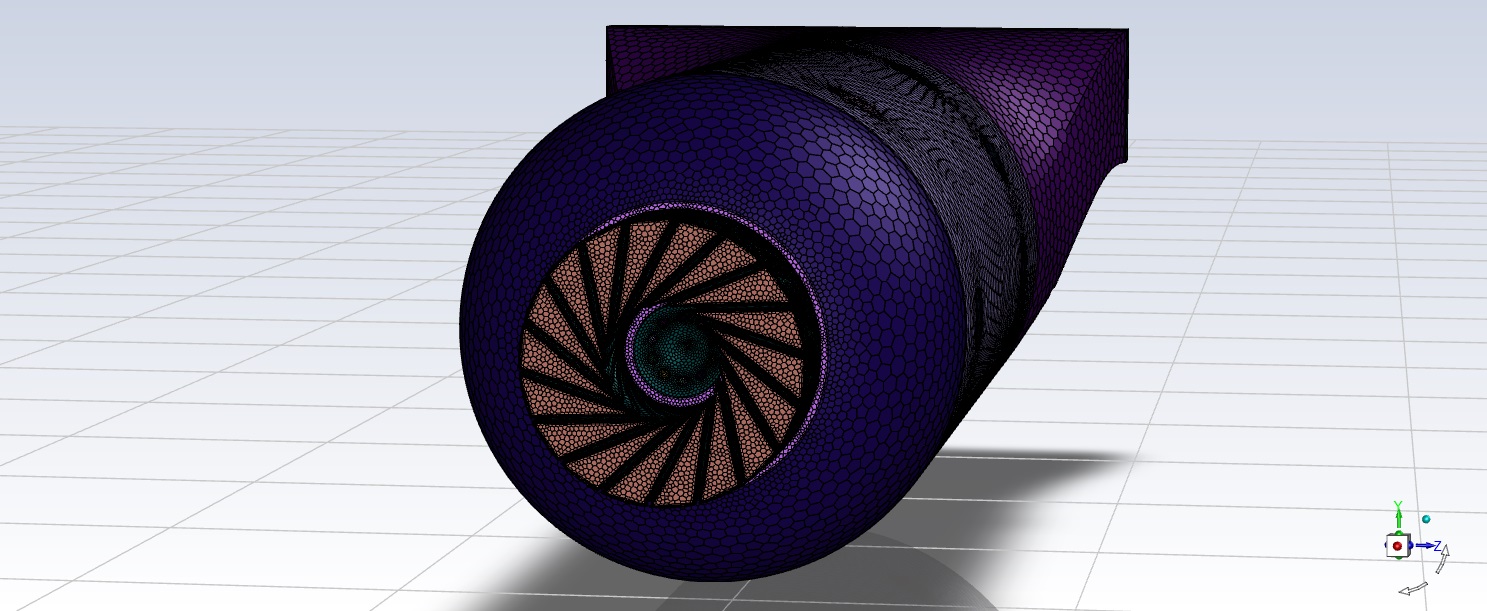}
		\caption{Front view}\label{fig:2a}
	\end{subfigure}
	
	\begin{subfigure}[b]{0.8\linewidth}
		\centering
		\includegraphics[width=\linewidth]{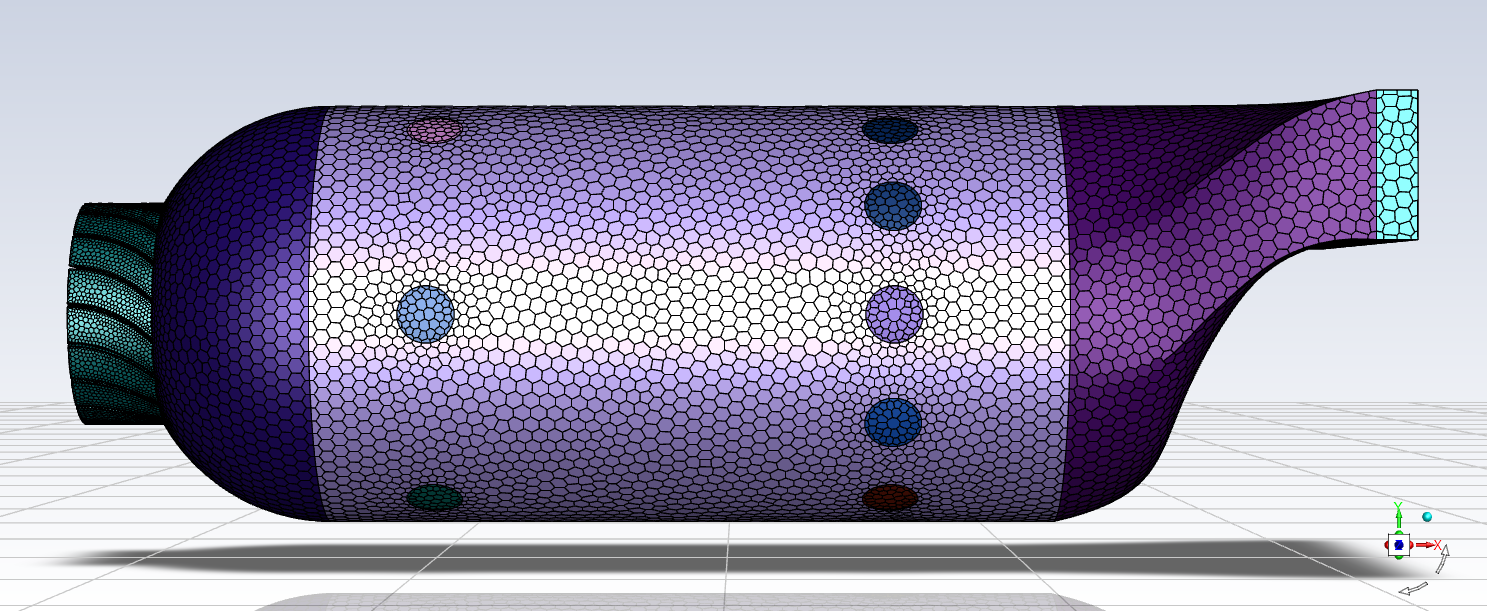}
		\caption{Side view}\label{fig:2b}
	\end{subfigure}
	
	\begin{subfigure}[b]{0.8\linewidth}
		\centering
		\includegraphics[width=\linewidth]{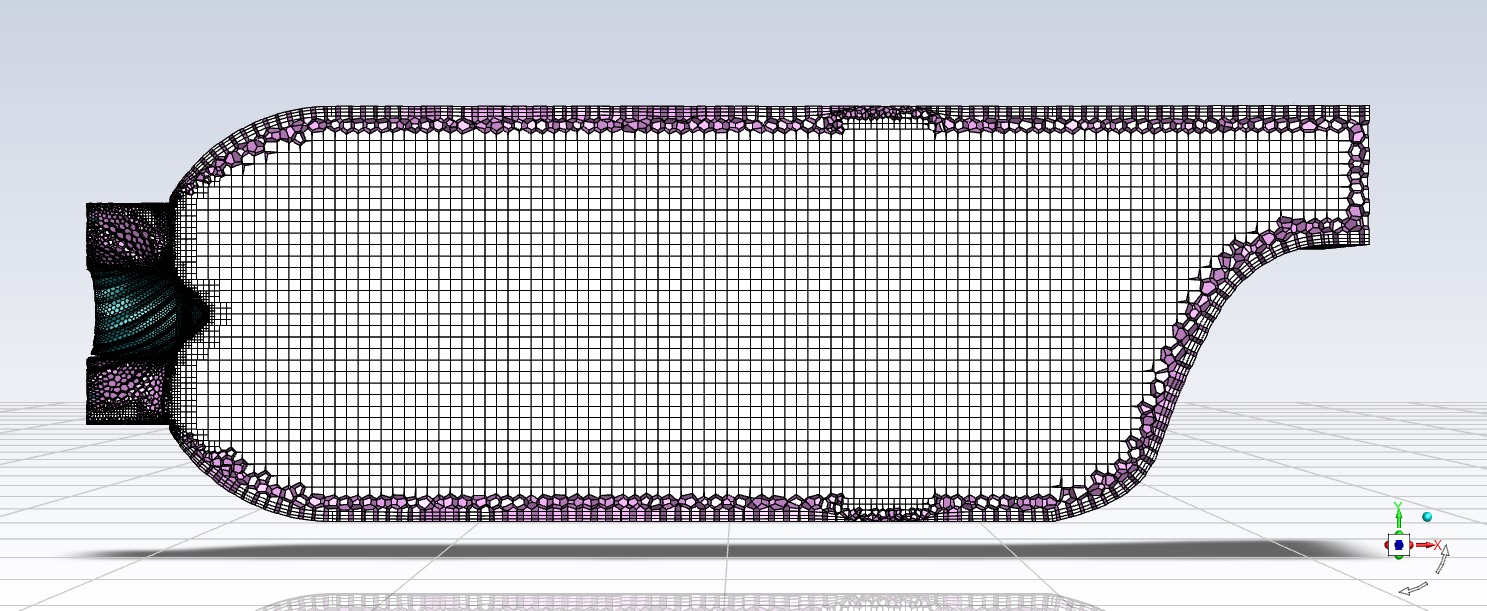}
		\caption{Sectional view}\label{fig:2c}
	\end{subfigure}
	\caption{Schematics of the computational mesh.}
	\label{fig:4}
\end{figure}
\newline
The computational domain is discretized using a \added{hybrid} unstructured, non-uniform \deleted{hybrid} poly-hexacore mesh, consisting of a \added{structured} hexahedral core surrounded by a polyhedral outer shell, as schematically illustrated in \fig\ref{fig:4}. Additionally, three layers of prism-type cells are generated near the walls to accurately resolve the boundary layer, as shown in \fig\ref{fig:2c}.
\begin{table}[!b]
	\centering
	\caption{Mesh characteristics ($N_e$: number of cells or control volumes, $h$: maximum cell length, $r$: grid refinement ratio, $l$: average cell length).}
	\label{tab:3}
	\begin{tabular}{|c|c|c|c|c|c|}
		\hline
		Mesh & $N_e$ & $h$ (mm) & $r$ (--) & $l$ (mm) & ${D_c}/l$ (--) \\
		\hline
		G1 & 347303 & 2.6 & -- & 1.363 & 55.01 \\
		G2 & 491403 & 2.0 & 1.3 & 1.214 & 61.75 \\
		G3 & 849619 & 1.4 & 1.428 & 1.011 & 74.12 \\
		\hline
	\end{tabular}
\end{table}
%
%\newline
Grid convergence analysis is performed to assess the influence of grid density on predictions of mean axial velocity, transverse velocity, turbulent kinetic energy and shear stress on the planes of primary holes (position $x=50$ mm) and dilution holes (position $x=130$ mm), and along the combustor axis (refer to \fig\ref{fig:1}). 
The mesh characteristics for the considered grids are specified in \tab\ref{tab:3}. The average cell size ($l$) is calculated as the cubic root of the ratio between the total fluid domain volume ($V_f=880069.13\text{ mm}^3$) and the total number of cells ($N_e$), i.e., $l=\sqrt[3]{V_f/N_e}$. The grid refinement ratio ($r = h_{\text{coarse}}/h_{\text{fine}}$) is maintained \citep{Celik2008} at $r \geq 1.3$. 
The computations are performed using a desktop computer equipped with an AMD Ryzen 3 processor (3.5 GHz, 4 cores) with an integrated Vega graphics card. Parallel processing was employed across all CPU cores to expedite computation. The total simulation time required on the dense grid (G3) is approximately 5 hours for two-equation turbulence models, and about 24 hours for the more complex seven-equation (LPS-RSM) model.
\section{Results and discussions}
In this section, present results are analyzed and discussed. Prior to the presentation of new results and the analysis of the accuracy of turbulence models at the reacting stage of the realistic can combustor, a grid independence study is presented to ensure the reliability of the present numerical results. Subsequently, the predictive ability of the turbulence models is analysed by comparing the predicted results with the experimental data \citep{heitor1985experiments, heitor1986velocity}. The velocity is normalized ($\mathbf{u}^\ast=\mathbf{u}/U_b$) using the bulk velocity ($U_b = 97$ \si{\meter\per\second}), the radial position is normalized ($r^\ast = r/R_c$) using the internal radius ($R_c = 37.5$ mm) , and the axial position is normalized ($x^* = x/L$) using the total axial length ($L = 182.80$ mm).
\subsection{Grid independence study}
\begin{figure}[!b]
	\centering
	\begin{subfigure}[t]{0.48\linewidth}
		\centering
		\includegraphics[width=\linewidth]{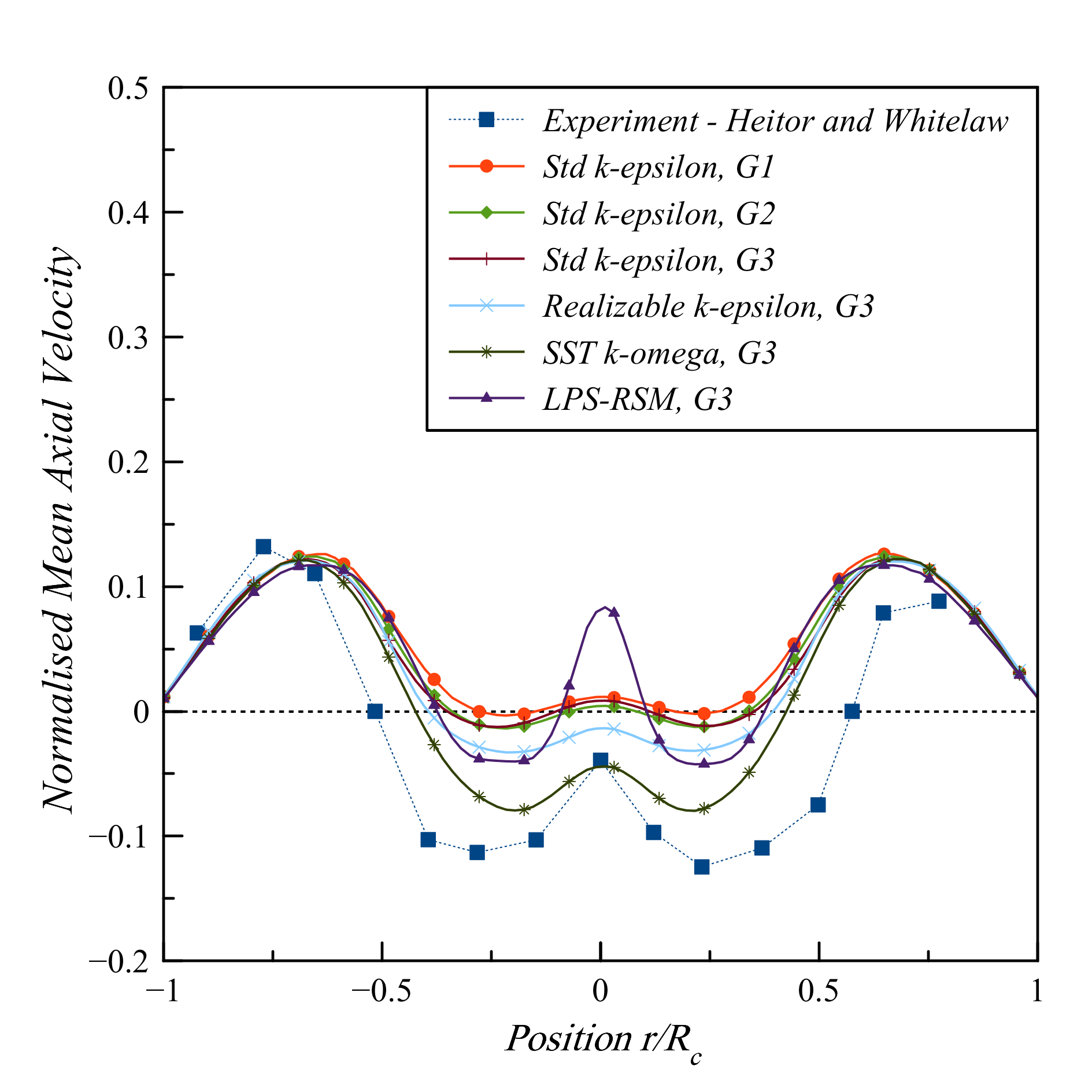}
		\caption{Normalized mean axial velocity ($\tilde{u}^\ast = {\tilde{u}}/{{U}_b}$)}
		\label{fig:5a}
	\end{subfigure}
	\hfill
	\begin{subfigure}[t]{0.48\linewidth}
		\centering
		\includegraphics[width=\linewidth]{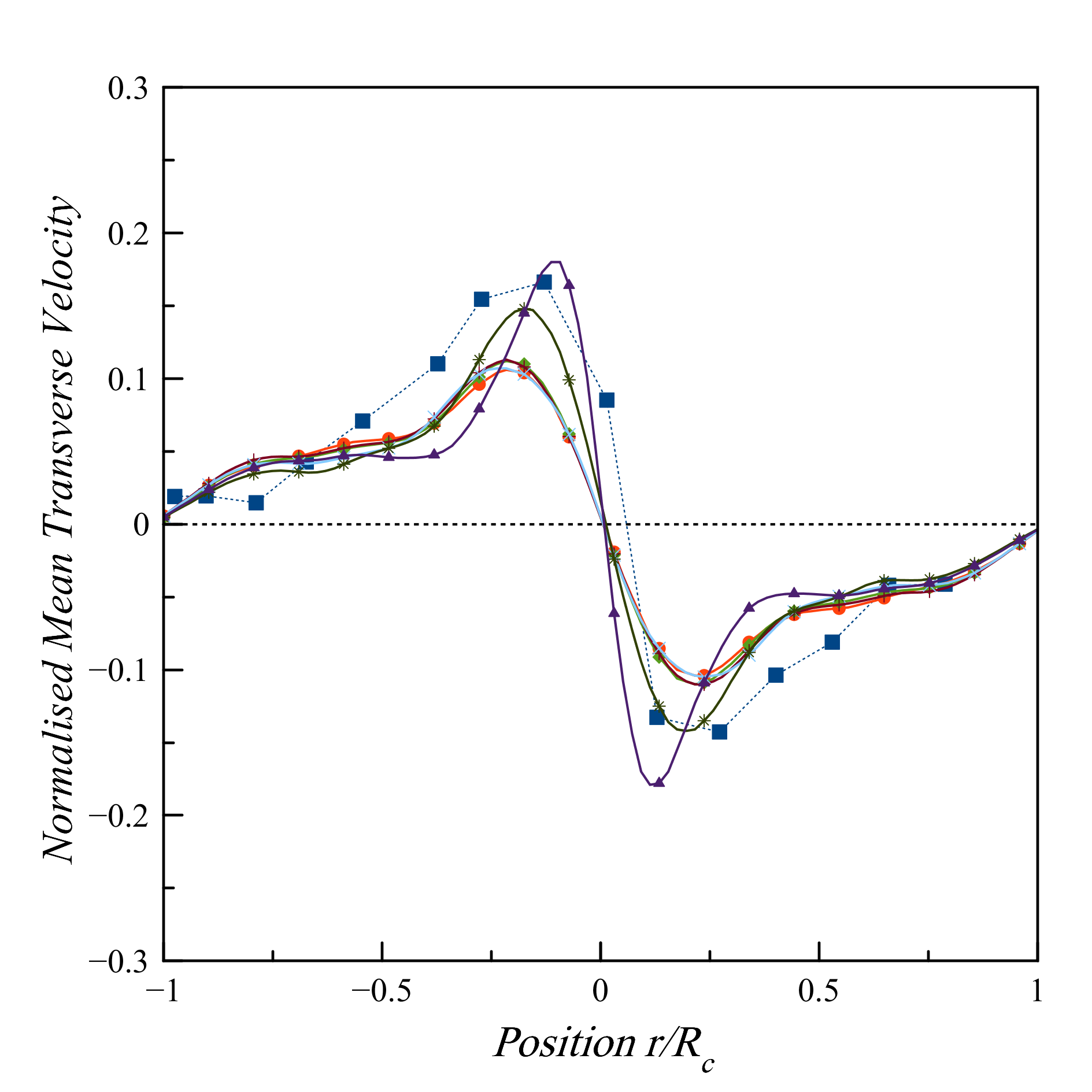}
		\caption{Normalized mean transverse velocity ($\tilde{v}^\ast = {\tilde{v}}/{{U}_b}$)}
		\label{fig:5b}
	\end{subfigure}
	\vspace{3mm}
	\begin{subfigure}[t]{0.48\linewidth}
		\centering
		\includegraphics[width=\linewidth]{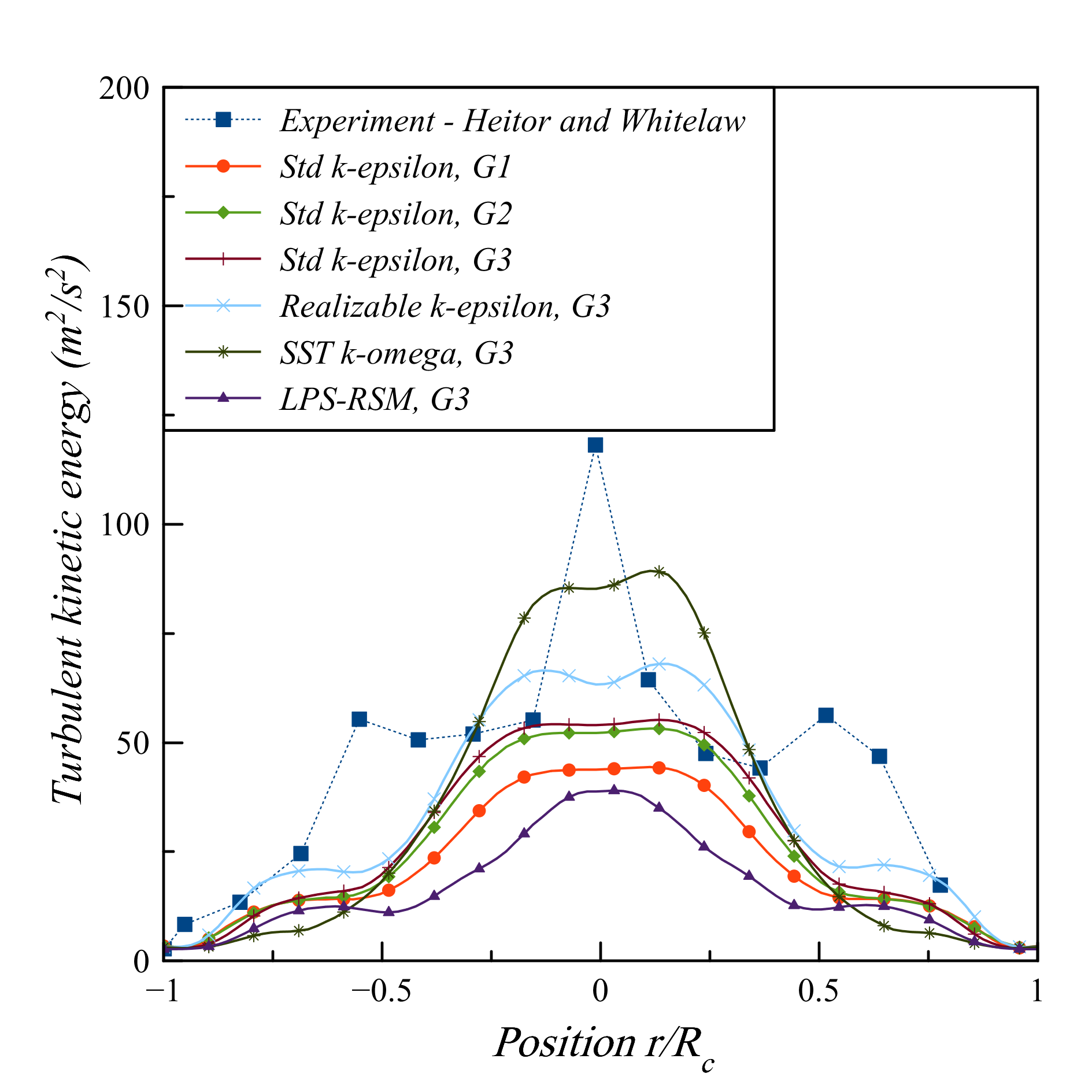}
		\caption{Turbulent kinetic energy, ${k} = \frac{3}{4} \left( \widetilde{{u^{\prime\prime}}^2} + \widetilde{{v^{\prime\prime}}^2} \right)$ \si{\meter\squared\per\second\squared}}
		\label{fig:5c}
	\end{subfigure}
	\hfill
	\begin{subfigure}[t]{0.48\linewidth}
		\centering
		\includegraphics[width=\linewidth]{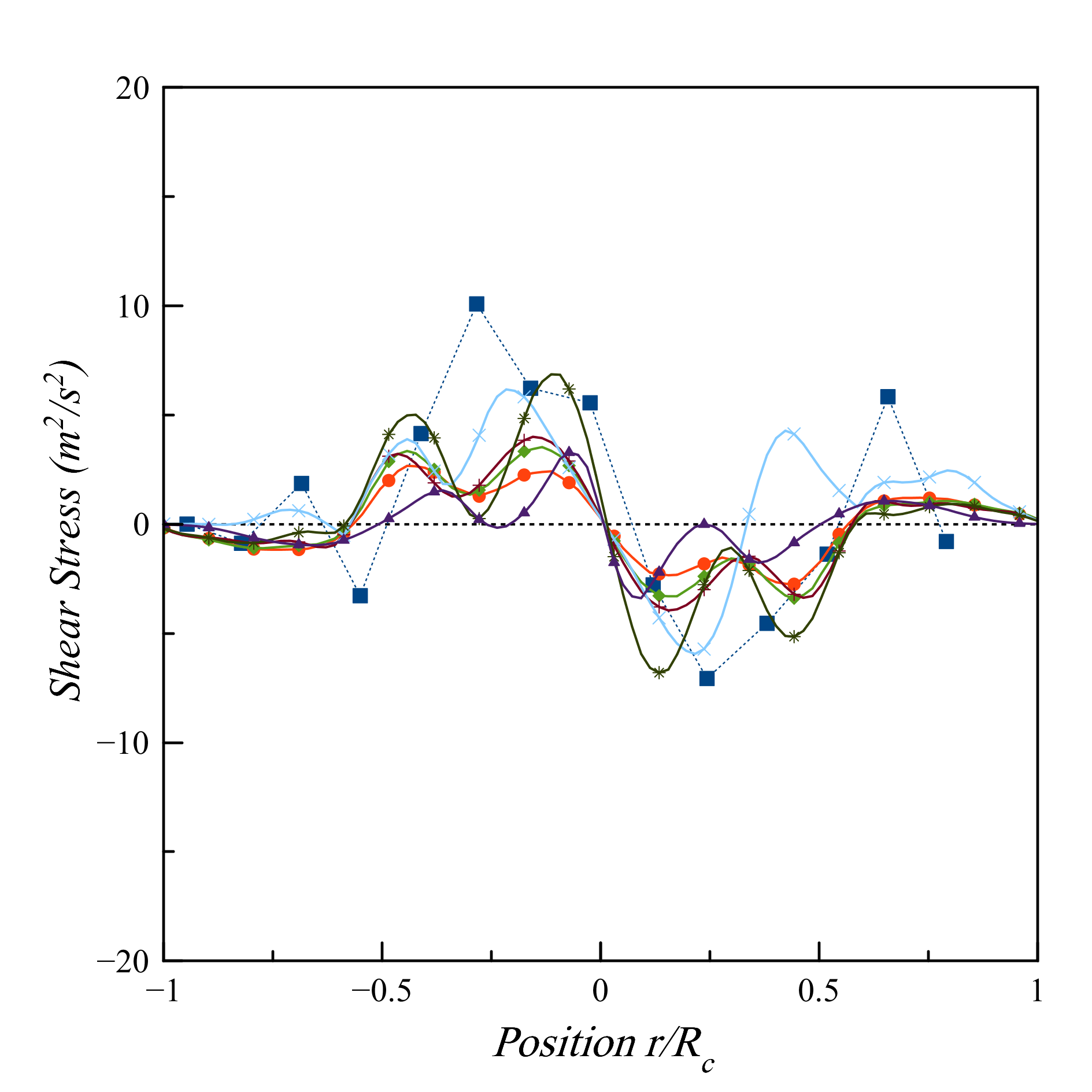}
		\caption{Shear stress, $-\widetilde{u^{\prime\prime}v^{\prime\prime}}$, \si{\meter\squared\per\second\squared}}
		\label{fig:5d}
	\end{subfigure}
	\caption{Velocity and turbulence characteristics at axial position $x = 50$ mm on the primary holes plane at reacting conditions (refer \tab\ref{tab:1}). The experimental data are from \cite{heitor1985experiments,heitor1986velocity} }
	\label{fig:5}
\end{figure}

\begin{figure}[!tb]
	\centering
	\begin{subfigure}[t]{0.48\linewidth}
		\centering
		\includegraphics[width=\linewidth]{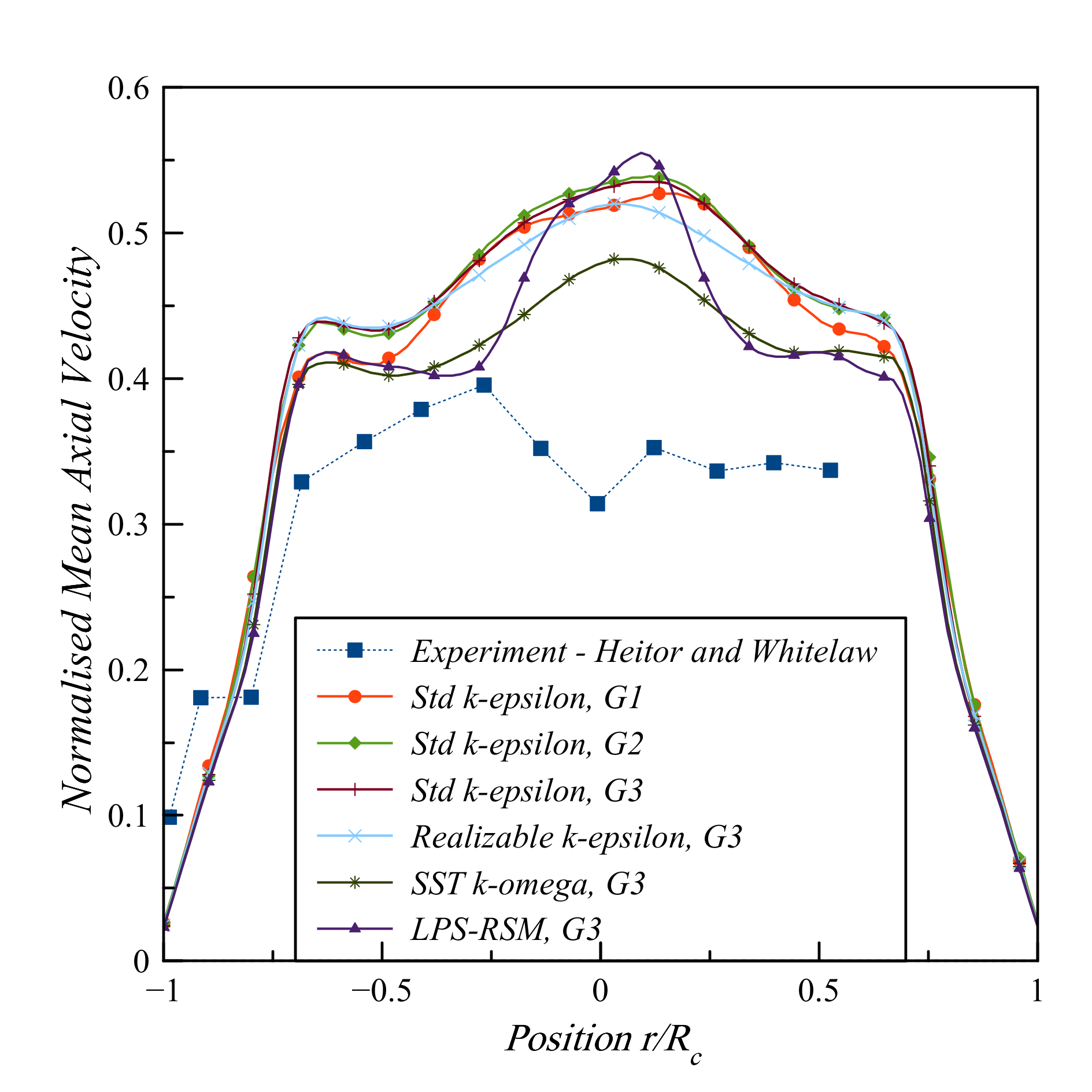}
		\caption{Normalized mean axial velocity ($\tilde{u}^\ast = {\tilde{u}}/{{U}_b}$)}
		\label{fig:6a}
	\end{subfigure}
	\hfill
	\begin{subfigure}[t]{0.48\linewidth}
		\centering
		\includegraphics[width=\linewidth]{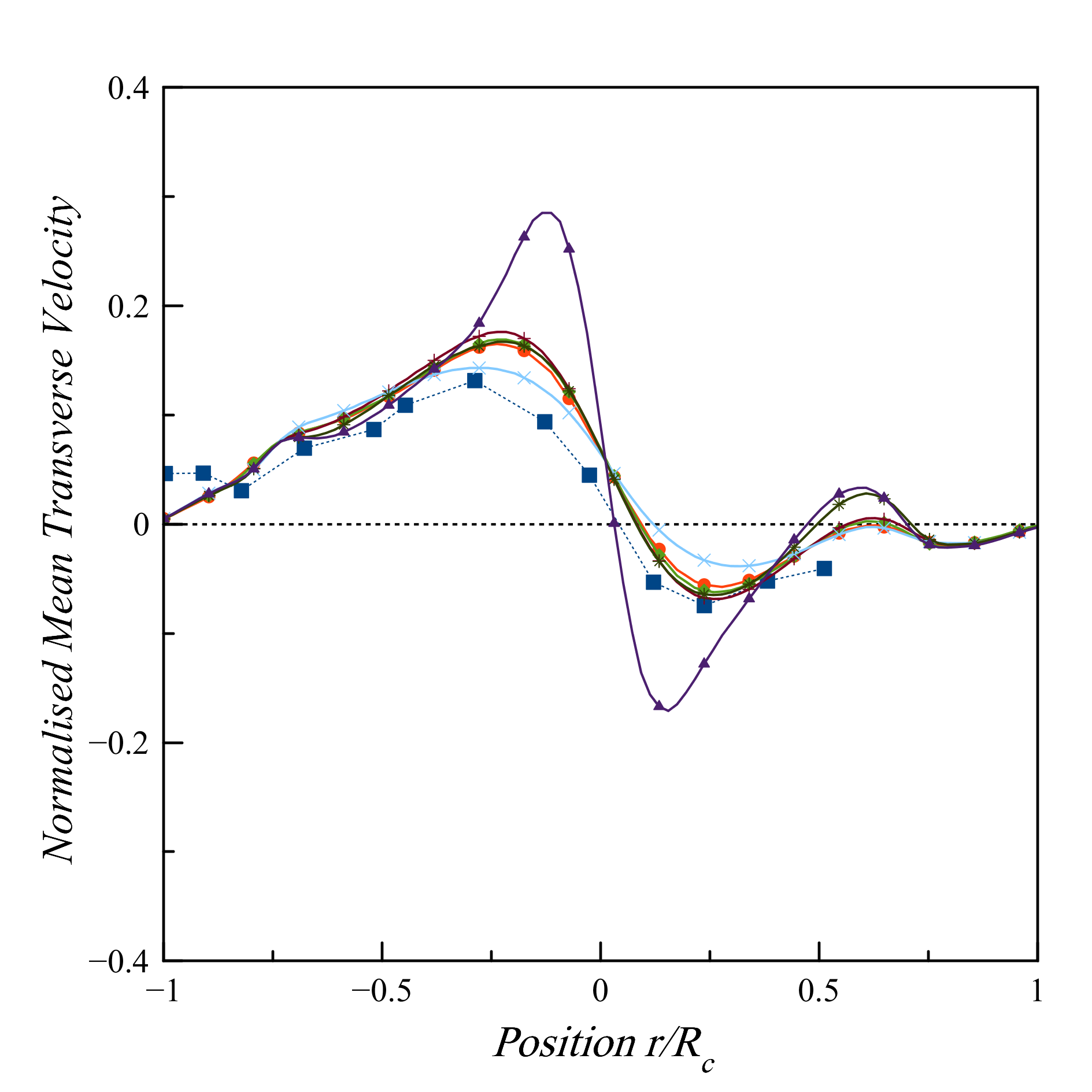}
		\caption{Normalized mean transverse velocity ($\tilde{v}^\ast = {\tilde{v}}/{{U}_b}$)}
		\label{fig:6b}
	\end{subfigure}
	\vspace{3mm}
	\begin{subfigure}[t]{0.48\linewidth}
		\centering
		\includegraphics[width=\linewidth]{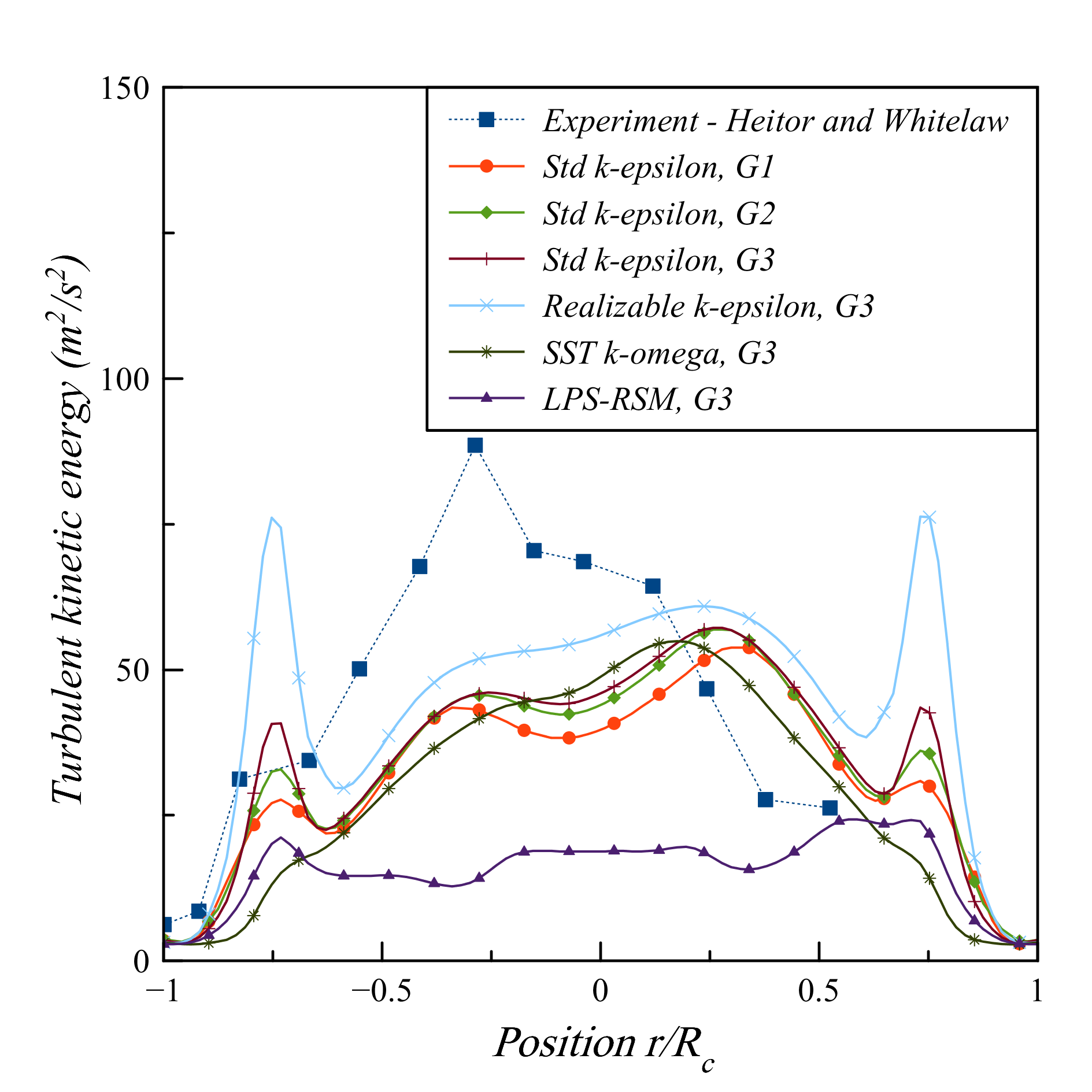}
		\caption{Turbulent kinetic energy, ${k} = \frac{3}{4} \left( \widetilde{{u^{\prime\prime}}^2} + \widetilde{{v^{\prime\prime}}^2} \right)$, \si{\meter\squared\per\second\squared}}
		\label{fig:6c}
	\end{subfigure}
	\hfill
	\begin{subfigure}[t]{0.48\linewidth}
		\centering
		\includegraphics[width=\linewidth]{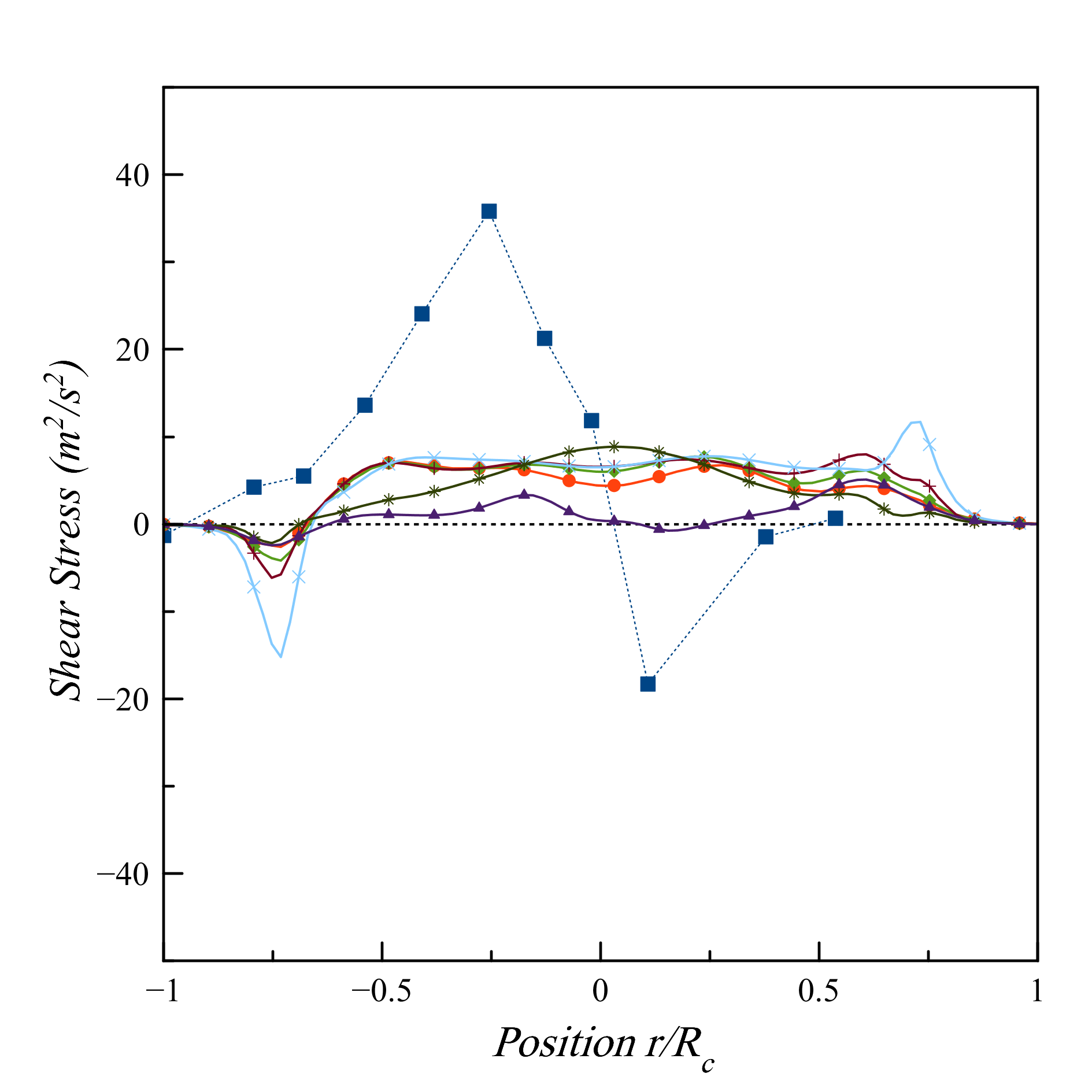}
		\caption{Shear stress, $-\widetilde{u^{\prime\prime}v^{\prime\prime}}$, \si{\meter\squared\per\second\squared}}
		\label{fig:6d}
	\end{subfigure}
	\caption{Velocity and turbulence characteristics at axial position $x = 130$ mm on the dilution holes plane at reacting conditions (refer \tab \ref{tab:1})). The experimental data are from \cite{heitor1985experiments,heitor1986velocity}}
	\label{fig:6}
\end{figure}

\begin{figure}[!tb]
	\centering
	\begin{subfigure}[t]{0.48\linewidth}
		\centering
		\includegraphics[width=\linewidth]{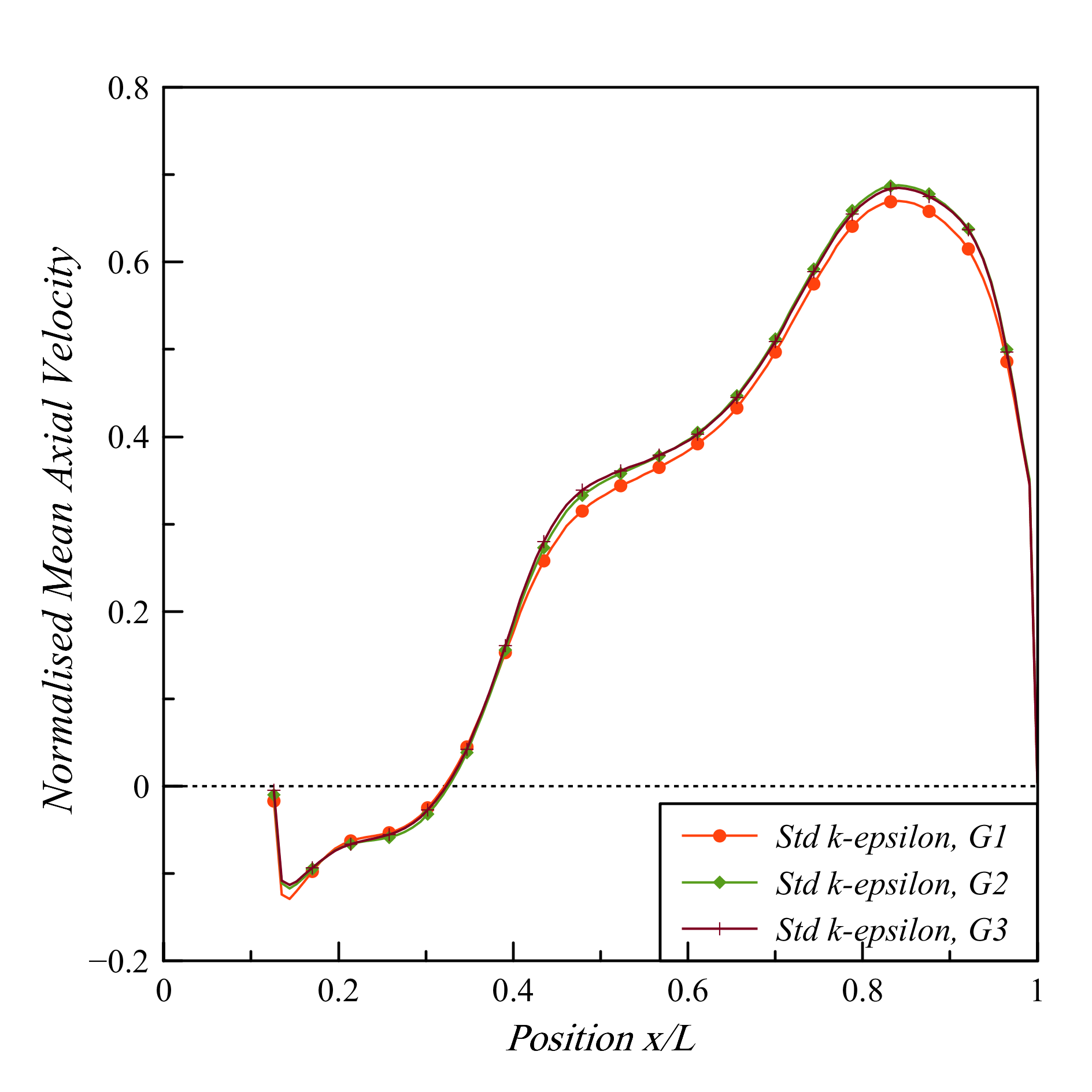}
		\caption{Normalized mean axial velocity ($\tilde{u}^\ast = {\tilde{u}}/{{U}_b}$)}
		\label{fig:7a}
	\end{subfigure}
	\hfill
	\begin{subfigure}[t]{0.48\linewidth}
		\centering
		\includegraphics[width=\linewidth]{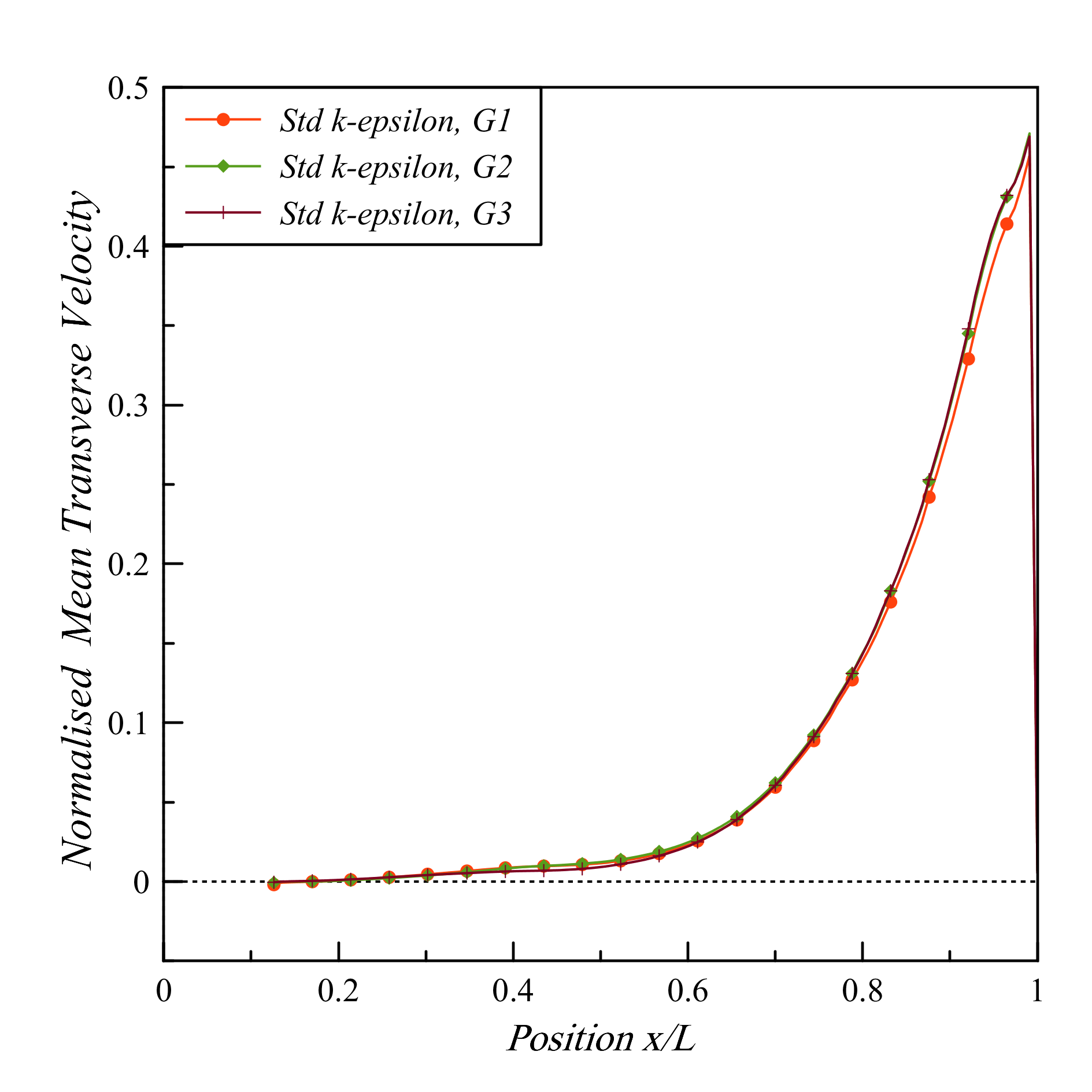}
		\caption{Normalized mean transverse velocity ($\tilde{v}^\ast = {\tilde{v}}/{{U}_b}$)}
		\label{fig:7b}
	\end{subfigure}
	\vspace{3mm}
	\begin{subfigure}[t]{0.48\linewidth}
		\centering
		\includegraphics[width=\linewidth]{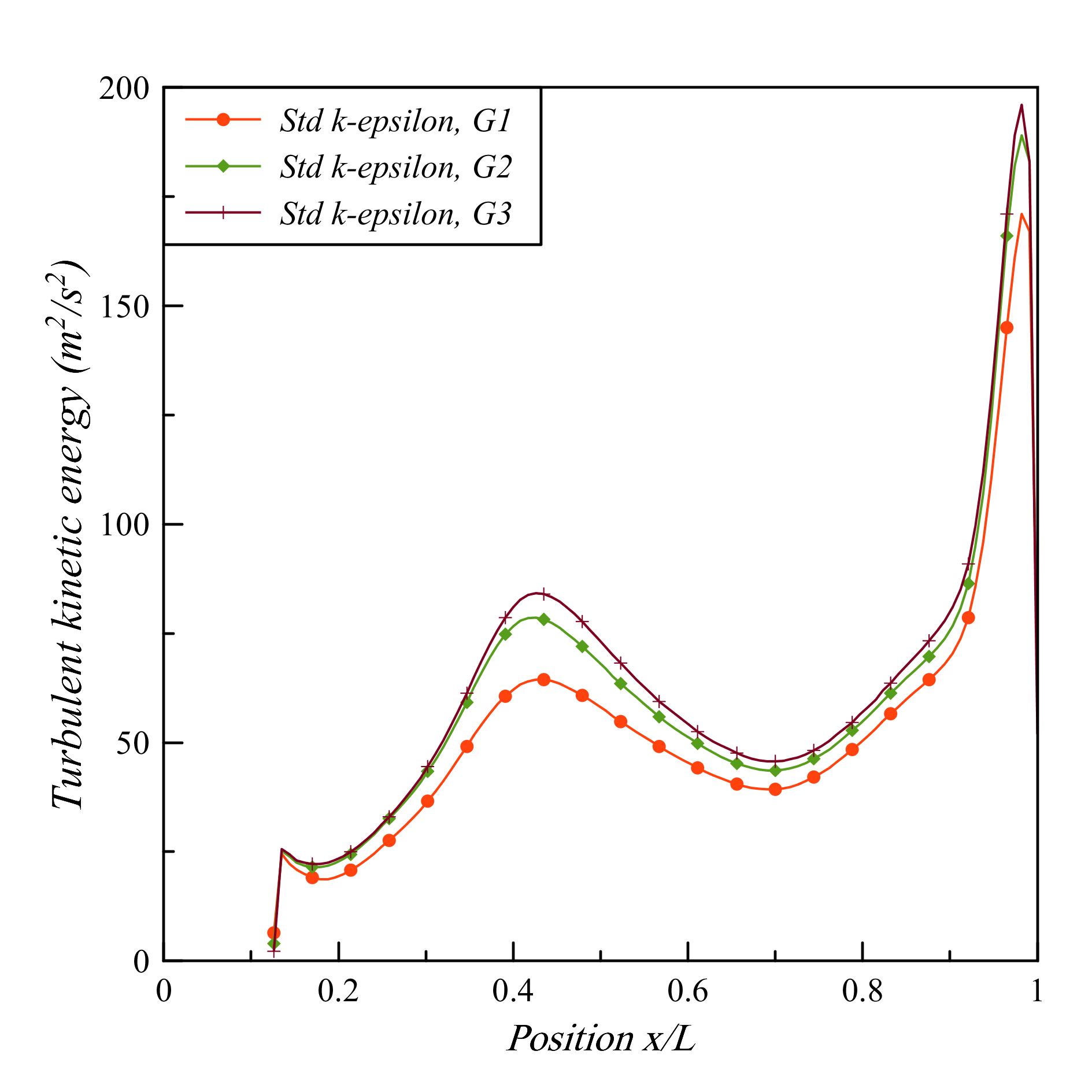}
		\caption{Turbulent kinetic energy, ${k} = \frac{3}{4} \left( \widetilde{{u^{\prime\prime}}^2} + \widetilde{{v^{\prime\prime}}^2} \right)$, \si{\meter\squared\per\second\squared}}
		\label{fig:7c}
	\end{subfigure}
	\hfill
	\begin{subfigure}[t]{0.48\linewidth}
		\centering
		\includegraphics[width=\linewidth]{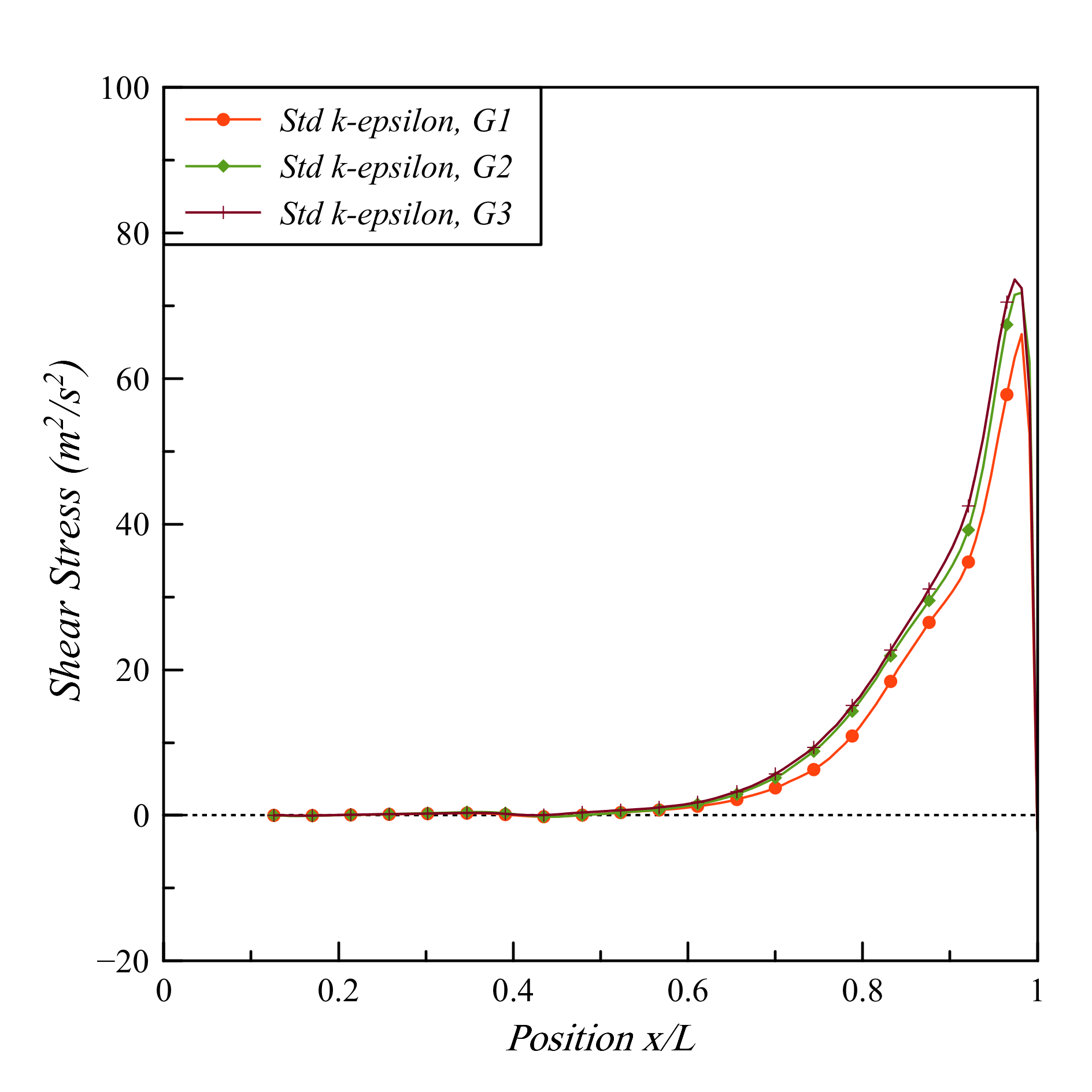}
		\caption{Shear stress, $-\widetilde{u^{\prime\prime}v^{\prime\prime}}$, \si{\meter\squared\per\second\squared}}
		\label{fig:7d}
	\end{subfigure}
	\caption{Velocity and turbulence characteristics variations over the length $x^* = x/L$ on the centreline ($r^* = 0$) at the reacting conditions  (refer \tab \ref{tab:1}). }
	\label{fig:7}
\end{figure}

The predicted velocity and turbulence characteristics obtained using three grids (G1, G2, and G3) at  primary holes (PH) plane ($x = 50$ mm) and dilution holes (DH) plane ($x = 130$ mm) are shown in \figs\ref{fig:5} - \ref{fig:7}. The grids are successively refined over the entire computational domain. \fig\ref{fig:5a} compares the normalized \added{mean} axial velocity profiles; a slight difference is observed between G1 and G2, whereas the discrepancy becomes negligible between G2 and G3. A similar trend is evident in \fig\ref{fig:5b} for the normalized \added{mean} transverse velocity at PH plane ($x = 50$ mm), with minimal variation between G2 and G3.
%
%\newline
On comparing the predicted turbulent kinetic energy ($k$) obtained using different grids (G1, G2, and G3) in \fig\ref{fig:5c}, it can be seen that the magnitude of $k$ increases due to improved resolution of mean velocity gradients with subsequent grid refinement (i.e., decreasing element size). This observations is consistent with previous studies \citep{kumar2017investigation, kumar2021predictions, kumar2022numerical, kumar2024assessment}.  A similar trend is observed for the predicted shear stress ($\widetilde{u^{\prime\prime}v^{\prime\prime}}$) in \fig\ref{fig:5d}, since the magnitude of shear stress  in two-equation turbulence models is directly proportional to the magnitude of mean velocity gradients (see \eqn\ref{viscous-stress-tensor}).
%\newline
On examining the influence of grid density (G1, G2, and G3) on the predicted velocity and turbulence fields at the DH plane ($x = 130$ mm), trends similar to those at the PH plane ($x = 50$ mm) are observed, i.e.,  a comparable mean axial and transverse velocity profiles are obtained using the three grids (G1, G2, and G3), with smaller variations between the predictions of G2 and G3, as illustrated in \figs\ref{fig:6a} and \ref{fig:6b}. The predicted turbulent kinetic energy ($k$) and shear stress ($\widetilde{u^{\prime\prime}v^{\prime\prime}}$) also exhibit similar grid-convergence behavior as observed at the PH plane.
\newline
Furthermore, the \added{normalized} mean axial and transverse velocities predicted on the axial centerline using grids G1, G2, and G3 (\figs\ref{fig:7a} and \fig\ref{fig:7b}) exhibit consistent trends, with the results from G2 and G3 showing closer agreement, indicating convergence with mesh refinement. The predicted turbulent kinetic energy along the axis (\fig\ref{fig:7c})   demonstrates a behavior consistent with that observed at the PH plane ($x = 50$ mm) and DH plane ($x = 130$ mm), showing a progressive increase in magnitude as the grid is refined. This systematic variation reflects improved resolution of turbulence structures with mesh refinement. Similarly, the predicted shear stress along the axis using G1, G2, and G3 (\fig\ref{fig:7d}) exhibits strong consistency and follows a trend analogous to that of the turbulent kinetic energy. 
Overall, these observations indicate that grid independence is being approached. Therefore, Grid 3 (refer \tab \ref{tab:3}) has been selected for subsequent simulations to ensure enhanced numerical accuracy and reliability of the results.
\subsection{Assessment of turbulence models}
In this section, the performance of various turbulence closure models, i.e., standard $k-\epsilon$, realizable $k-\epsilon$, SST $k-\omega$, and LPS-RSM,  in predicting the mean axial velocity, mean transverse velocity, turbulent kinetic energy, and shear stress, as illustrated in \fig\ref{fig:5} and \ref{fig:6}, is systematically evaluated. 
%\newline 
The present numerical results are compared with experimental measurements \citep{heitor1985experiments,heitor1986velocity} to assess their predictive capability of turbulence models for combustion flow in a realistic gas turbine combustor (see \fig \ref{fig:1s}). 
Additionally, the sensitivity of predicted temperature fields and scalar transport characteristics to the choice of turbulence model is examined to quantify model-dependent variations in reactive-thermofluidic flow behavior.
%\deleted[id=AK]{Combustion reactants and products, including the mole fractions of C$_3$H$_8$, O$_2$, H$_2$, CO$_2$ and CO (as illustrated in Figs.~3--5), as well as velocity (Figs.~6 and 9) and temperature fields (Figs.~10 and 11), are assessed by comparing the present computational results with experimental studies to determine their suitability for accurately simulating combustion flow in a realistic gas turbine combustor (see Fig.~1). Experimental measurements of mean temperature and mean species concentration are available in the literature and are used for model validation. However, velocity measurements for the reacting flow in the current geometry are not available. Velocity data exist only for a water model of the Tay combustor under different inflow conditions. Consequently, this study validates only the predicted mean temperature and species concentrations against the available experimental data, while the velocity field is examined qualitatively due to the lack of corresponding measurements. The CO$_2$ levels are underpredicted, as shown in Fig.~9c, consistent with the underprediction of temperature in Fig.~3f. The CO$_2$ and temperature predictions also deviate from experimental measurements, showing agreement only at $r^\ast \approx \pm 0.5$.} 
%
\subsubsection{Standard $k-\epsilon$ model predictions}
\begin{figure}[!b]
	\centering
	\begin{subfigure}[t]{0.48\linewidth}
		\centering
		\includegraphics[width=\linewidth]{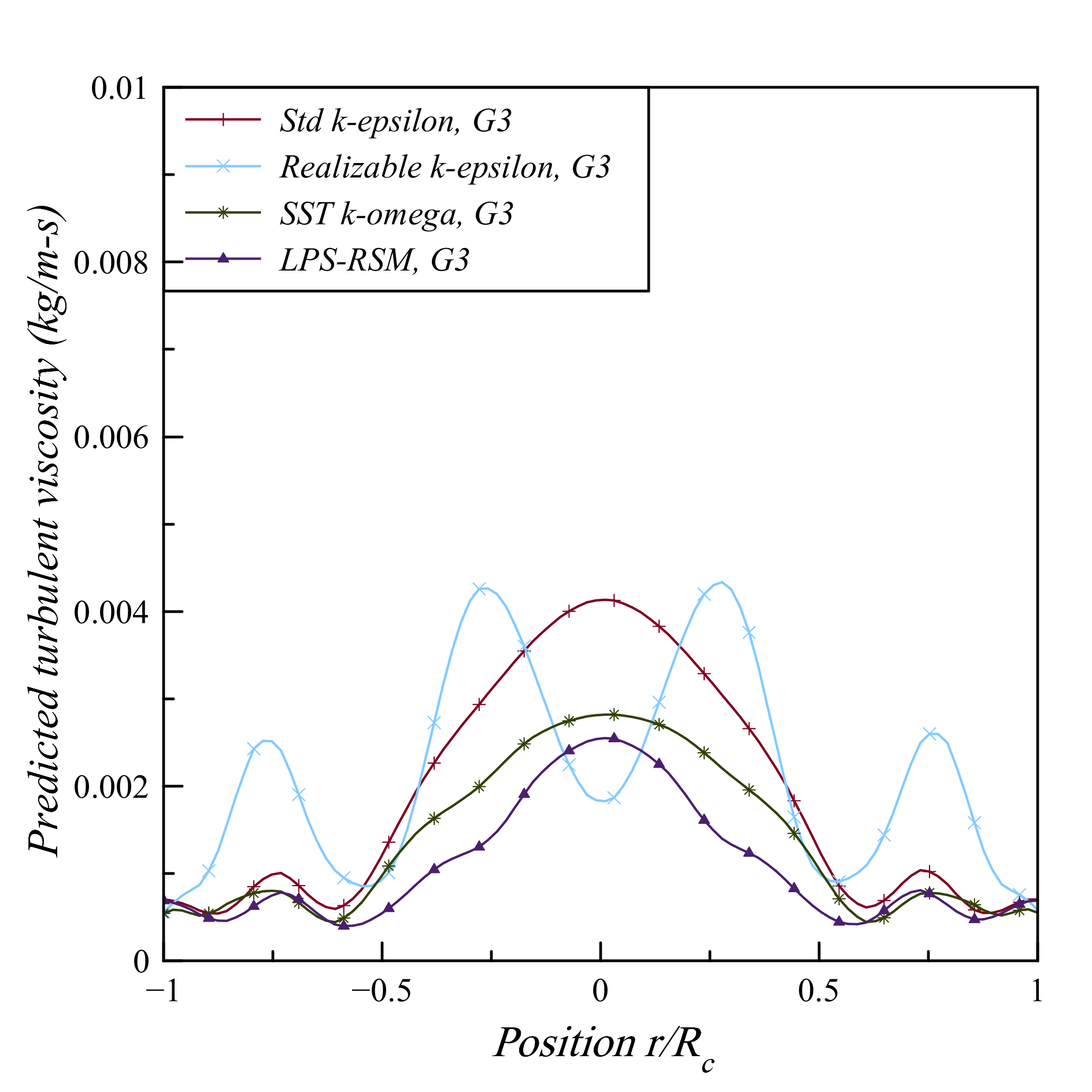}
		\caption{At $x = 50$ mm on the primary holes plane}
		\label{fig:8a}
	\end{subfigure}
	\hfill
	\begin{subfigure}[t]{0.48\linewidth}
		\centering
		\includegraphics[width=\linewidth]{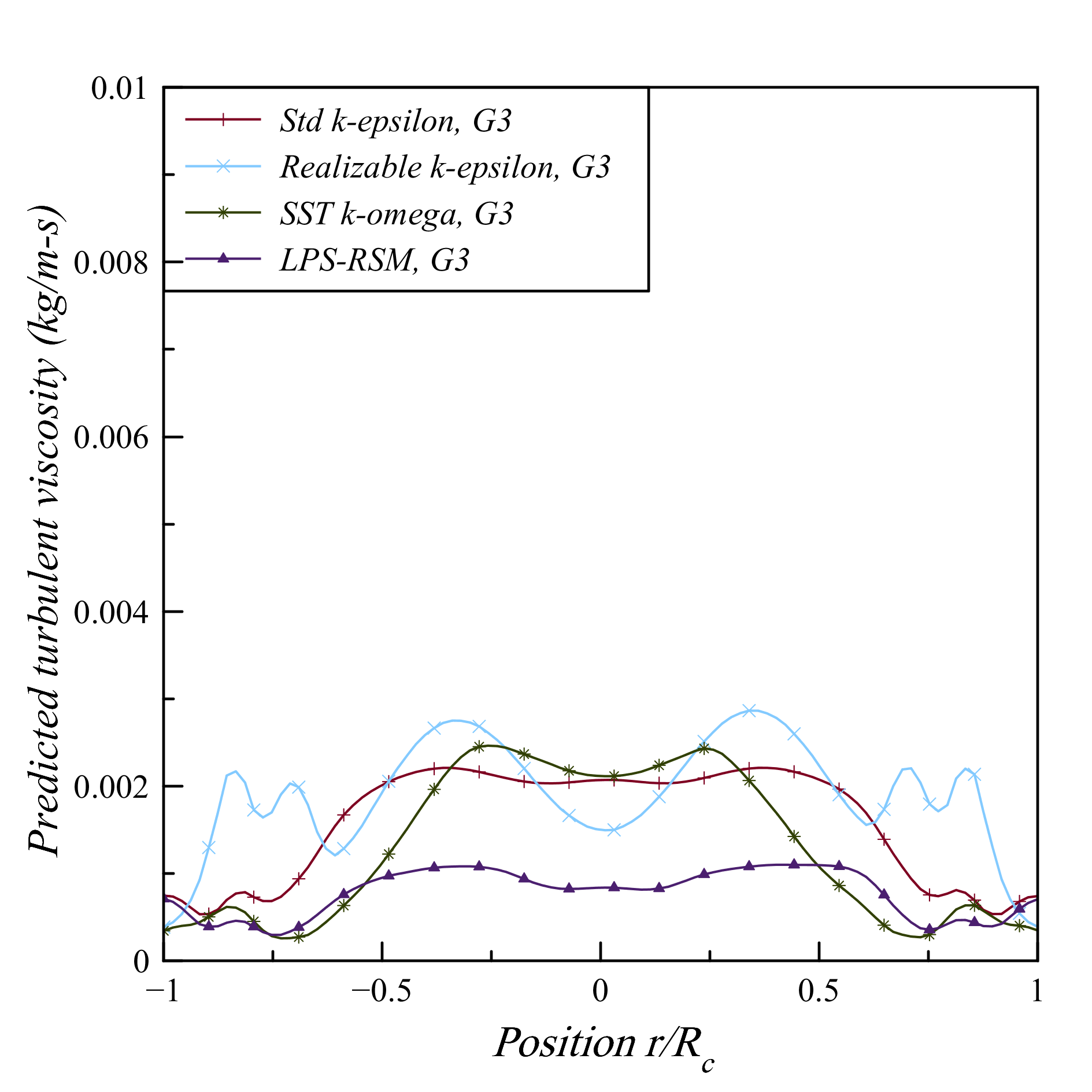}
		\caption{At $x = 130$ mm on the  dilution holes plane}
		\label{fig:8b}
	\end{subfigure}
	\vspace{3mm}
	
	\caption{Predicted turbulent viscosity ($\mu_t$, \si{kg \per \meter \per\second} kg m$^{-1}$s$^{-1}$) on primary and dilution holes plane at \deleted{at} the reacting conditions  (refer \tab\ref{tab:1}). }
	\label{fig:8}
\end{figure}
\begin{figure}[!b]
	\centering
	\begin{subfigure}{0.48\linewidth}
		\includegraphics[width=\linewidth]{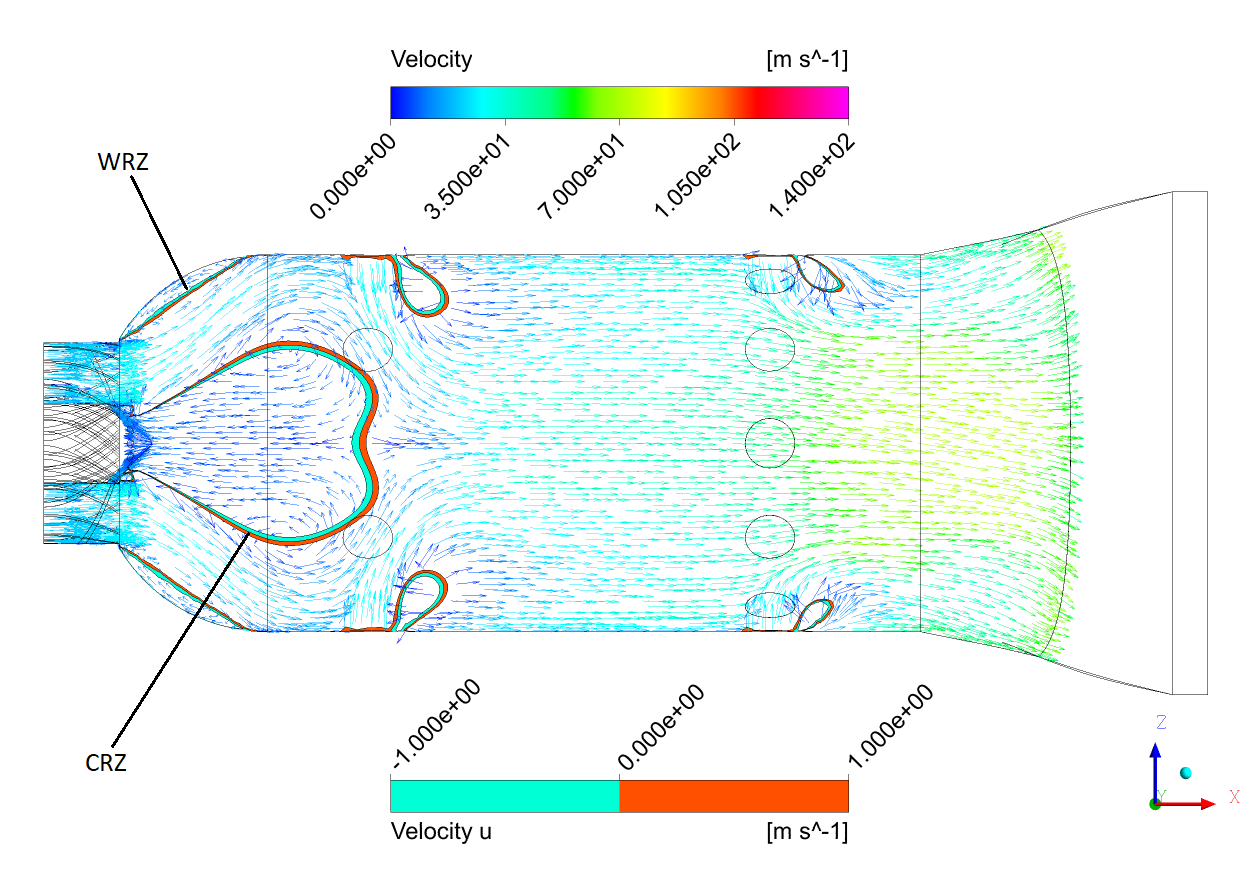}
		\caption{Standard $k-\epsilon$}
		\label{fig:9a}
	\end{subfigure}
	\begin{subfigure}{0.48\linewidth}
		\includegraphics[width=\linewidth]{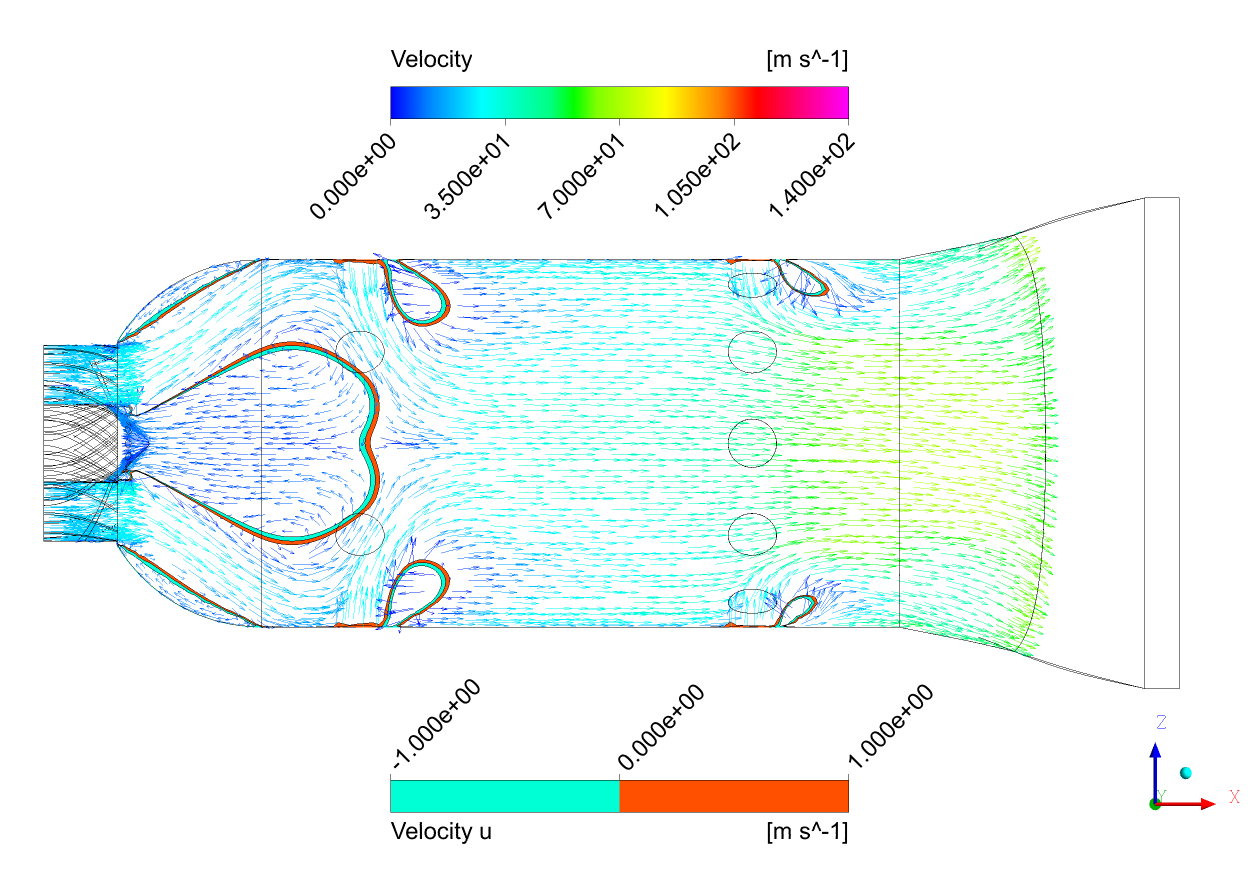}
		\caption{Realizable $k-\epsilon$}
		\label{fig:9b}
	\end{subfigure}
	\begin{subfigure}{0.48\linewidth}
		\includegraphics[width=\linewidth]{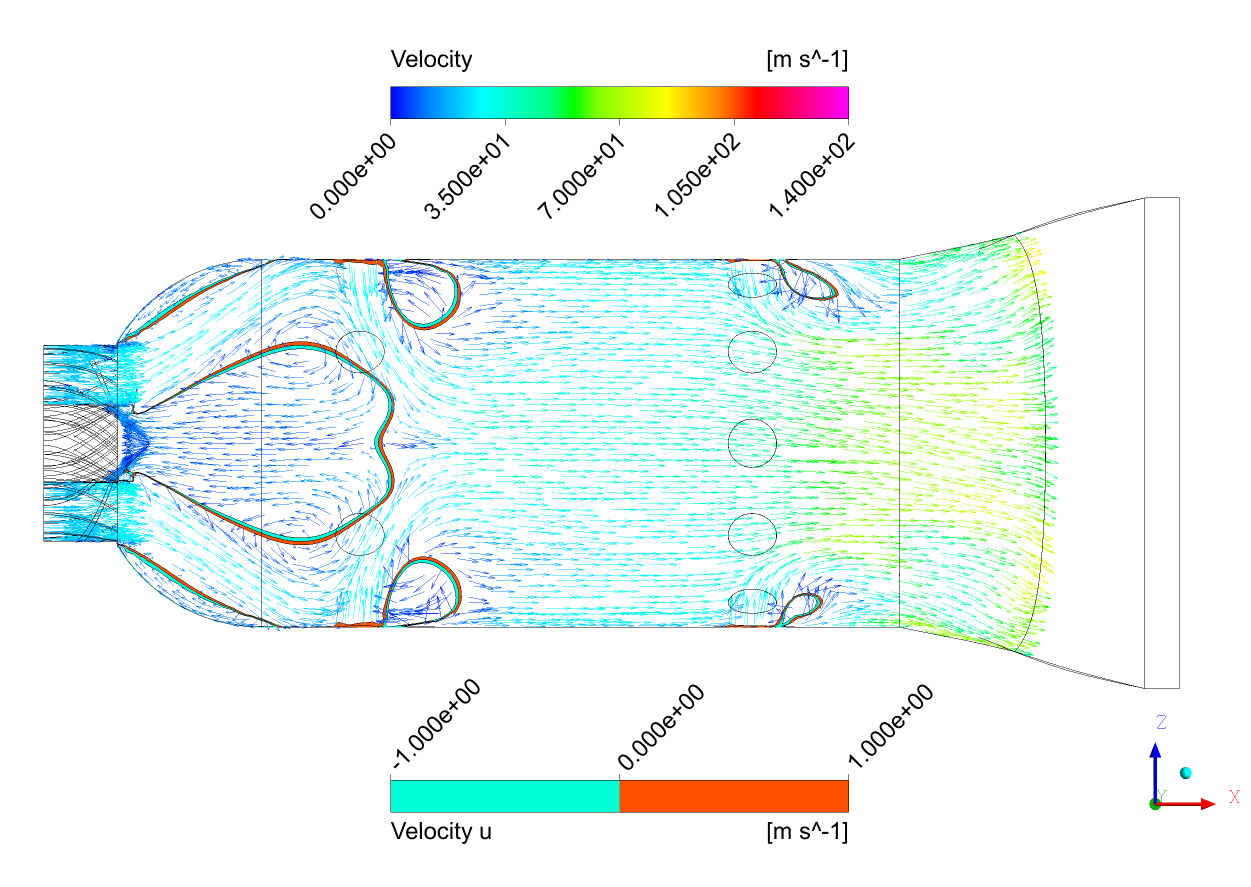}
		\caption{SST $k-\omega$}
		\label{fig:9c}
	\end{subfigure}
	\begin{subfigure}{0.48\linewidth}
		\includegraphics[width=\linewidth]{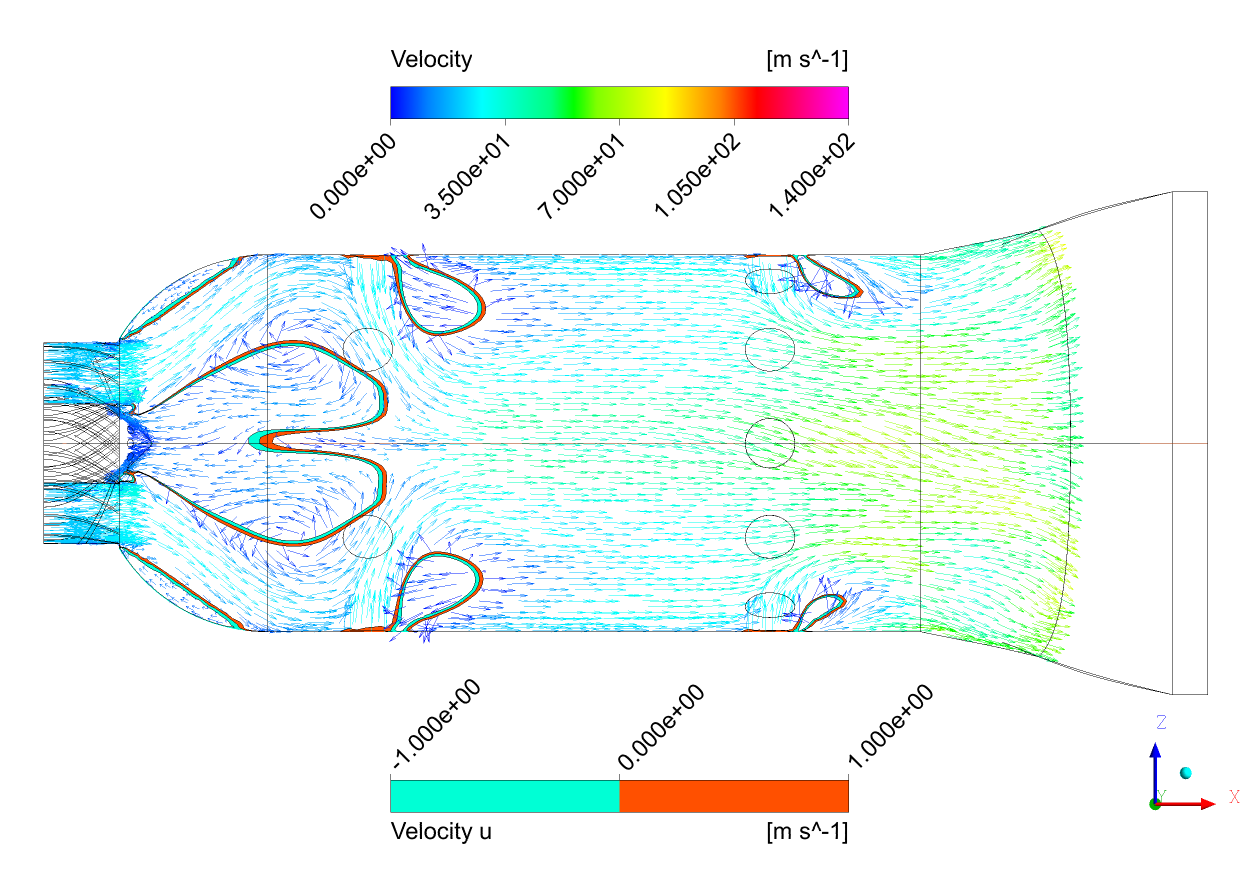}
		\caption{LPS-RSM}
		\label{fig:9d}
	\end{subfigure}
	\caption{Comparison of velocity vectors predicted on the horizontal (Z-Y) plane using various turbulence models.}
	\label{fig:9}
\end{figure}
\begin{figure}[!b]
	\centering
	\begin{subfigure}{0.48\linewidth}
		\includegraphics[width=\linewidth]{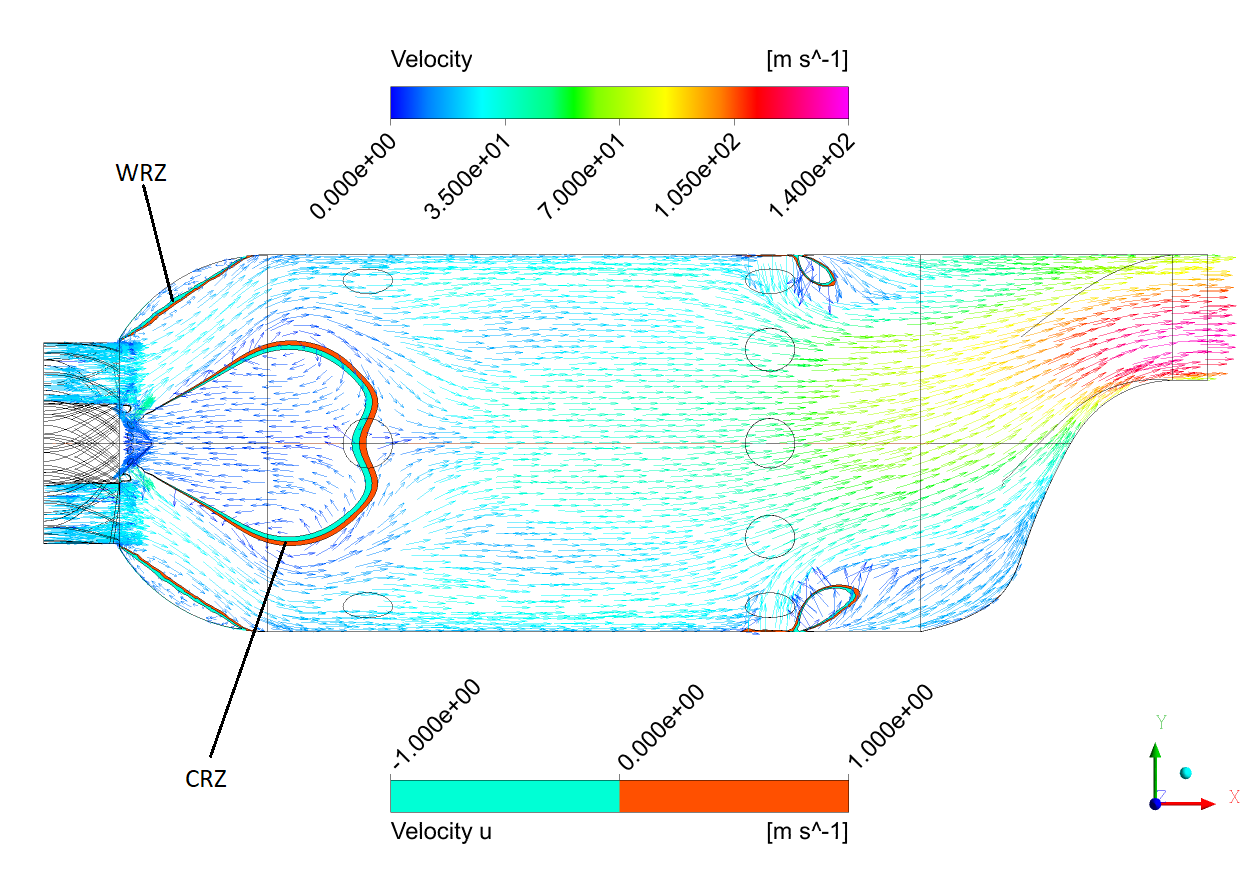}
		\caption{Standard $k-\epsilon$}
		\label{fig:10a}
	\end{subfigure}
	\begin{subfigure}{0.48\linewidth}
		\includegraphics[width=\linewidth]{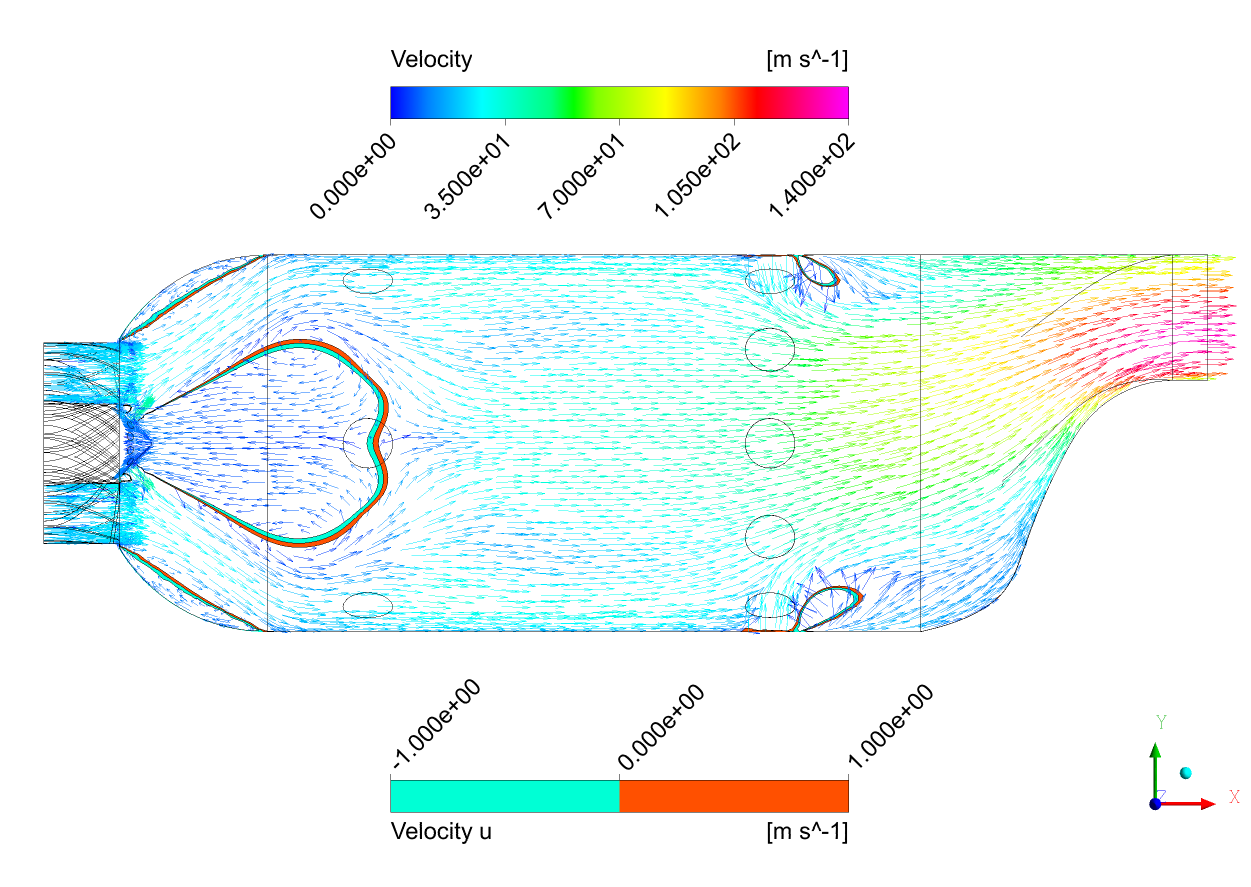}
		\caption{Realizable $k-\epsilon$}
		\label{fig:10b}
	\end{subfigure}
	\begin{subfigure}{0.48\linewidth}
		\includegraphics[width=\linewidth]{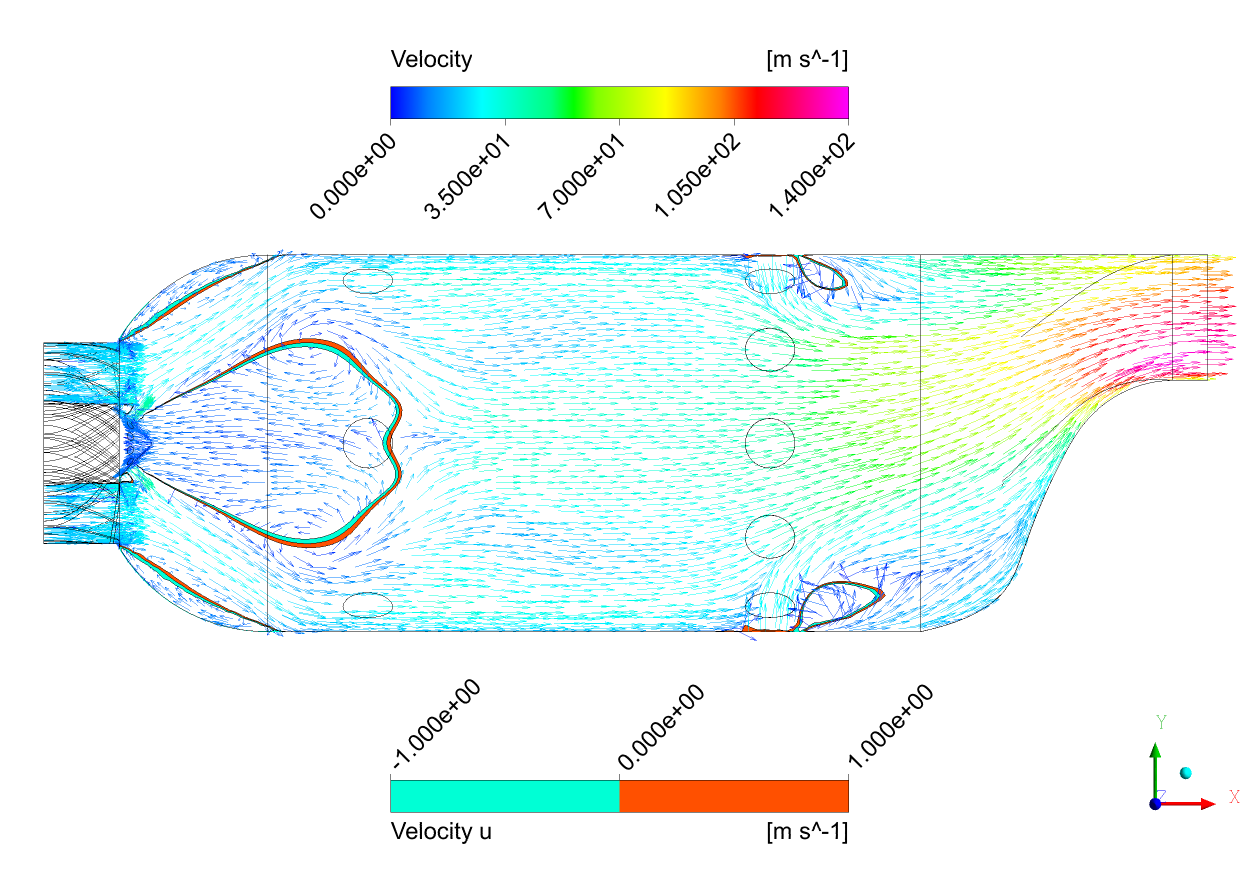}
		\caption{SST $k-\omega$}
		\label{fig:10c}
	\end{subfigure}
	\begin{subfigure}{0.48\linewidth}
		\includegraphics[width=\linewidth]{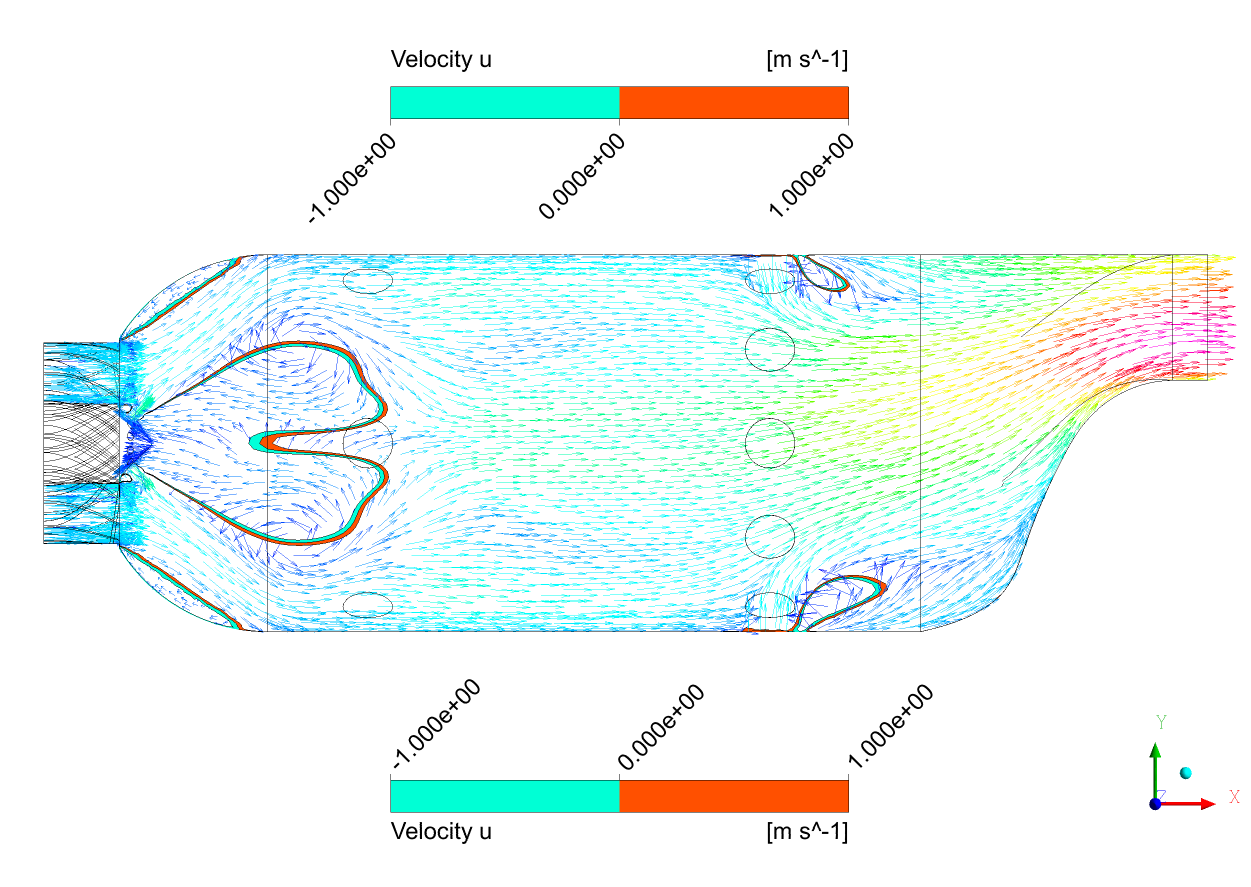}
		\caption{LPS-RSM}
		\label{fig:10d}
	\end{subfigure}
	\caption{Comparison of velocity vectors predicted on the front (X-Y) plane using various turbulence models.}
	\label{fig:10}
\end{figure}
\begin{figure}[!b]
	\centering
	\begin{subfigure}{0.48\linewidth}
		\includegraphics[width=\linewidth]{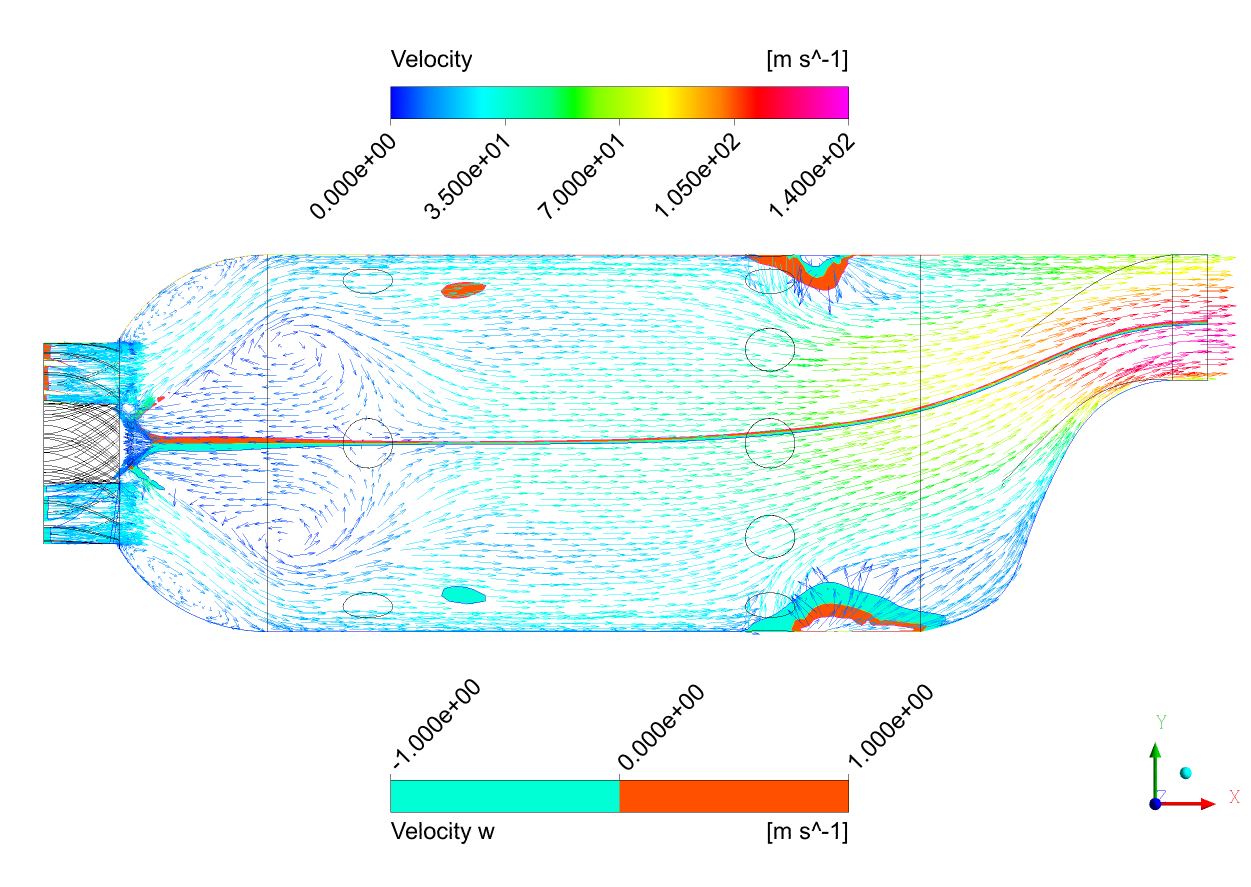}
		\caption{Standard $k-\epsilon$}
		\label{fig:11a}
	\end{subfigure}
	\begin{subfigure}{0.48\linewidth}
		\includegraphics[width=\linewidth]{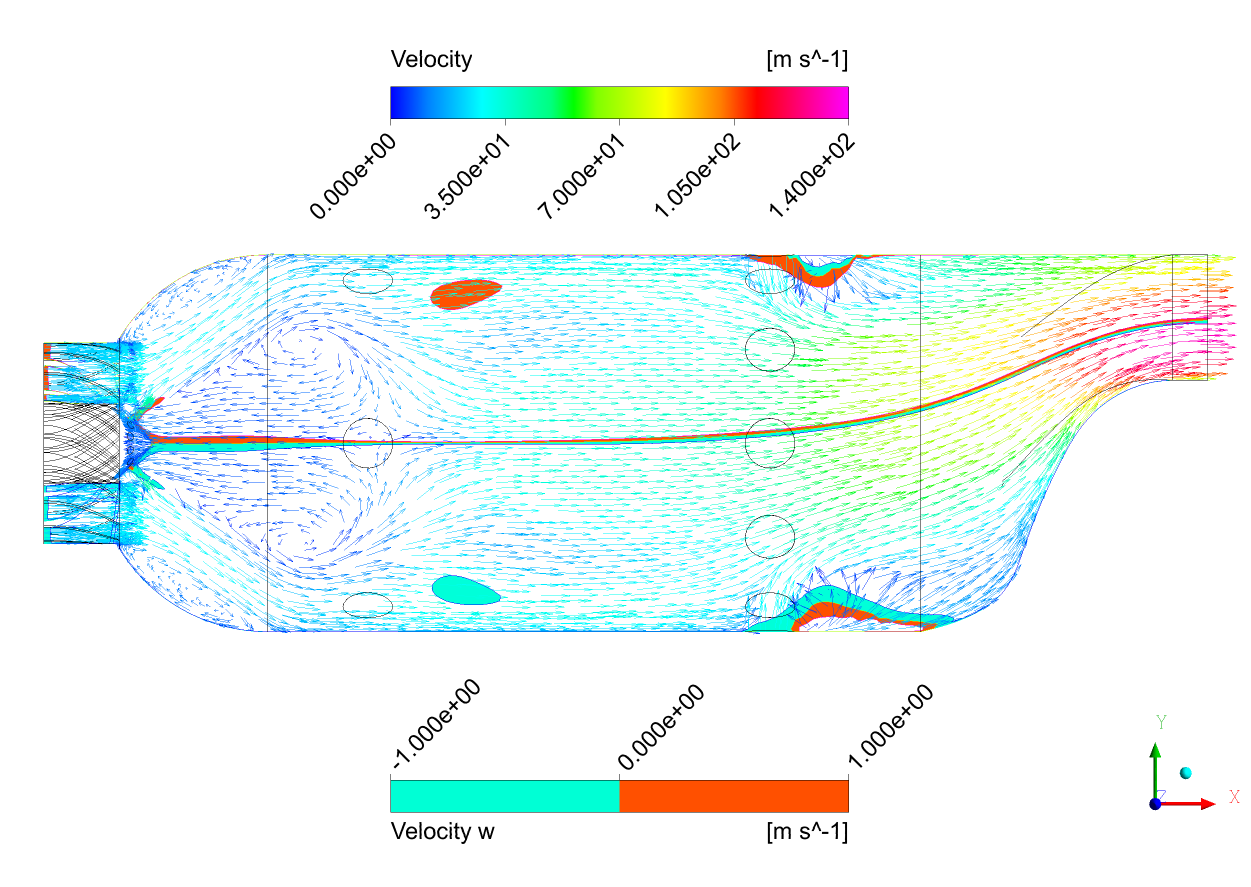}
		\caption{Realizable $k-\epsilon$}
		\label{fig:11b}
	\end{subfigure}
	\begin{subfigure}{0.48\linewidth}
		\includegraphics[width=\linewidth]{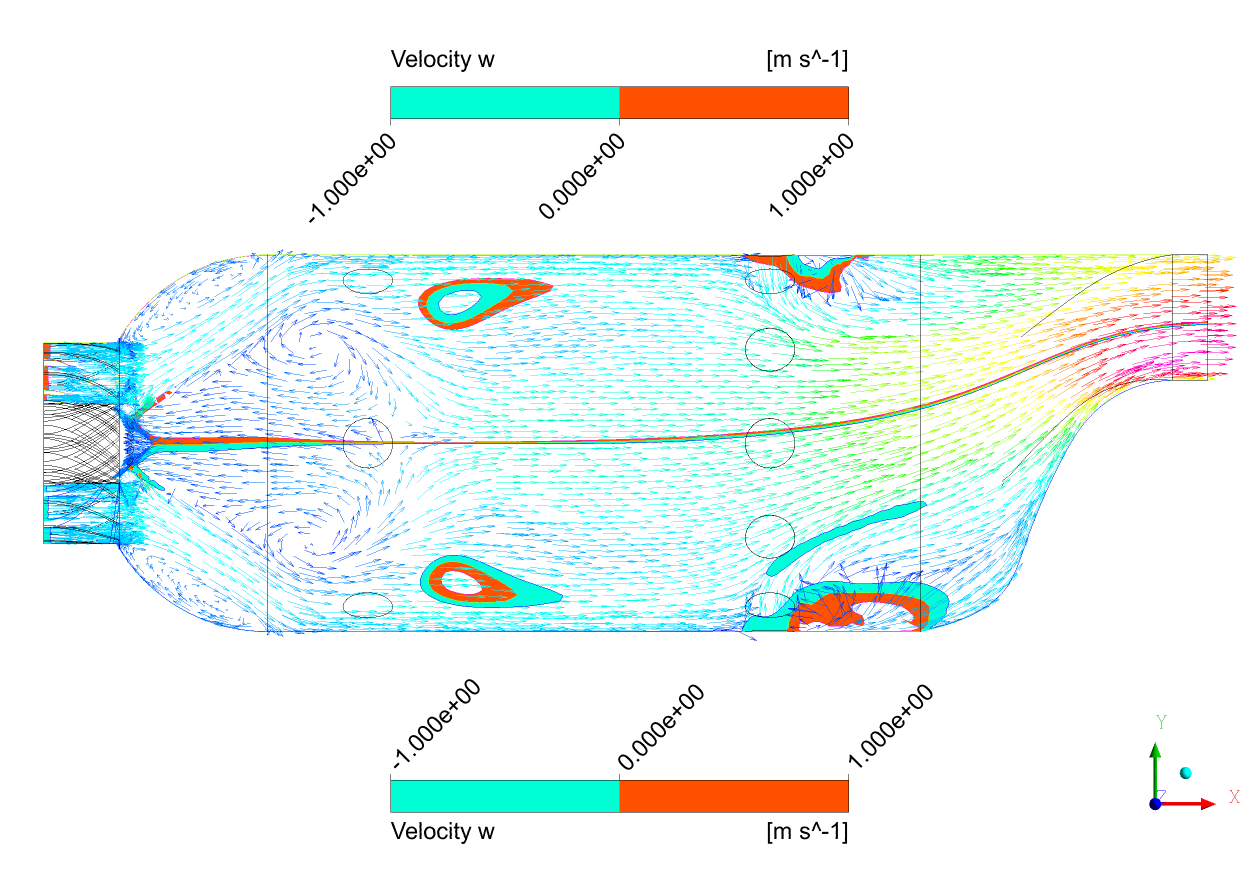}
		\caption{SST $k-\omega$}
		\label{fig:11c}
	\end{subfigure}
	\begin{subfigure}{0.48\linewidth}
		\includegraphics[width=\linewidth]{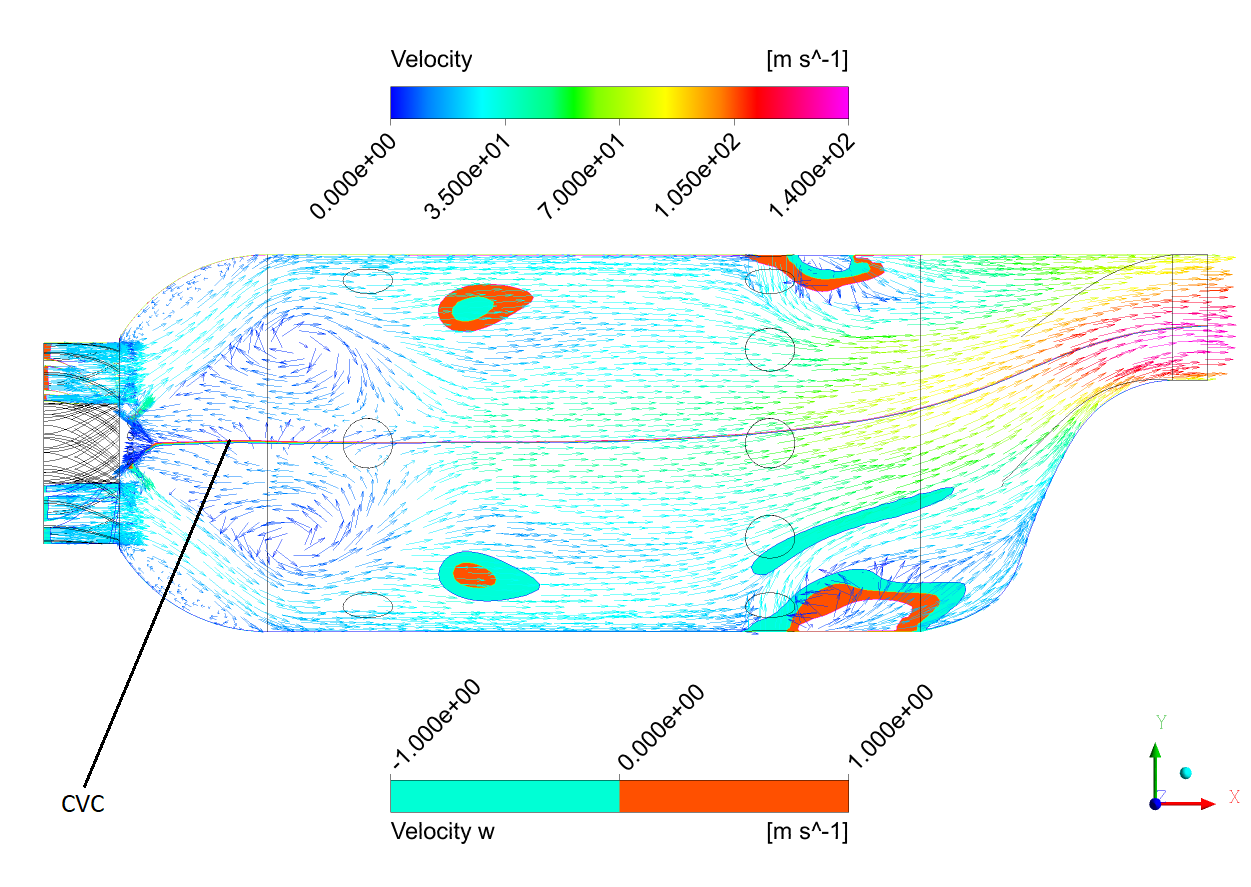}
		\caption{LPS-RSM}
		\label{fig:11d}
	\end{subfigure}
	\caption{Comparison of velocity vectors predicted on the  front (X-Y)  plane using various turbulence models.}
	\label{fig:11}
\end{figure}

\figs\ref{fig:5}–\ref{fig:6} present the predictions of the standard $k-\epsilon$ model compared with the experimental findings \citep{heitor1985experiments,heitor1986velocity}.  The normalized mean axial velocity ($\tilde{u}^\ast = \tilde{u}/{U}_b$) profiles (\fig\ref{fig:5a}) clearly exhibit the characteristic features of swirling flow inside the combustor.
At the outer radial locations ($0.8\le |r^\ast|  \le 1$), the axial velocity remains positive due to the forward momentum imparted by the incoming primary jets. As the radial position moves toward the core region ($|r^\ast| \approx 0.5$), the axial velocity gradually decreases and becomes negative, indicating the formation of the central recirculation zone (CRZ) resulting from the interaction between the swirling main flow and the cross-jets issued from the primary holes. At the centreline ($r^\ast = 0$), the axial velocity remains negative but with reduced magnitude, further confirming the presence of a stabilized CRZ. It should be noted that the experimental data of the remaining radial position ($\approx 0.75 \leq r^\ast \leq 1$) are not available in the literature \citep{heitor1985experiments,heitor1986velocity}  for the mean axial velocity, transverse velocity, turbulent kinetic energy (TKE), and shear stress in both primary holes (PH) and dilution holes (DH) planes (see \fig\ref{fig:5} to \fig\ref{fig:6}). 
\newline
On comparing the predicted \added{normalized} mean axial velocity obtained using the standard $k-\epsilon$ model with the experimental data, it is observed that the model shows good agreement in the outer radial region ($-1 \le r^\ast \le -0.65$). In the intermediate region ($-0.65 \le r^\ast \le -0.35$), the predictions follow the experimental trend, although quantitative deviations begin to emerge. Further inward ($-0.35 \le r^\ast < 0$), the model correctly predicts negative axial velocities, indicating the presence of a central recirculation zone (CRZ). However, the magnitude of the predicted negative velocity is substantially smaller than the measured values, suggesting that the model underpredicts the strength of the CRZ. This behaviour is further illustrated in \figs\ref{fig:9a} and \ref{fig:10a}, which present \deleted{azimuthal} \added{axial} velocity \deleted{($\tilde{w}$)} \added{($\tilde{u}$)} contours within the range \deleted{$-1 < \tilde{w} < 1$} \added{$-1 < \tilde{u} < 1$} m/s. The contours confirm a relatively weak CRZ and also reveal the presence of a wall recirculation zone (WRZ).  On the positive radial side, the axial velocity becomes slightly positive in the region $0 \le r^\ast \le 0.25$, then turns negative briefly, and subsequently increases to positive values in the range $0.25 \le r^\ast \le 0.5$. Beyond this location ($0.5 \le r^\ast \le 1$), the axial velocity gradually decreases toward zero.  
\newline 
Overall, the predicted axial velocity distribution is symmetric about the centreline and captures the general trend of the experimental data; however, it consistently underpredicts the CRZ intensity. The observed discrepancies can be attributed to the inherent limitations of the standard $k-\epsilon$ model. The model relies on the Boussinesq hypothesis, which assumes an isotropic eddy viscosity and relates the Reynolds stresses linearly to the mean strain rate, thereby limiting its ability to capture turbulence anisotropy in complex flows, i.e., locally isotropic turbulent viscosity, implying that turbulent transport behaves similarly in all directions at a given point \cite{tang2022approach}. This assumption restricts its ability to represent turbulence anisotropy, streamline curvature effects, and strong swirl-strain interactions. Consequently, turbulent momentum redistribution and secondary flow structures are inadequately resolved, leading to inaccuracies in predicting complex swirling and reacting flow patterns.
\newline
\fig\ref{fig:5b} presents the normalized mean transverse velocity distribution ($\tilde{v}^\ast = \tilde{v}/{U}_b$) as a function of the radial position ($r^\ast$). As the radial position approaches the centreline, the transverse velocity decreases sharply and changes sign, becoming negative. This behaviour is characteristic of swirling flows and reaches its maximum negative magnitude at $r^\ast \approx 0.15$.  Moving further toward the centre, the transverse velocity gradually recovers and its magnitude decreases, although it remains negative. 
A comparison of the predicted mean transverse velocity obtained using the standard $k-\epsilon$ model with the experimental data (\fig\ref{fig:5b}) shows that the numerical results generally follow the experimental trend. In the outer region ($-1 < r^\ast < -0.65$), the transverse velocity is overpredicted, whereas in the region $-0.65 < r^\ast < -0.25$ it is underpredicted. From this location onward, the predicted velocity sharply decreases and becomes negative, indicating the presence of swirl, and reaches its peak negative value at approximately $r^\ast \approx 0.235$. Although still slightly underpredicted, the transverse velocity profile subsequently begins to recover, matching the experimental value near $r^\ast \approx 0.65$, and gradually approaches zero toward the outer radial boundary. The discrepancies in the predicted mean transverse velocity can primarily be attributed to the isotropic nature of the standard $k-\epsilon$ model, which limits the model’s ability to represent turbulence anisotropy in strongly swirling flows. Nevertheless, while the transverse velocity profiles indicate the presence of swirl, the azimuthal velocity component ($\tilde{w}$) confirms (\fig\ref{fig:11a}) the formation of a central vortex core (CVC). The contour of $\tilde{w}$ within the range $-1 < \tilde{w} < 1$ m/s, presented together with velocity vectors in \fig\ref{fig:11a}, clearly indicates the formation of a CVC originating from the swirler exit and extending downstream through the combustor.
\newline 
On examining the turbulent kinetic energy (TKE) distribution shown in \fig\ref{fig:5c}, the TKE is very small near the left boundary ($r^\ast \approx -1$), remaining below 10 m$^2$/s$^{2}$, which is consistent with the low velocity gradients adjacent to the combustor liner. In the region ($-0.8 \le r^\ast \le -0.5$) corresponding to the shear layer where the primary hole (PH) jets interact with the circulating flow, the TKE increases steadily to approximately 55 m$^2$/s$^{2}$. The TKE remains elevated between $r^\ast \approx -0.55$ and $r^\ast \approx -0.15$, before increasing sharply to a pronounced peak at the centreline ($r^\ast \approx 0$). This peak indicates the presence of the central recirculation zone (CRZ), where opposing axial momentum  generates strong velocity fluctuations and consequently enhance turbulence \deleted{intensity} \added{levels}. Beyond the centreline, the TKE gradually decreases to approximately 40 - 45 m$^2$/s$^{2}$ within the shear layer, where the primary jets again interact with the swirling flow, and subsequently decays rapidly toward the right boundary. A comparison with the TKE predicted using the standard $k-\epsilon$ model on the primary hole (PH) plane (\fig\ref{fig:5c}) shows that the model underpredicts the turbulence \deleted{intensity} \added{levels} across most of the radial span. The standard $k-\epsilon$ predicted TKE profile intersects the experimental values only at isolated radial locations (i.e., at $r^\ast \approx -0.15$, $\approx 0.25$, $\approx 0.35$). The numerical prediction increases gradually from the left side beginning at $r^\ast \approx -0.5$, followed by a steep rise to a peak at $r^\ast \approx -0.15$, where it coincides with the experimental value. Moving further inward, the predicted TKE increases again and reaches a maximum near $r^\ast \approx 0.2$, after which it begins to decrease and intersects the experimental profile near $r^\ast \approx 0.25$. Beyond this location, the predicted TKE continues to decline toward the outer radial region and gradually approaches zero toward the right boundary.
\newline 
In two-equation turbulence models, the eddy viscosity plays a crucial role in determining the TKE and overall turbulence characteristics. Consistent with the predicted TKE behaviour using the standard $k-\epsilon$ mode, the turbulent viscosity distribution shown in \fig\ref{fig:8a} exhibits a plateau in the same radial region where the TKE curve also becomes nearly flat. This indicates that the eddy viscosity predicted by the standard $k-\epsilon$ model is not sufficiently sensitive to local flow variations in the present configuration. Moreover, eddy-viscosity models are known to be overly dissipative \cite{wilcox1998turbulence}, which can lead to excessive damping of turbulence fluctuations and smoothing of mean velocity gradients.
\newline
A comparison of the predicted Reynolds shear stress obtained using the standard $k-\epsilon$ model with the experimental data is presented in \fig\ref{fig:5d}. In the left outer region ($-1 \le r^\ast \le -0.8$), the shear stress remains close to zero, indicating weak turbulence activity and relatively small velocity gradients near the combustor liner. Moving inward, a small positive peak appears around $r^\ast \approx -0.7$, followed by a negative dip near $r^\ast \approx -0.55$, reflecting alternating turbulent fluctuations associated with local shear-layer interactions. Further inward, the shear stress increases sharply and reaches a prominent positive peak at approximately $r^\ast \approx -0.3$, marking a region of strong velocity gradients and intensified turbulence. Beyond this location, the shear stress gradually decreases but remains positive until about $r^\ast \approx 0.1$, indicating continued but weakening turbulent mixing. At $r^\ast \approx 0.1$, the shear stress transitions, i.e., changes sign and becomes negative, reaching a minimum near $r^\ast \approx 0.25$. This negative shear stress region corresponds to the inner portion of the central recirculation zone (CRZ), where the direction of turbulent momentum transport reverses. Moving further outward, the shear stress transitions back toward positive values, attaining another peak near $r^\ast \approx 0.65$. Beyond this point, it decreases again and becomes negative around $r^\ast \approx 0.8$, suggesting another reversal in turbulent momentum transfer associated with the outer shear-layer region.
\newline
On comparison, in the left outer region ($-1 \le r^\ast \le -0.7$), the model predicts small negative shear stress values that follow the experimental trend, although their magnitude is smaller than the measured values. In the shear-layer region ($-0.7 \le r^\ast \le -0.3$), the experimental data exhibit a pronounced drop followed by a sharp rise in shear stress. The numerical results obtained using the standard $k-\epsilon$ model significantly underpredict this behaviour, capturing only a mild increase and failing to reproduce the sharp experimental peak. In the central region, the experimental data show strong variations in shear stress, decreasing rapidly around $r^\ast \approx -0.25$ and becoming negative, with a minima occurring near $r^\ast \approx 0.25$. Although the standard $k-\epsilon$ model captures the overall trend, it considerably underpredicts the magnitude of the shear stress in this region. Further outward ($0.3 \le r^\ast \le 0.7$), the experimental results indicate that the shear stress returns to positive values before decreasing again and becoming negative near $r^\ast \lesssim 0.8$, suggesting the presence of another shear-layer interaction. The numerical predictions again underrepresent this behaviour.  Toward the outer boundary ($0.8 \lesssim r^\ast \lesssim 1$), the shear stress gradually approaches zero. The overall underprediction of Reynolds shear stress across the primary hole (PH) radial plane can be attributed to the isotropic limitations of the standard $k-\epsilon$ model, which overly dissipative representation of turbulence. Consequently, the model struggles to capture the strong turbulence anisotropy and shear-layer dynamics characteristic of confined swirling flows in can-type and cannular-type combustors.
\newline
The normalized axial velocity ($\tilde{u}^\ast$) distribution on the dilution hole (DH) plane is shown in \fig\ref{fig:6a}. The experimental profile exhibits a steep rise in normalized axial velocity from the left side ($r^\ast=-1$ to $r^\ast \approx -0.7$), after which the velocity continues to increase more gradually and reaches a peak near $r^\ast \approx -0.25$. Moving further toward the centre, the mean axial velocity decreases in the region $-0.25 \lesssim r^\ast \le 0$, followed by a slight increase up to about $r^\ast \approx 0.125$. Beyond this location, the velocity profile remains nearly constant, forming a plateau over the range $0.125 \le r^\ast \le 0.55$. The numerical results obtained using the standard $k-\epsilon$ model overpredict the axial velocity across most of the radial plane. The predicted mean axial velocity increases rapidly from the left side ($r^\ast=-1$) and forms a plateau between $r^\ast \approx -0.65$ to $\approx 0.65$. From this point, the velocity continues to increase and reaches a peak near $r^\ast \approx 0.135$. Thereafter, the predicted velocity gradually decreases toward the outer radial region and eventually approaches zero near the right side ($r^\ast \approx 1$). Overall, the predicted axial velocity profile deviates from the experimental trend and fails to capture the mild asymmetry observed in the measurements. These discrepancies can primarily be attributed to the limitations of the standard $k-\epsilon$ model, i.e., isotropic turbulent viscosity and inherently dissipative nature. The elevated eddy viscosity predicted in this region (see \fig\ref{fig:8b}) leads to excessive smoothing of velocity gradients and consequently an overprediction of the axial velocity.
\newline 
The \added{normalized} mean transverse velocity ($\tilde{v}^\ast$)  distribution on the dilution hole (DH) plane is presented in \fig\ref{fig:6b}. Starting from the left side ($r^\ast=-1$) and moving toward the centreline, the experimental transverse velocity remains positive and reaches a peak in the shear-layer region at $r^\ast \approx -0.25$. Beyond this location, the transverse velocity decreases, becoming negative near the centreline, and reaches a negative peak within the opposite shear-layer region at $r^\ast \approx 0.25$. Further outward, the transverse velocity gradually increases in magnitude toward zero while remaining negative. The standard $k$–$\epsilon$ model captures the transverse velocity distribution reasonably well, reproducing the overall profile shape and the sign change across the radial plane, with positive values on the left side, near-zero values around the centreline, and negative values on the right side. The model slightly overpredicts the positive peak on the left side, and the numerical profile appears somewhat smoother than the experimental data. However, the negative peak on the right side is reproduced fairly well, with the predicted values coinciding with the experimental data near $r^\ast \approx 0.235$ and $r^\ast \approx 0.385$. The simulation also captures the zero-crossing location with a small radial shift of approximately $r^\ast \approx 0.05$.
\newline
The turbulent kinetic energy (TKE) distribution on the dilution hole  (DH) plane is shown in \fig\ref{fig:6c}. The experimental measurements indicate that the TKE gradually increases from the left boundary ($r^\ast \approx -1$). A rapid rise in TKE occurs in the region $-0.9 \lesssim r^\ast \lesssim -0.8$, primarily due to the influence of the dilution jets, after which the profile begins to form a plateau. Further inward, the TKE increases again from $r^\ast \approx -0.5$, associated with the development of a shear layer, and reaches a pronounced peak near $r^\ast \approx -0.25$. Beyond this location, the TKE decreases rapidly up to $r^\ast \approx -0.15$, followed by a more gradual decline up to $r^\ast \approx 0.15$. This region corresponds to the transition across the central part of the flow toward the opposite shear layer. Moving further toward the right side, the experimental TKE distribution again approaches a plateau. The standard $k-\epsilon$ model consistently underpredicts the TKE across most radial locations, except near the DH regions on both sides ($r^\ast \approx \pm 0.75$), where the interaction between the convected swirling flow originating from the swirler and the dilution jets enhances turbulence levels. Moreover, the overall shape of the predicted TKE profile differs significantly from the experimental measurements, with the predicted peaks appearing at radial locations opposite to those observed experimentally. In general, the standard $k-\epsilon$ model fails to accurately reproduce the turbulence \deleted{intensity} \added{levels} generated by the strong shear associated with confined swirling flows and their interaction with dilution jets in can-type and can-annular combustors.
\newline 
The Reynolds shear stress distribution on the dilution hole (DH) plane is shown in\fig\ref{fig:6d}. Near the left side ($r^\ast \approx -1$), the shear stress is a small negative value. Moving inward, it gradually increases up to $r^\ast \approx -0.8$, beyond which the profile exhibits a short plateau. Further inward, the shear stress rises sharply and reaches a pronounced positive peak at  $r^\ast \approx -0.25$, indicating the presence of a strong shear-layer interaction. Subsequently, the shear stress decreases rapidly, crossing zero near the centreline and becoming negative, with a minimum occurring at $r^\ast \approx 0.1$. Beyond this point, the shear stress gradually recovers while remaining negative over a finite radial region, again suggesting the influence of a shear layer. Moving further toward the right side, the shear stress continues to recover and eventually becomes positive. In contrast, the numerical predictions obtained using the standard $k-\epsilon$ model show significant deviations from the experimental measurements across the radial plane. The predicted shear stress begins close to zero at the left ($r^\ast \approx -1$), decreases slightly to negative values at $r^\ast \approx -0.8$, and reaches a small negative peak near $r^\ast \approx -0.75$. Beyond this location, the shear stress begins to increase, becoming positive around $r^\ast \approx -0.65$ and forming a weak plateau near $r^\ast \approx -0.7$. However, the positive shear stress is substantially underpredicted in the region $-0.8 \le r^\ast \le 0$. Moreover, in the radial region where the experimental measurements show negative shear stress ($0 < r^\ast < 0.5$), the model instead predicts positive values. These discrepancies further highlight the inability of the standard $k-\epsilon$ model to accurately capture the shear-layer dynamics and associated turbulent momentum exchange occurring in the swirling dilution-jet interaction region of confined swirling combustor flows.
\subsubsection{Realizable $k-\epsilon$ model predictions}
The predicted \added{normalized} mean axial velocity ($\tilde{u}^\ast$) obtained using the realizable $k-\epsilon$ model is compared with the experimental data in \fig\ref{fig:5a}. Overall, the numerical predictions follow the experimental trend reasonably well. Compared with the standard $k-\epsilon$ model, the realizable $k-\epsilon$ model predicts a stronger recirculation region within $-0.4 \le r^\ast \le 0.4$, indicating a more pronounced central recirculation zone (CRZ). These differences can be attributed to the modified formulation of the dissipation-rate transport equation (\eqn\ref{realizable-favre-epsilon}) and the variable turbulent viscosity coefficient $C_\mu$ (\eqn\ref{eq:Cmu_favre}) employed in the realizable $k-\epsilon$ model. The predicted CRZ structure using the realizable $k-\epsilon$ model is further illustrated in \figs\ref{fig:9b} and \ref{fig:10b}.
\newline
The predicted \added{normalized} transverse velocity ($\tilde{v}^\ast$) distributions are compared with the experimental measurements in \fig\ref{fig:5b}. The realizable $k-\epsilon$ model generally captures the overall trend of the experimental data; however, the transverse velocity is underpredicted across most of the radial plane. A comparison between the realizable and standard $k-\epsilon$ model predictions in \fig\ref{fig:5b} indicates that both models produce very similar transverse velocity profiles on the primary hole (PH) plane. Since both turbulence models reproduce the sign change in transverse velocity, indicating the presence of swirl, the azimuthal velocity ($w$) component is further examined to confirm the formation of the central vortex core (CVC). The azimuthal velocity ($w$) contours illustrating the predicted CVC are presented in \fig\ref{fig:11b}.
\newline
The predicted turbulent kinetic energy (TKE) using the realizable $k-\epsilon$ model is compared with the experimental data in \fig\ref{fig:5c}. The predicted TKE gradually increases toward the left side of the radial plane and then rises sharply from $r^\ast \approx -0.5$, reaching a peak near $r^\ast \approx -0.175$. Moving toward the centreline, the TKE slightly decreases around $r^\ast \approx 0$ and increases again, forming a symmetric peak at $r^\ast \approx 0.175$. Beyond this point, the TKE decreases sharply until approximately $r^\ast \approx 0.5$ and then gradually approaches zero toward the outer region. 
Compared with the experimental measurements, the predicted TKE is generally underpredicted across the radial plane. However, good agreement is observed at several locations, particularly around $r^\ast \approx -0.3$, $r^\ast \approx 0.10$, $r^\ast \approx 0.365$, and $r^\ast \approx 0.75$. While the overall trend is captured, a discrepancy occurs near the centreline where the experimental TKE exhibits a peak, whereas the prediction shows a slight dip.
Compared with the standard $k-\epsilon$ model, the realizable $k-\epsilon$ model predicts slightly higher TKE within $-0.5 \le r^\ast \le 0.5$, with symmetrical peaks around $r^\ast \approx \pm 0.25$. These differences arise from the modified formulation of the dissipation-rate transport equation (\eqn\ref{realizable-favre-epsilon}) and the variable turbulent viscosity coefficient $C_{\mu}$ (\eqn\ref{eq:Cmu_favre}) used in the realizable $k-\epsilon$ model. The modified $C_{\mu}$ ensures realizability by maintaining the positivity of the normal Reynolds stresses $\widetilde{{u^{\prime\prime}\alpha}^2}$. Since $C_{\mu}$ directly influences the turbulent viscosity ($\mu_t$), differences in $\mu_t$ affect the overall flow field, as shown in \fig\ref{fig:8a}.
The turbulent viscosity predicted by the realizable $k-\epsilon$ model exhibits distinct spatial variations across the radial plane. It initially increases from a small value on the left side, forming a minor peak around $r^\ast \approx -0.75$, reflecting the interaction between primary jets and the swirling flow. Moving inward, $\mu_t$ decreases near $r^\ast \approx -0.5$ and then rises sharply to a pronounced peak around $r^\ast \approx -0.25$, indicating strong turbulence generated by the interaction between the outer swirl and the central vortex core (CVC). Toward the centreline, $\mu_t$ decreases, followed by a primary peak near $r^\ast \approx 0.25$. Further outward, $\mu_t$ decreases around $r^\ast \approx 0.5$, forms a secondary peak near $r^\ast \approx 0.75$, and finally drops toward the right side.  The TKE distribution, therefore, shows a clear spatial correlation with the turbulent viscosity field. Regions with higher $\mu_t$ correspond to enhanced turbulence production and mixing, particularly near the central recirculation zone (CRZ) and the surrounding shear layers.
\newline
The predicted shear stress using the realizable $k-\epsilon$ model is compared with the experimental values in \fig\ref{fig:5d}. The comparison shows a reasonable agreement in terms of overall trend. The predicted shear stress begins with minimum values in the left outer region around $r^\ast \approx -1$ and increases gradually from about $r^\ast \approx -0.8$, forming a small peak near $r^\ast \approx -0.75$. This behavior indicates sensitivity to the shear layer generated by the interaction between the swirling flow and the primary jets. Thereafter, the shear stress decreases slightly until approximately $r^\ast \approx -0.6$, after which it increases again and forms an intermediate peak around $r^\ast \approx -0.4$, where it closely matches the experimental value. In the region $-0.5 \le r^\ast \le 0$, the experimental shear stress attains locally high values, peaking near $r^\ast \approx -0.285$, which is associated with the shear layer formed due to the interaction between the swirling flow and the central vortex core (CVC). The realizable $k-\epsilon$ model captures the trend in this region but underpredicts the magnitude. Compared with the standard $k-\epsilon$ model, the realizable model predicts slightly higher shear stress values within this range. Moving further inward, the experimental shear stress changes sign near the centreline ($r^\ast \approx 0$), becoming negative and reaching a minimum near $r^\ast \approx 0.25$. It then gradually recovers and becomes positive again around $r^\ast \approx 0.5$, indicating the opposite side of the inner shear layer generated by the interaction of swirl and the CVC. The realizable $k-\epsilon$ model captures this overall behaviour; however, the sign change occurs slightly closer to the centreline. This shift can be attributed to the displacement of turbulent viscosity peaks predicted by the model, which influences the radial distribution of shear stress. Furthermore, the isotropic eddy-viscosity assumption limits the model's ability to fully represent the anisotropic shear layers produced by the swirl-CVC interaction. Beyond this region, the predicted shear stress forms a local peak near $r^\ast \approx 0.45$, consistent with the experimental trend in the outer shear layer region, and then gradually decreases toward a minimum value near $r^\ast \approx 1$.
\newline
On comparing the predicted \added{normalized} mean axial velocity with the experimental values across the dilution hole (DH) plane in \fig\ref{fig:5a}, it can be observed that the realizable $k-\epsilon$ model overpredicts the mean axial velocity across the radial plane. The model predicts a higher centreline velocity and steeper velocity gradients in the shear-layer region, and fails to reproduce the experimentally observed flattened plateau in the right-side shear layer region $(0.15 \le r^\ast \le 0.5)$. A comparison with the predictions obtained using the standard $k-\epsilon$ model indicates a slight improvement with the realizable $k-\epsilon$ model, which can be attributed to the modified model formulation. However, inherent limitations of the realizable $k-\epsilon$ model, particularly the isotropic eddy-viscosity assumption, restrict its ability to accurately capture the anisotropic shear-layer behaviour present in this flow field at the dilution hole (DH) plane.
\newline
The predicted \added{normalized} transverse velocity using the realizable $k-\epsilon$ model across the dilution hole (DH) plane is compared with the experimental values in \fig\ref{fig:5b}. The model captures the overall trend, including the sign change across the radial plane, with positive values on the left side, near-zero values around the centreline, and negative values on the right side. Comparison with the standard $k-\epsilon$ model shows very similar transverse velocity profiles, with only minor differences arising from variations in model formulation. The predicted turbulent kinetic energy (TKE) across the dilution hole (DH) plane is shown in \fig\ref{fig:5c}. In the region ($r^\ast \approx \pm 0.5$) where the dilution jets interact with the swirling flow, the TKE is significantly overpredicted, consistent with the elevated turbulent viscosity predicted in the same region (\fig\ref{fig:8b}). 
The overall predicted TKE distribution does not match the experimental profile, where the peak occurs on the opposite side around $r^\ast \approx 0.25$. 
\newline
Similar to the standard $k-\epsilon$ model, the realizable $k-\epsilon$ model fails to reproduce the turbulence \deleted{intensity} \added{levels} generated by the strong shear arising from the interaction between confined swirl and dilution jets. The predicted shear stress also shows notable discrepancies with the experimental data. The profile begins with negative values on the left side ($r^\ast$), reaching a minimum near $r^\ast \approx -0.75$, which is consistent with the peaks in turbulent viscosity (see \fig\ref{fig:8b}) corresponding to the region of strong dilution jet-swirl interaction. Moving inward, the shear stress becomes positive around $r^\ast \approx -0.65$ and forms a plateau near $r^\ast \approx -0.7$. The model significantly under-predicts the positive shear stress in the range $-0.8 \le r^\ast \le 0$ and incorrectly predicts positive shear in the region $0 < r^\ast < 0.5$, where the experimental data indicate negative values. These discrepancies highlight the limitations of the realizable $k-\epsilon$ model in accurately capturing the shear-layer dynamics associated with swirl–jet interaction.
\subsubsection{SST $k-\omega$ model predictions}
The comparison of \added{normalized} mean axial velocity ($u^\ast$) predicted using using the SST $k-\omega$ model with the experimental values on the primary holes (PH) plane is presented in \fig\ref{fig:5a}. The predictions are consistent with the experimental values and, when compared with those obtained using the standard and realizable $k-\epsilon$ models, show better agreement with the experiment. The negative mean axial velocity values observed in the region $-0.5 \leq r^\ast \leq 0.5$ indicate the presence of a central recirculation zone (CRZ), which is stronger CRZ compared to the both $k-\epsilon$ models. The predicted CRZ is illustrated using the mean axial velocity contours in \figs\ref{fig:9c} and \ref{fig:10c}.
\newline
The predicted \added{normalized} mean transverse velocity ($v^\ast$) using the SST $k-\omega$ model on the primary holes (PH) plane is compared with the experimental data in \fig\ref{fig:5b}, showing overall good agreement. When compared with the predictions of the standard and realizable $k-\epsilon$ models and the experimental values of transverse velocity in \fig\ref{fig:5b}, the SST $k-\omega$ model provides better overall agreement with the experimental data. The counter-rotating vortex core (CVC) is predicted and mapped using the azimuthal component of velocity ($\tilde{w}$) in \fig\ref{fig:11c}.
\newline
The turbulent kinetic energy (TKE) predicted using the SST $k-\omega$ model on the primary holes (PH) plane is presented in \fig\ref{fig:5c}. Although the TKE is generally under-predicted across the plane, except at positions $-0.25 \lesssim r^\ast \lesssim -0.10$ and $0.10 \lesssim r^\ast \lesssim 0.35$, the predicted TKE does not attain a peak near the central position ($r^\ast \approx 0$) as observed in the experimental data. Nevertheless, the TKE predicted using the SST $k-\omega$ model follows the experimental trend more closely than the predictions obtained using the standard and realizable $k-\epsilon$ models. As aforementioned, in two-equation turbulence models, the turbulent viscosity strongly governs both TKE and the overall turbulence behaviour. By examining the TKE profile alongside the turbulent viscosity distribution of the SST $k-\omega$ model, it is observed that the spatial variation of TKE closely follows the turbulent viscosity distribution. The TKE predicted by the SST $k-\omega$ model is higher in the central region ($-0.5 \lesssim r^\ast \lesssim 0.5$) compared with the predictions from the standard and realizable $k-\epsilon$ models. Consequently, both standard and realizable $k$–$\epsilon$ models predict higher turbulent viscosity in this region, which corresponds to lower turbulence levels and weaker mixing. The relatively lower turbulent viscosity predicted by the SST $k-\omega$ model can be attributed to its turbulent viscosity formulation. The turbulent viscosity expression (\eqn\ref{turbulent-viscosity-SST-k-omega}) in the SST $k-\omega$ model includes a strain-dependent term ($\tilde{S}_{ij}$) and a blending function ($F_2$), which acts as a viscosity limiter \cite{Menter1994}. By limiting the turbulent viscosity in shear-dominated regions, the model prevents excessive turbulence dissipation. This mechanism allows higher turbulence production to be sustained, ultimately leading to an increased level of TKE prediction in these regions. The predicted shear stress is compared with the experimental values in \fig\ref{fig:5d}. The predictions follow the experimental trend, including the alternating positive and negative shear stress distribution and the major sign change around the central position observed experimentally. However, the model under-predicts both the positive and negative peak magnitudes near $r^\ast \approx \pm 0.3$. This under-prediction occurs partly due to model formulation and the use of Favre averaging, which smooths turbulent fluctuations and suppresses the peak shear stress values.
\newline 
The \added{normalized mean} axial velocity predicted using the SST $k-\omega$ model on the dilution holes (DH) plane is shown in \fig\ref{fig:6a}. The axial velocity is generally over-predicted across the plane; however, the agreement with experimental data is still better than that obtained using the standard and realizable $k-\epsilon$ models. Similar to the $k-\epsilon$ models, the SST $k-\omega$ model overestimates the axial velocity in the core region $-0.5 \lesssim r^\ast \lesssim 0.5$ and fails to capture the experimentally observed flattened plateau in the rightward shear layer.
The predicted \added{normalized mean} transverse velocity using the SST $k-\omega$ model is compared with experimental values in \fig\ref{fig:6b}, showing overall agreement. The predicted TKE is compared with experimental data in \fig\ref{fig:6c}. Although the overall shape of the TKE distribution is reasonably captured, the dominant peak location is shifted: the experimental peak at approximately $r^\ast \approx -0.3$ is predicted on the right side at about $r^\ast \approx 0.2$.
The predicted TKE is also compared with the turbulent viscosity distribution of the SST $k-\omega$ model (see \fig\ref{fig:8b}). The TKE generally follows the spatial trend of turbulent viscosity; however, the TKE peak occurs near $r^\ast \approx 0.2$, while the turbulent viscosity peak appears around $r^\ast \approx -0.3$. The relatively lower turbulent viscosity limits excessive dissipation and sustains turbulence production in this region. Nevertheless, due to the isotropic eddy-viscosity formulation, the model does not accurately distribute turbulence generation and dissipation across the plane. Despite these limitations, the SST $k-\omega$ model predicts TKE more accurately than the standard and realizable $k-\epsilon$ models (see \fig\ref{fig:6c}), particularly near the outer shear layer ($r^\ast \approx \pm 0.5$), which forms due to the interaction of swirl with the dilution jets. This indicates improved handling of adverse pressure gradients and shear-layer effects by the SST $k-\omega$ model.
The predicted shear stress using the SST $k-\omega$ model on the dilution holes (DH) plane is shown in \fig\ref{fig:6d}. The shear stress is generally under-predicted and fails to reproduce the sign reversal near the centre observed experimentally. On the left side ($-1 \lesssim r^\ast \lesssim 0$), the shear stress is significantly under-predicted, while on the right side ($0 \lesssim r^\ast \lesssim 1$) the predicted values become positive, whereas the experimental values remain negative up to approximately $r^\ast \approx 0.4$. As discussed earlier, the SST $k-\omega$ model relies on an isotropic eddy-viscosity assumption and therefore cannot fully capture the highly anisotropic turbulence and rapid strain-rate variations generated by the interaction of swirl and dilution jets.
\subsubsection{LPS-RSM predictions}
The LPS-RSM predictions for \added{normalized} mean axial velocity are compared with experimental data in \fig\ref{fig:5a}. The LPS-RSM model captures the overall axial velocity trend, reproducing high axial velocity in the shear layer region ($|r^\ast| \gtrsim 0.5$). However, in the central region ($|r^\ast| \lesssim 0.25$) the model departs from the measurements, predicting a pronounced axial velocity peak that indicates excessive re-acceleration of the axial flow and consequently a weaker recirculation region. This behaviour is illustrated in \figs\ref{fig:9d} and \ref{fig:10d}. Overall, the LPS-RSM captures the trend better than the standard and realizable $k-\epsilon$ models, although the SST $k-\omega$ model shows closer agreement with the experimental data.
The improved performance of the SST $k-\omega$ model can be attributed to its blended $k-\epsilon$ and $k-\omega$ formulation, which better handles adverse pressure gradients and flow separation, leading to improved prediction of swirl-induced recirculation and shear layer dynamics. Although the LPS-RSM accounts for turbulence anisotropy, it \deleted{does} \added{did} not capture the anisotropic behaviour accurately in the shear-layer interaction zone, resulting in excessive core re-acceleration and weaker shear layer prediction compared with the experiment \added{in the present case}.
Fairly good agreement with experimental data is obtained for the \added{normalized mean} transverse velocity (see \fig\ref{fig:5b}), particularly within the shear layer ($-0.5 \lesssim r^\ast \lesssim 0.5$). However, the transverse velocity magnitude is slightly over-predicted in the central region ($-0.10 \lesssim r^\ast \lesssim 0.10$). While the SST $k-\omega$ model remains closest to the experimental measurements, the LPS-RSM predictions are still more accurate than those obtained using the standard and realizable $k-\epsilon$ models. The \deleted{counter-rotating} \added{central} vortex core (CVC) predicted using the LPS-RSM model is identified using the azimuthal velocity component ($\tilde{w}$) in \fig\ref{fig:11d}.
\newline
\fig\ref{fig:5c} shows that TKE is significantly under-predicted across the primary holes (PH) plane. Although the spatial trend is qualitatively captured, the magnitude of TKE is considerably lower than the experimental measurements and less accurate than the predictions obtained using the standard $k-\epsilon$, the realizable $k-\epsilon$, and the SST $k-\omega$ models. In the LPS-RSM formulation, the overall pressure-strain term ($\tilde{\phi}_{ij}$, \eqn\ref{lps-pressure-strain}) is expressed as the sum of the slow and rapid pressure-strain terms (\eqns\ref{slow-pressure-strain} and \ref{rapid-pressure-strain}) and the wall-reflection contribution, where the slow and rapid terms redistribute TKE among the Reynolds stress components. The slow pressure-strain term ($\tilde{\phi}_{ij,1}$) contains the Reynolds stress anisotropy tensor ($\tilde{b}{ij}$, \eqn\ref{anisotropy-tensor}) and represents the tendency of the Reynolds stress tensor to relax toward isotropy by redistributing energy from components with higher TKE to those with lower TKE. The rapid pressure-strain term ($\tilde{\phi}_{ij,2}$) represents the interaction between pressure fluctuations and mean velocity gradients and accounts for the rapid distortion of turbulence. Although symmetric, its unequal components can sustain anisotropy. Since these pressure-strain terms primarily redistribute rather than generate turbulent energy, they limit the buildup of TKE in regions with strong swirl and shear. Consequently, the production term ($\tilde{P}_{ij}$ in \eqn\ref{rsm-equation}) does not fully convert mean flow gradients into TKE, as continuous redistribution among the normal stress components counterbalances turbulence production, resulting in lower predicted TKE compared with two-equation models.
\newline 
The predicted shear stress using LPS-RSM is compared with experimental values in \fig\ref{fig:5d}. Although LPS-RSM captures the overall experimental shear stress profile, its magnitude is significantly under-predicted and remains inferior to the predictions of the standard and realizable $k-\epsilon$ models and the SST $k-\omega$ model. The discrepancy between experimental values and LPS-RSM prediction of shear stress can be attributed to the strong action of the pressure-strain correlation ($\tilde{\phi}{ij}$), which redistributes the Reynolds stresses and suppresses the shear component $-\widetilde{u^{\prime\prime}v^{\prime\prime}}$. This suppression reduces the turbulent production ($\tilde{P}{ij}$) and weakens shear layer development. Although LPS-RSM does not primarily rely on the eddy-viscosity hypothesis, the reduced TKE production leads to a lower modeled effective turbulent viscosity based on \eqn\eqref{turbulent-viscosity}. The lower viscosity weakens the diffusion term ($\tilde{D}_{T,ij}$, \eqn\ref{rsm-equation}), which governs the redistribution and mixing of shear stress across the shear layer. Consequently, less momentum is transported by turbulence, resulting in a weaker shear layer and reduced shear stress magnitude compared with both the experimental data and the predictions of the standard and realizable $k-\epsilon$ and SST $k-\omega$ models.
\newline 
The predicted \added{normalized} mean axial velocity at dilution holes (DH) plane using LPS-RSM is shown in \fig\ref{fig:6a}. It is over-predicted across the plane, with pronounced over-prediction near the center ($r/R_c \approx \pm 0.25$). This behavior is attributed to the pressure-strain terms ($\tilde{\phi}{ij,1}$ and $\tilde{\phi}{ij,2}$), as discussed above, which redistribute Reynolds stresses and suppress shear components, thereby reducing turbulent production and inhibiting shear layer growth and mixing. Consequently, outward transport of axial momentum is limited, leading to excessive core acceleration and over-prediction of axial velocity. In contrast, the SST $k-\omega$ model predicts the axial velocity more accurately due to its shear-sensitive production term ($P_k$) and turbulent viscosity formulation (\eqn\ref{turbulent-viscosity-SST-k-omega}), which enhance turbulence generation in high-strain regions and improve momentum transport across the shear layer. The \added{normalized} mean transverse velocity is also overpredicted near the center ($r/R_c \approx \pm 0.25$) by LPS-RSM (\fig\ref{fig:6b}) for the same reasons.
\newline
The TKE predicted by LPS-RSM in \fig\ref{fig:6c} is significantly under-predicted across the dilution holes (DH) plane compared to the standard and realizable $k-\epsilon$ and SST $k-\omega$ models. This deficiency is primarily due to the pressure-strain correlation ($\tilde{\phi}{ij}$), which redistributes energy and controls anisotropy but appears to over-suppress shear stress components. This is consistent with the strong under-prediction of shear stress in \fig\ref{fig:6d}. Since TKE production ($\tilde{P}{ij}$, \eqn\ref{rsm-equation}) is largely driven by shear stress, its suppression leads to a significant reduction in TKE. Consequently, the flow lacks sufficient turbulent energy for realistic transport, resulting in a globally reduced TKE magnitude.
\subsection{Scalar Field Analysis}
%
%\added[id=AK]{The authors also examined the impacts of turbulence models on the scalar fields such as temperature distribution, species concentration ($C_3 H_8$, $CO_2$, $CO$), mixture fraction, and progress variables. The temperature profile on the primary hole plane predicted using the standard $k-\epsilon$ model is shown in \fig\ref{fig:12a}. The temperature increases sharply from the ambient value (318$K$) leftward side $r/R_c =-1$, to peak approx $r/R_c =-0.55$, from which it sharply decreases and starts to form a plateau at $r/R_c =-0.35$, which continues till $r/R_c =0.35$ from where it symmetrically rises again, peaking about $r/R_c =0.55$ and sharply decreases to ambient value (318$K$) at rightward side $r/R_c =1$. }
%
\begin{figure}[!tb]
	\centering
	\begin{subfigure}[t]{0.48\linewidth}
		\centering
		\includegraphics[width=\linewidth]{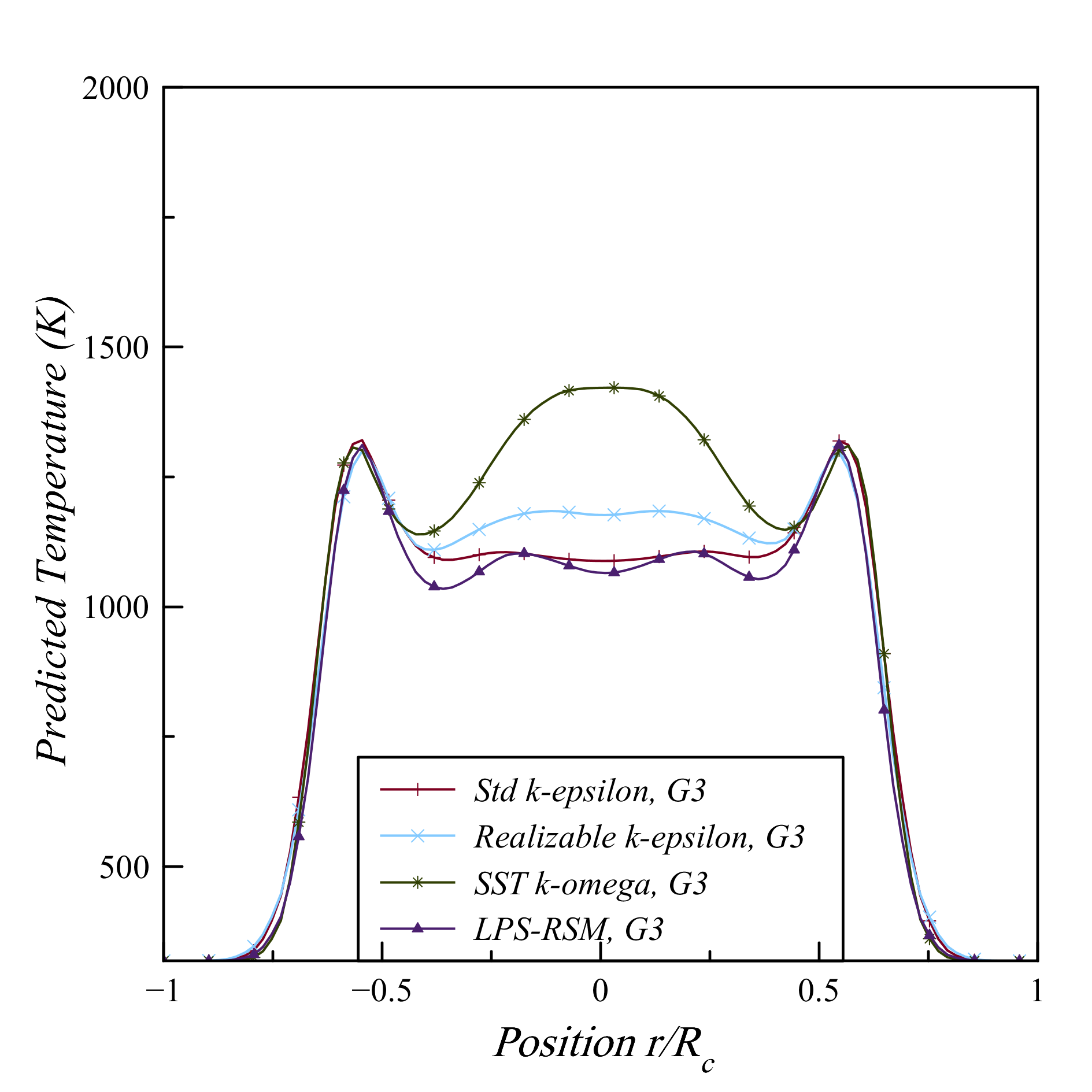}
		\caption{At $x = 50$ mm on the primary holes plane}
		\label{fig:12a}
	\end{subfigure}
	\hfill
	\begin{subfigure}[t]{0.48\linewidth}
		\centering
		\includegraphics[width=\linewidth]{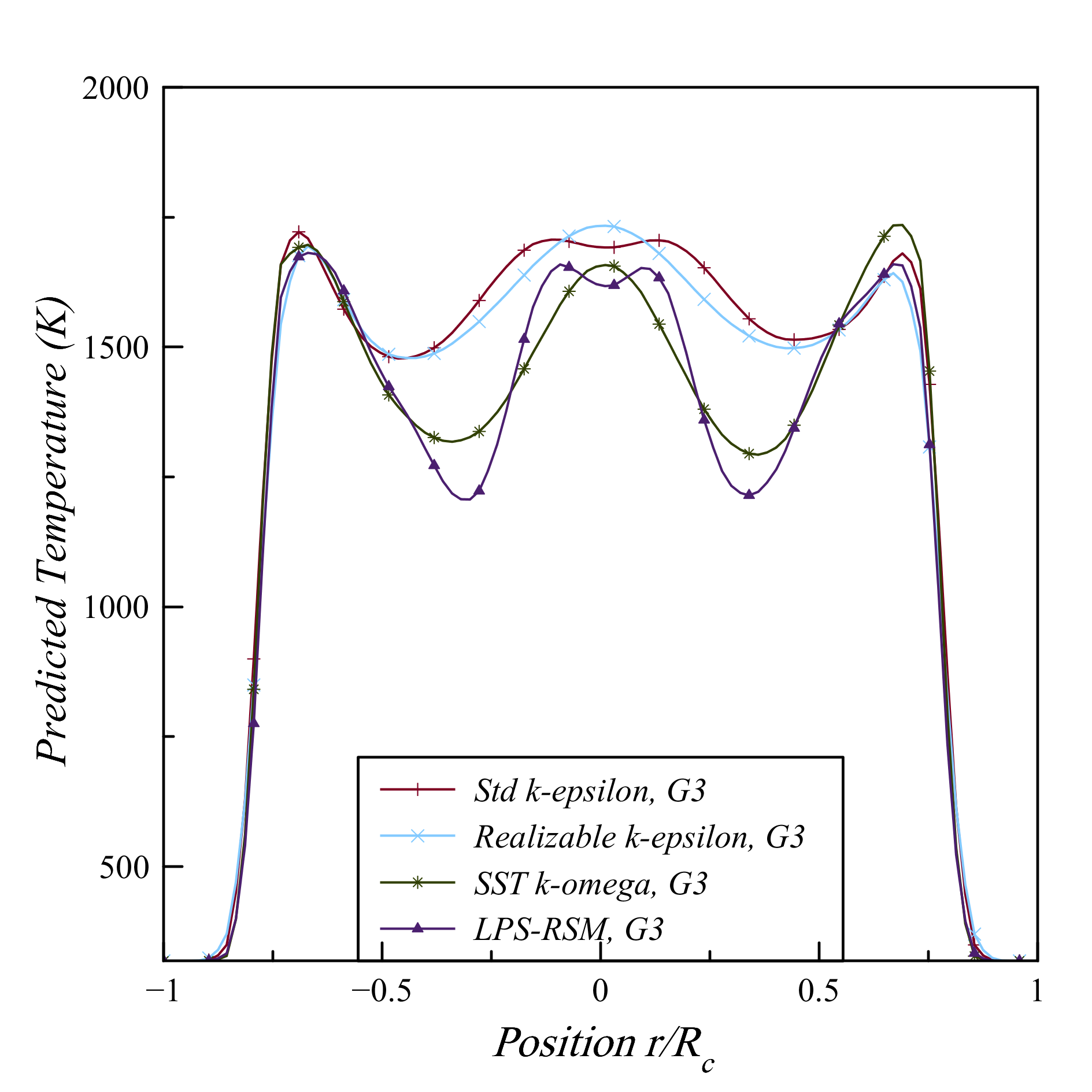}
		\caption{At $x = 130$ mm on the dilution holes plane}
		\label{fig:12b}
	\end{subfigure}
	\vspace{3mm}
	\caption{Predicted Favre-averaged temperature, $\tilde{T} = (\overline{\rho T}/\overline{\rho})$ (\si{K}), on primary and dilution holes plane at \deleted{ at} the reacting conditions (refer \tab \ref{tab:1}). }
	\label{fig:12}
\end{figure}

\begin{figure}[!tb]
	\centering
	\begin{subfigure}[t]{0.48\linewidth}
		\centering
		\includegraphics[width=\linewidth]{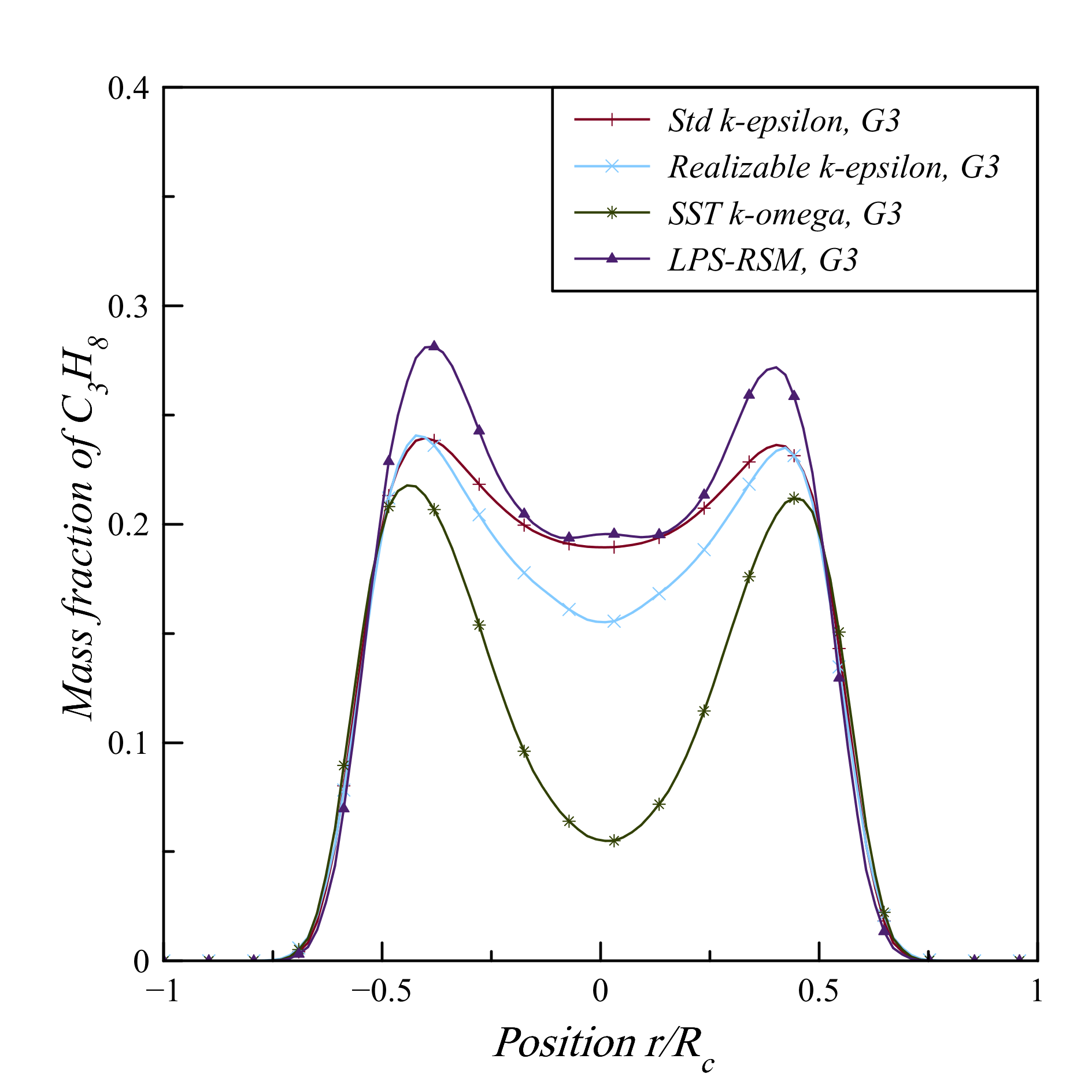}
		\caption{At $x = 50$ mm on the primary holes plane.}
		\label{fig:13a}
	\end{subfigure}
	\hfill
	\begin{subfigure}[t]{0.48\linewidth}
		\centering
		\includegraphics[width=\linewidth]{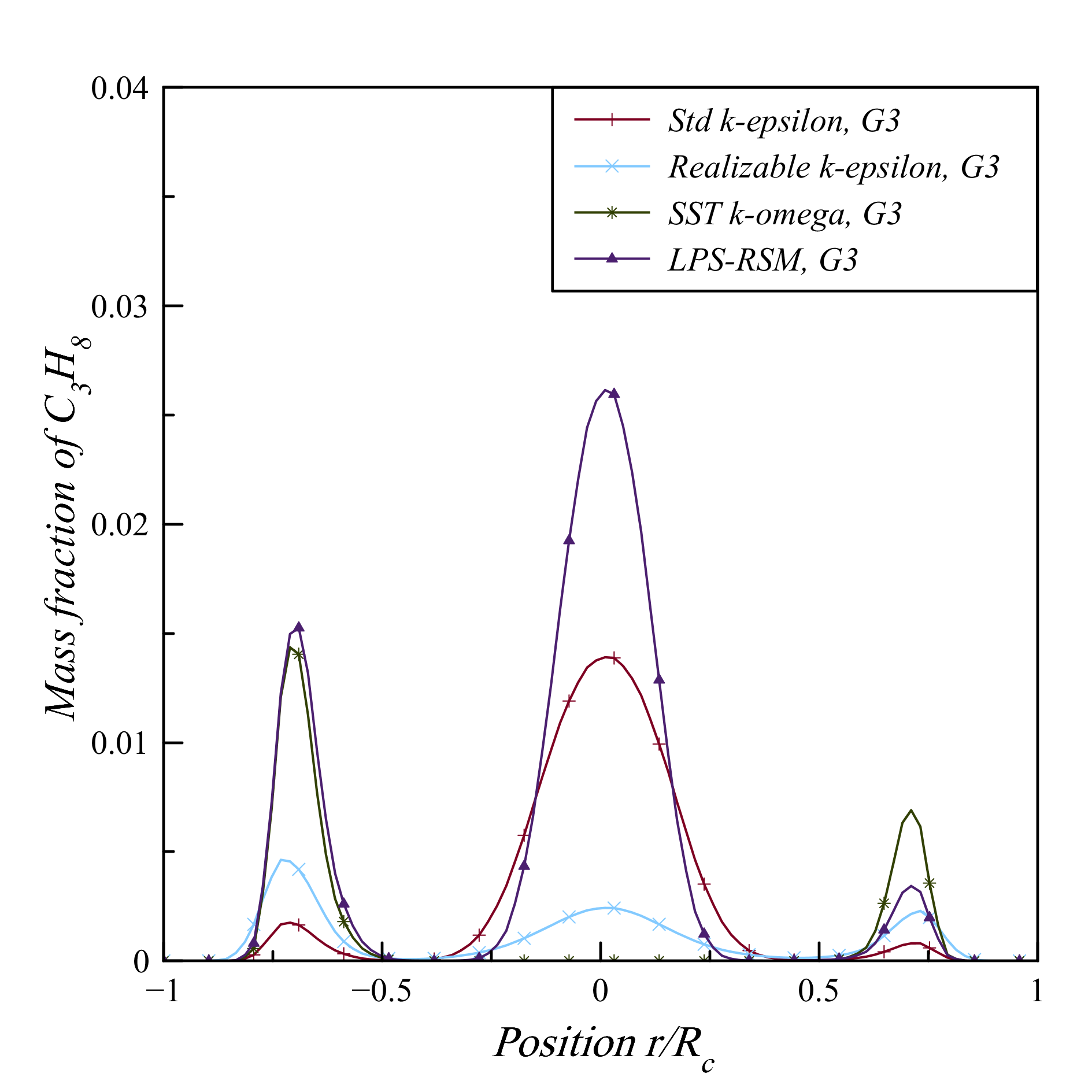}
		\caption{At $x = 130$ mm on the dilution holes plane.}
		\label{fig:13b}
	\end{subfigure}
	\vspace{3mm}
	
	\caption{Predicted Favre-averaged mass fraction of ${C_3H_8}$ ($\tilde{Y}_{C_3H_8} = {\overline{\rho Y_{C_3H_8}}}/{\overline{\rho}}$), on the primary and dilution holes planes under reacting conditions (refer \tab~\ref{tab:1}).}
	\label{fig:13}
\end{figure}

\begin{figure}[!tb]
	\centering
	\begin{subfigure}[t]{0.48\linewidth}
		\centering
		\includegraphics[width=\linewidth]{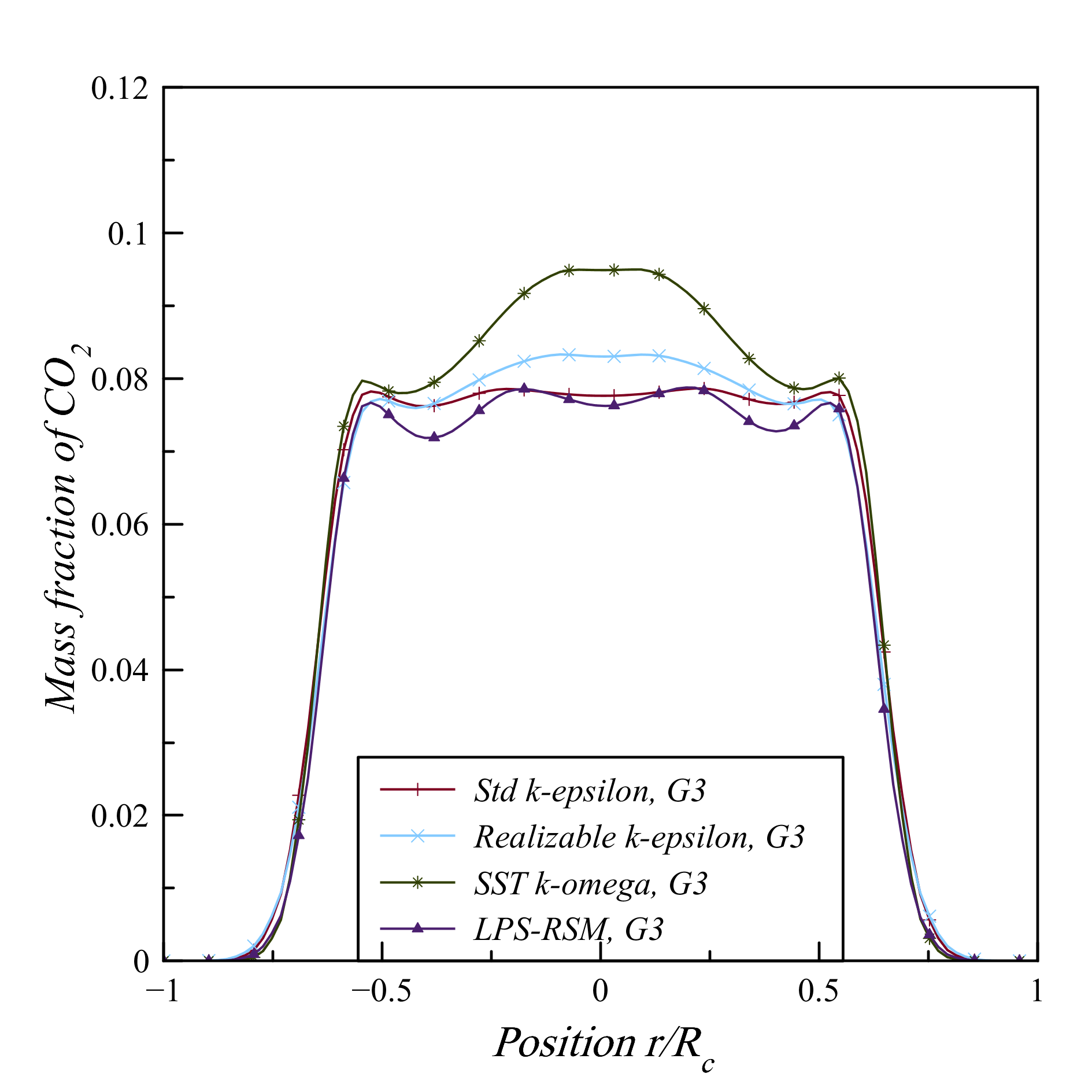}
		\caption{At $x = 50$ mm on the primary holes plane.}
		\label{fig:14a}
	\end{subfigure}
	\hfill
	\begin{subfigure}[t]{0.48\linewidth}
		\centering
		\includegraphics[width=\linewidth]{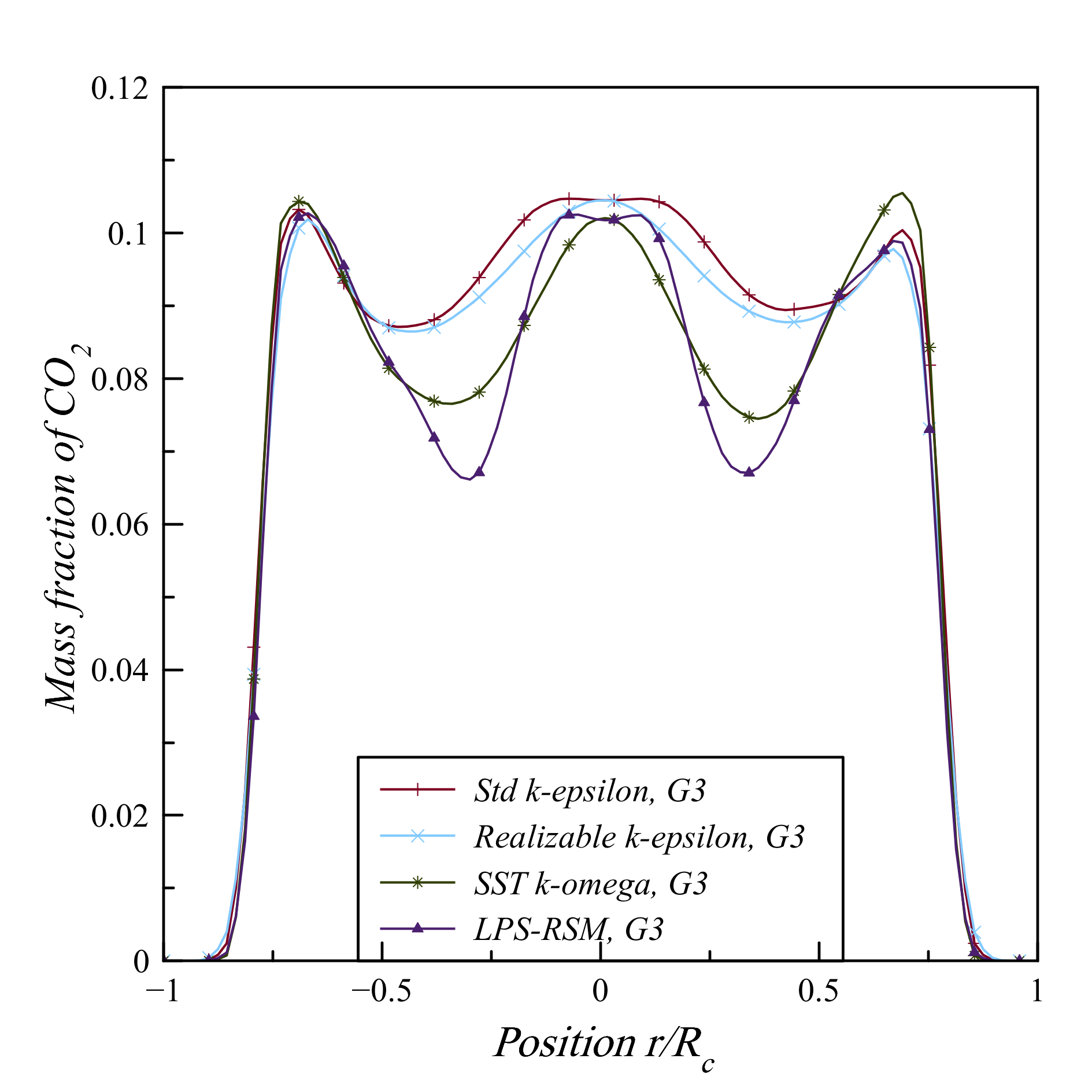}
		\caption{At $x = 130$ mm on the dilution holes plane.}
		\label{fig:14b}
	\end{subfigure}
	\vspace{3mm}
	
	\caption{Predicted Favre-averaged mass fraction of ${CO_2}$ ($\tilde{Y}_{CO_2} = {\overline{\rho Y_{CO_2}}}/{\overline{\rho}}$) on the primary and dilution holes planes under reacting conditions (refer \tab~\ref{tab:1}).}
	\label{fig:14}
\end{figure}

\begin{figure}[!tb]
	\centering
	\begin{subfigure}[t]{0.48\linewidth}
		\centering
		\includegraphics[width=\linewidth]{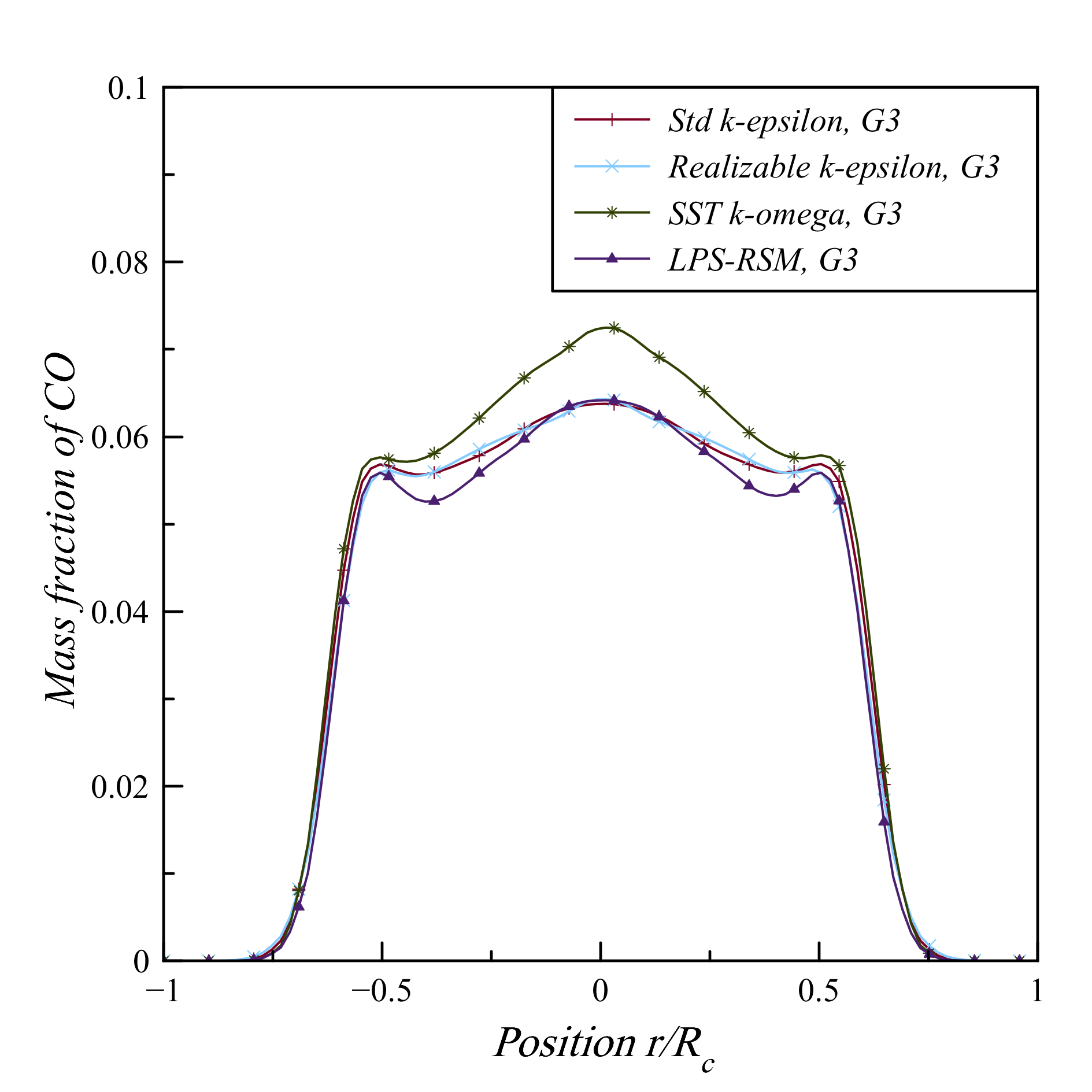}
		\caption{At $x = 50$ mm on the primary holes plane.}
		\label{fig:15a}
	\end{subfigure}
	\hfill
	\begin{subfigure}[t]{0.48\linewidth}
		\centering
		\includegraphics[width=\linewidth]{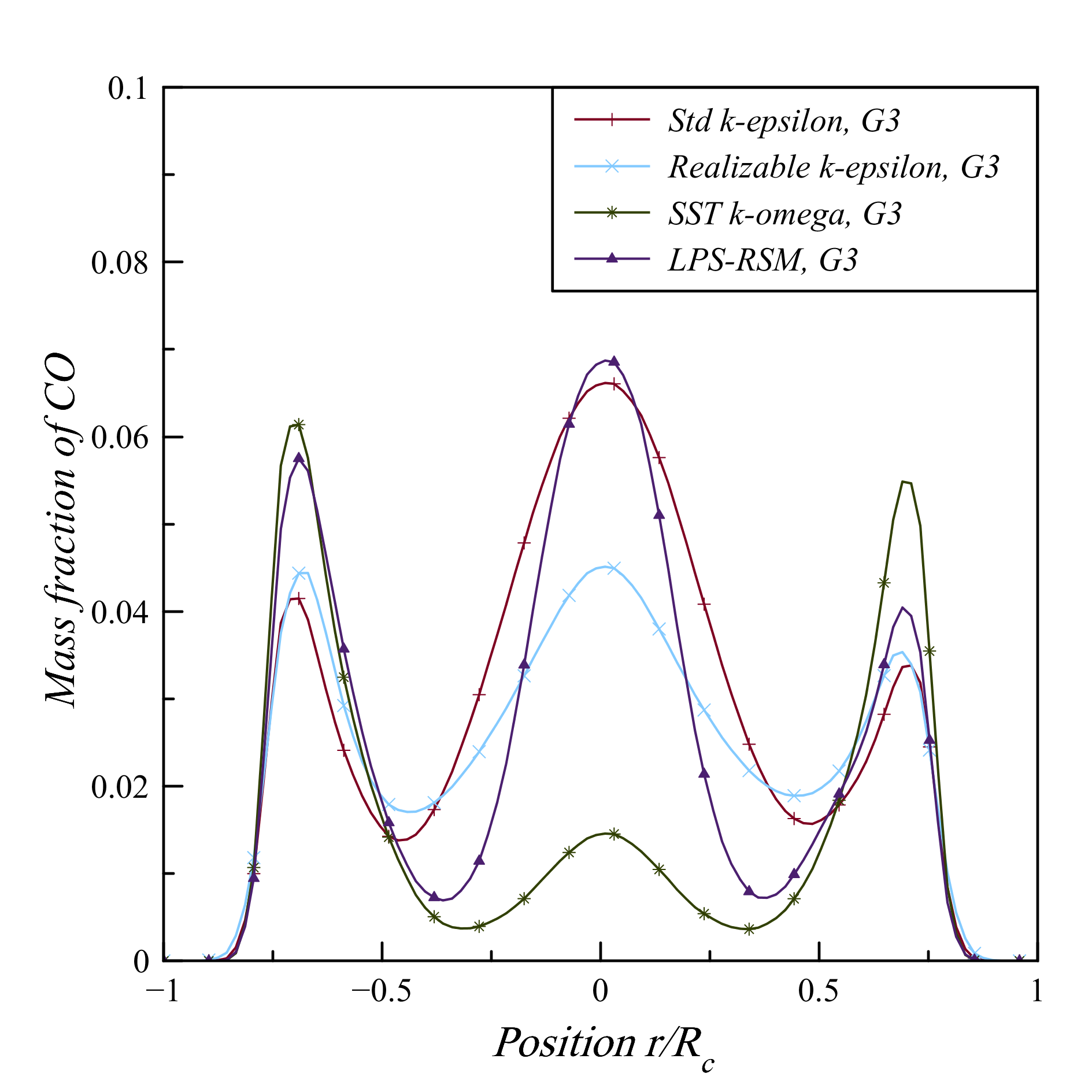}
		\caption{At $x = 130$ mm on the dilution holes plane.}
		\label{fig:15b}
	\end{subfigure}
	\vspace{3mm}
	
	\caption{Predicted Favre-averaged mass fraction of ${CO}$ ($\tilde{Y}_{CO} = {\overline{\rho Y_{CO}}}/{\overline{\rho}}$) on the primary and dilution holes planes under reacting conditions (refer \tab~\ref{tab:1}).}
	\label{fig:15}
\end{figure}

\begin{figure}[!tb]
	\centering
	\begin{subfigure}[t]{0.48\linewidth}
		\centering
		\includegraphics[width=\linewidth]{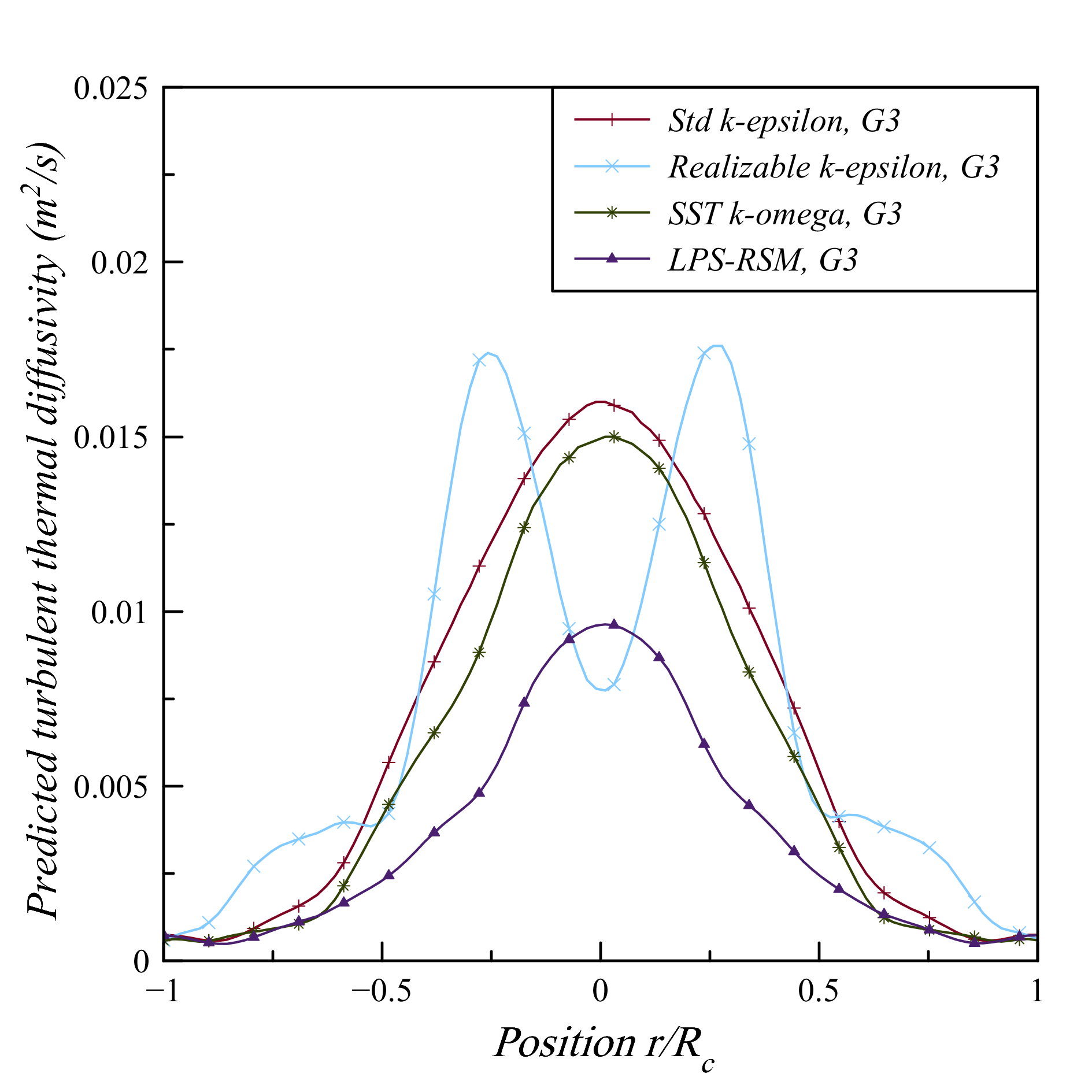}
		\caption{At $x = 50$ mm on the primary holes plane.}
		\label{fig:16a}
	\end{subfigure}
	\hfill
	\begin{subfigure}[t]{0.48\linewidth}
		\centering
		\includegraphics[width=\linewidth]{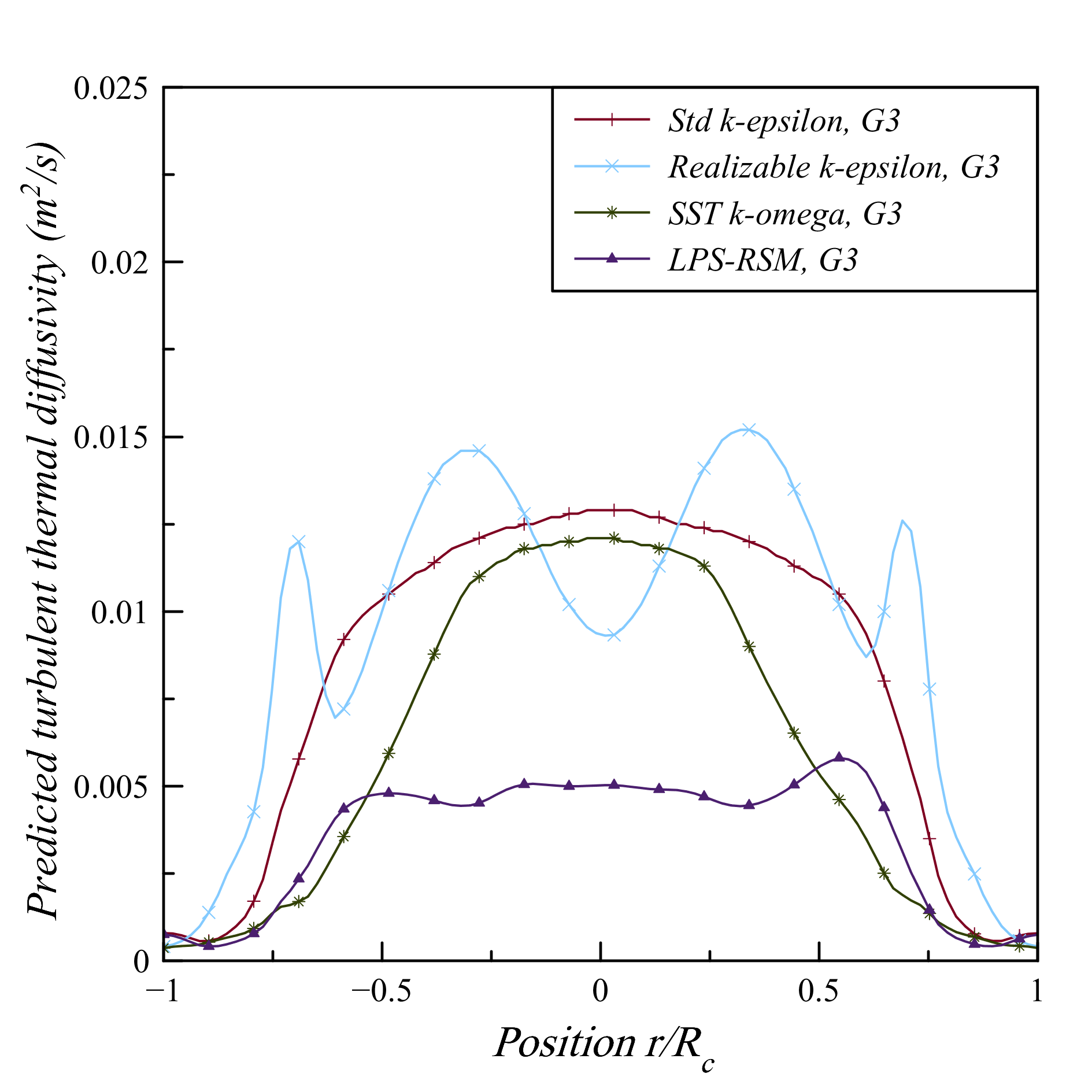}
		\caption{At $x = 130$ mm on the dilution holes plane.}
		\label{fig:16b}
	\end{subfigure}
	\vspace{3mm}
	
	\caption{Predicted turbulent thermal diffusivity ($\alpha_t = {\mu_t}/{(\bar{\rho} \, Pr_t)}$, \si{\meter\squared\per\second}) on the primary and dilution holes planes under reacting conditions (refer \tab~\ref{tab:1}).}
	\label{fig:16}
\end{figure}
\begin{figure}[!b]
	\centering
	\begin{subfigure}{0.48\linewidth}
		\includegraphics[width=\linewidth]{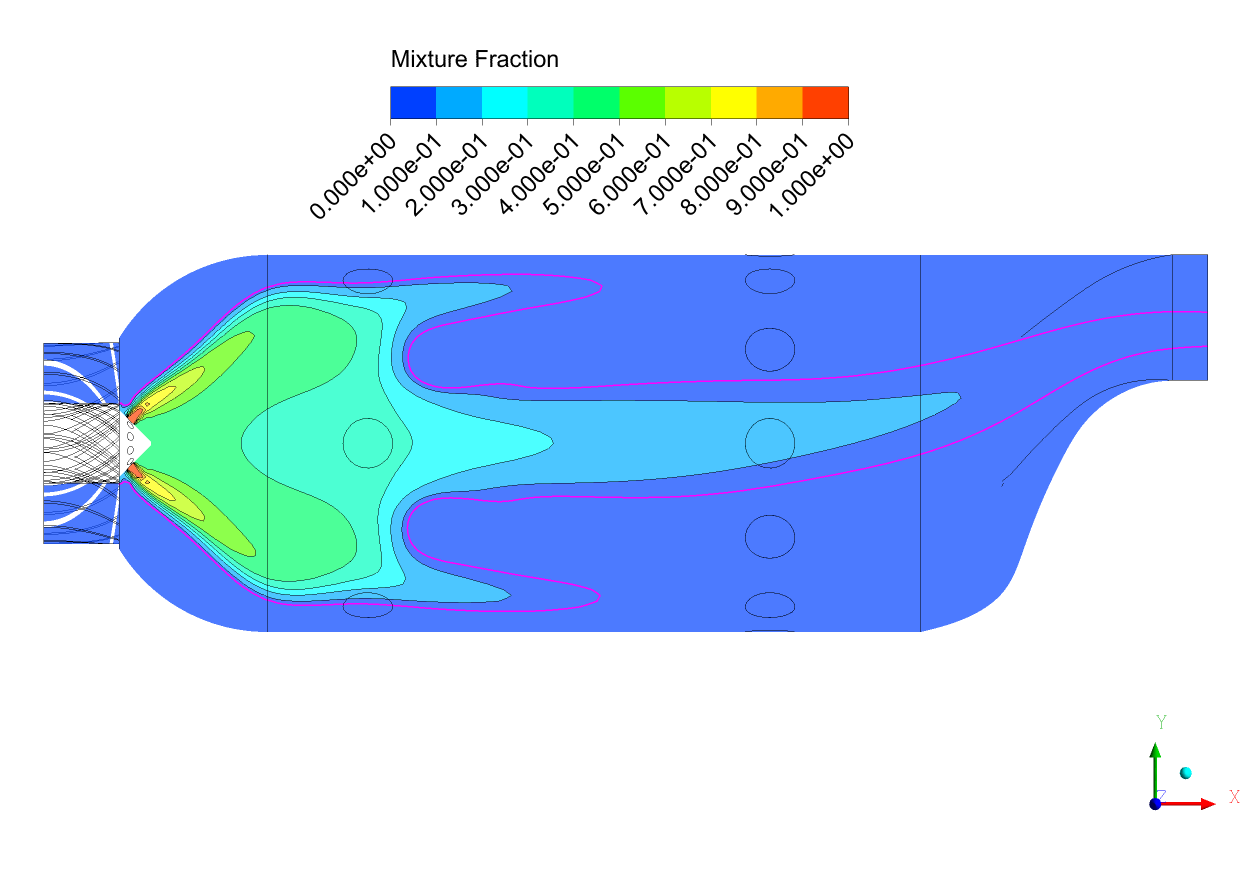}
		\caption{Standard $k$–$\epsilon$}
		\label{fig:18a}
	\end{subfigure}
	\begin{subfigure}{0.48\linewidth}
		\includegraphics[width=\linewidth]{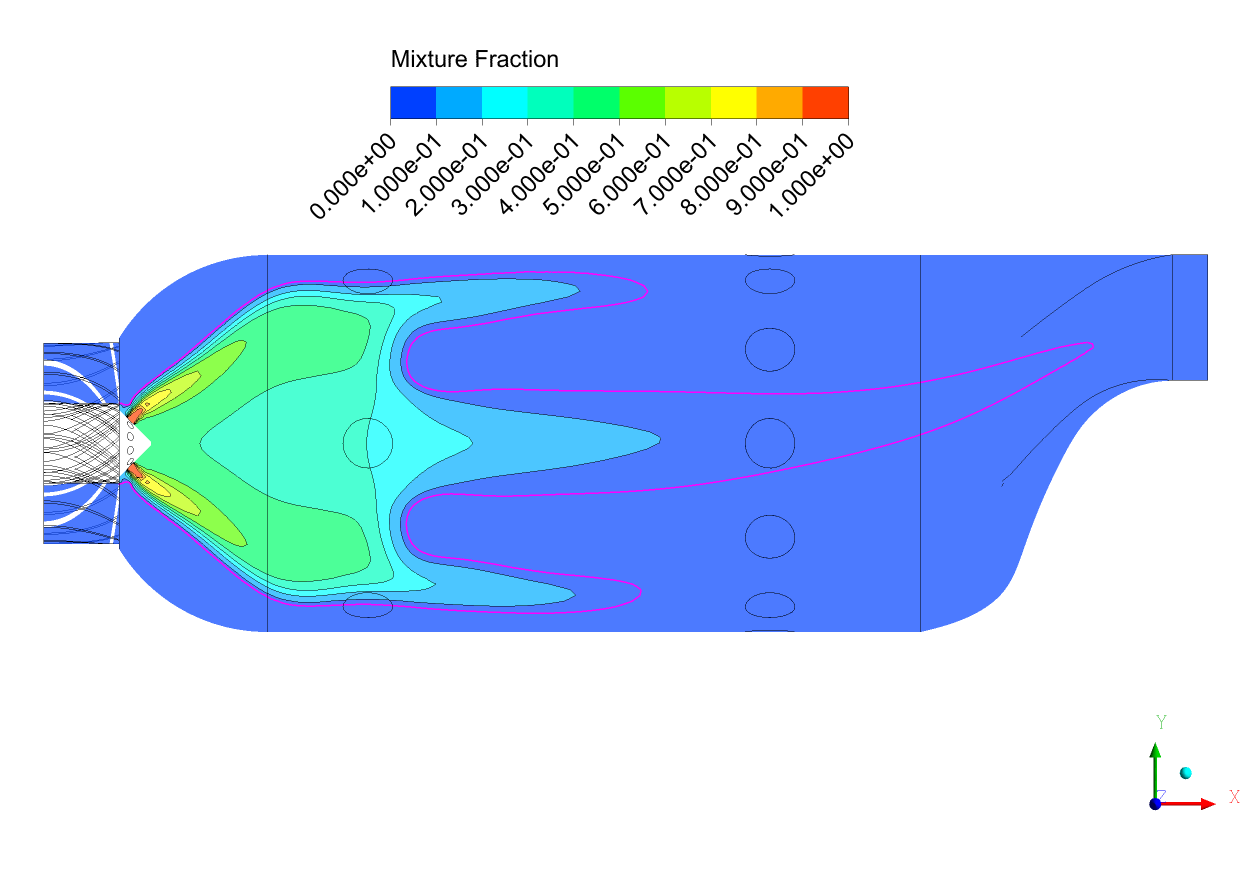}
		\caption{Realizable $k$–$\epsilon$}
		\label{fig:18b}
	\end{subfigure}
	\begin{subfigure}{0.48\linewidth}
		\includegraphics[width=\linewidth]{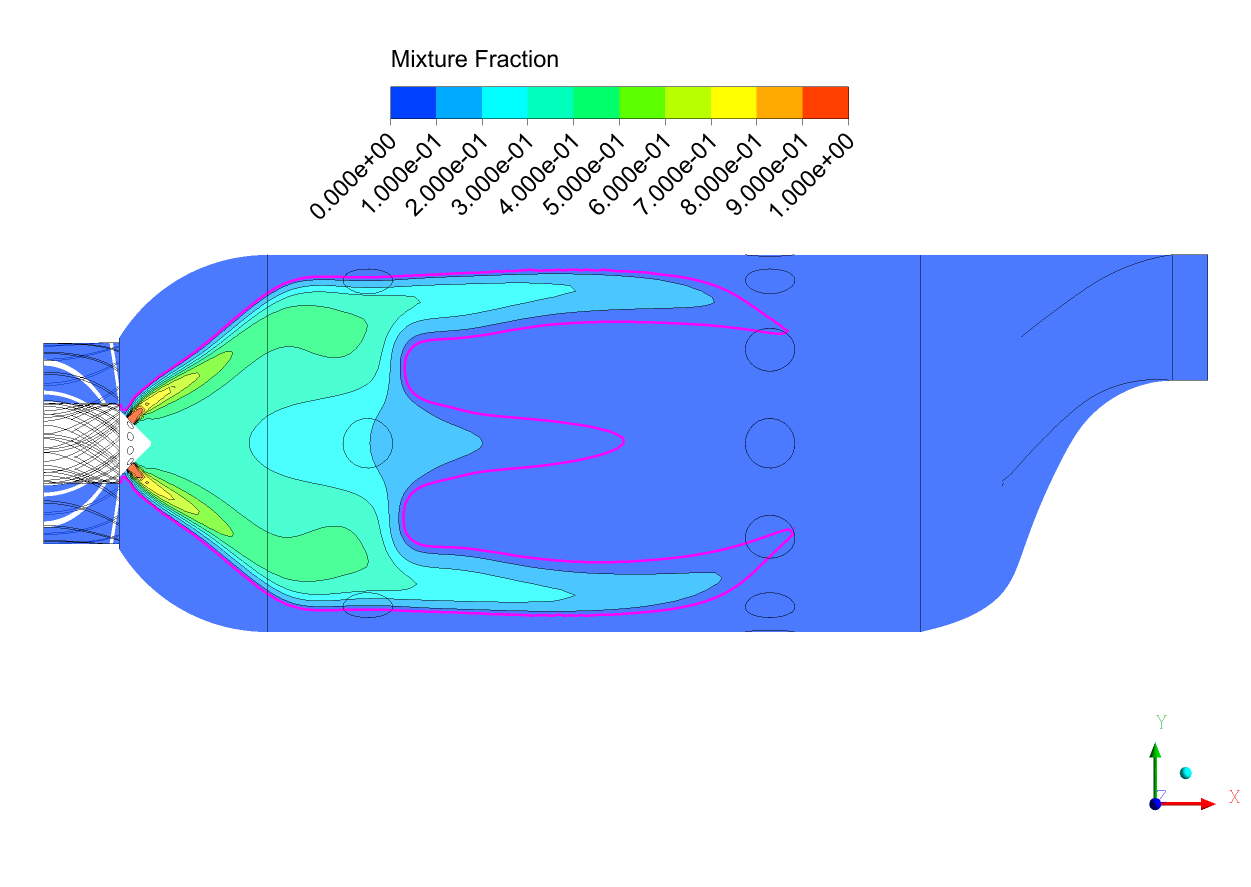}
		\caption{SST $k$–$\omega$}
		\label{fig:18c}
	\end{subfigure}
	\begin{subfigure}{0.48\linewidth}
		\includegraphics[width=\linewidth]{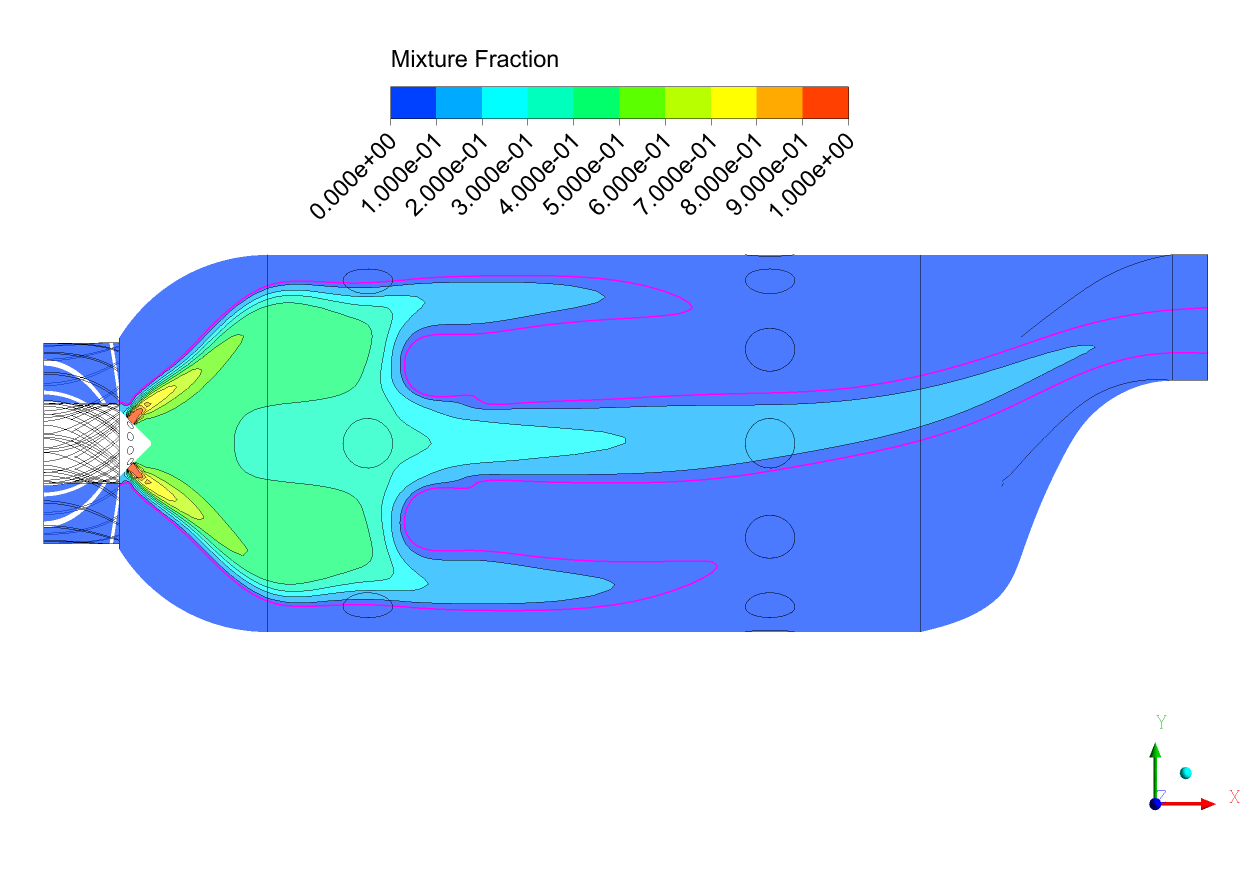}
		\caption{LPS–RSM}
		\label{fig:18d}
	\end{subfigure}
	\caption{Comparison of mixture fraction ($Z$) predicted on the X-Y (front) plane using various turbulence models.The stoichiometric  mixture fraction $Z =  0.06$ is depicted by the pink line.}
	\label{fig:18}
\end{figure}
\begin{figure}[!b]
	\centering
	\begin{subfigure}{0.48\linewidth}
		\includegraphics[width=\linewidth]{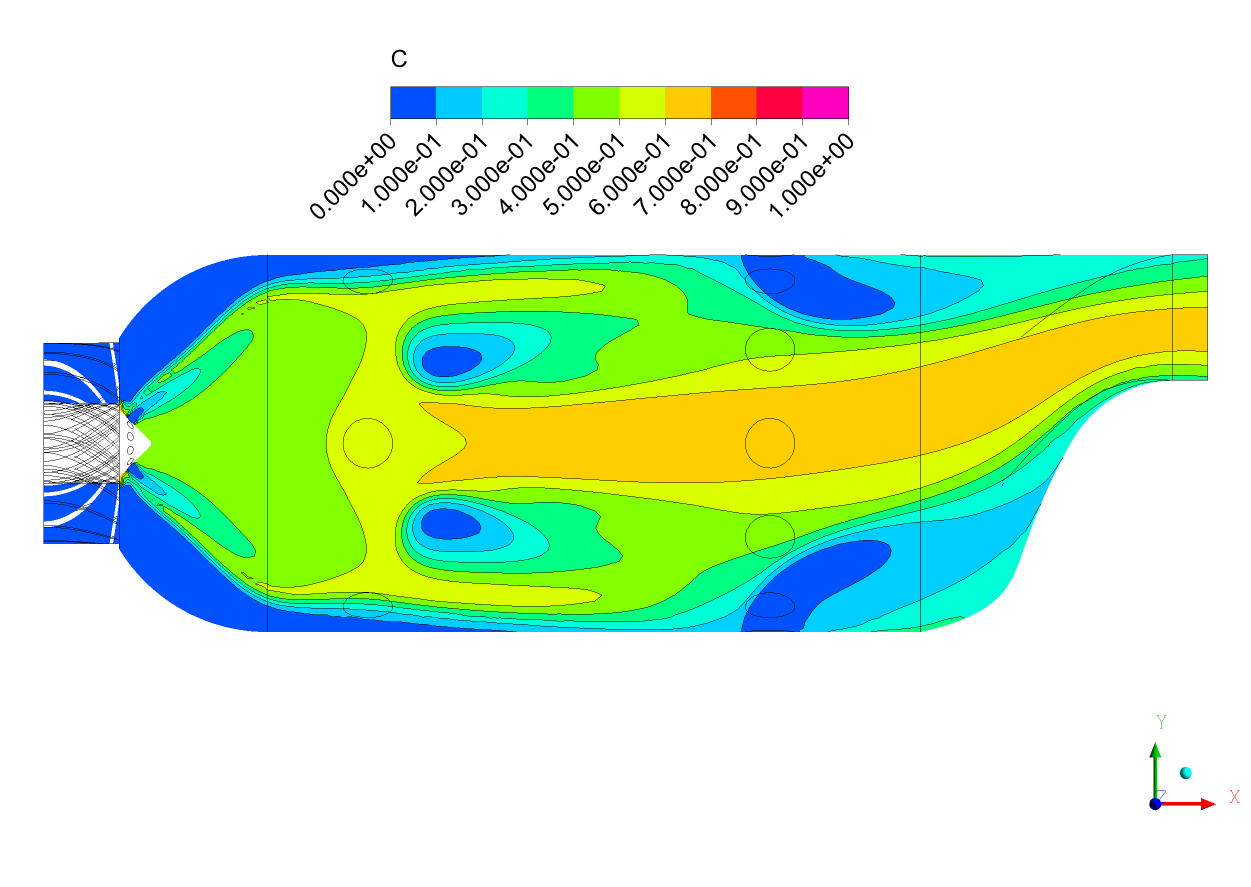}
		\caption{Standard $k$–$\epsilon$}
		\label{fig:20a}
	\end{subfigure}
	\begin{subfigure}{0.48\linewidth}
		\includegraphics[width=\linewidth]{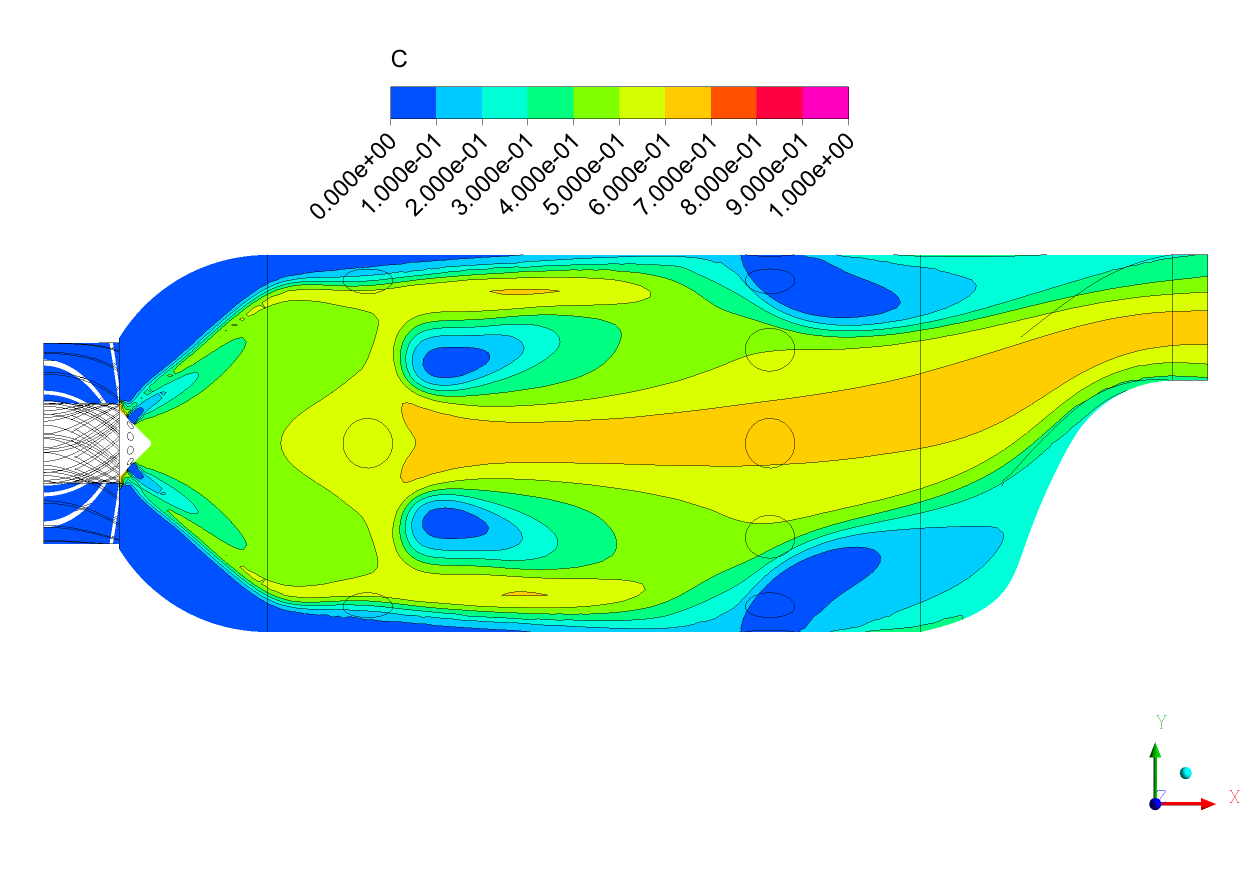}
		\caption{Realizable $k$–$\epsilon$}
		\label{fig:20b}
	\end{subfigure}
	\begin{subfigure}{0.48\linewidth}
		\includegraphics[width=\linewidth]{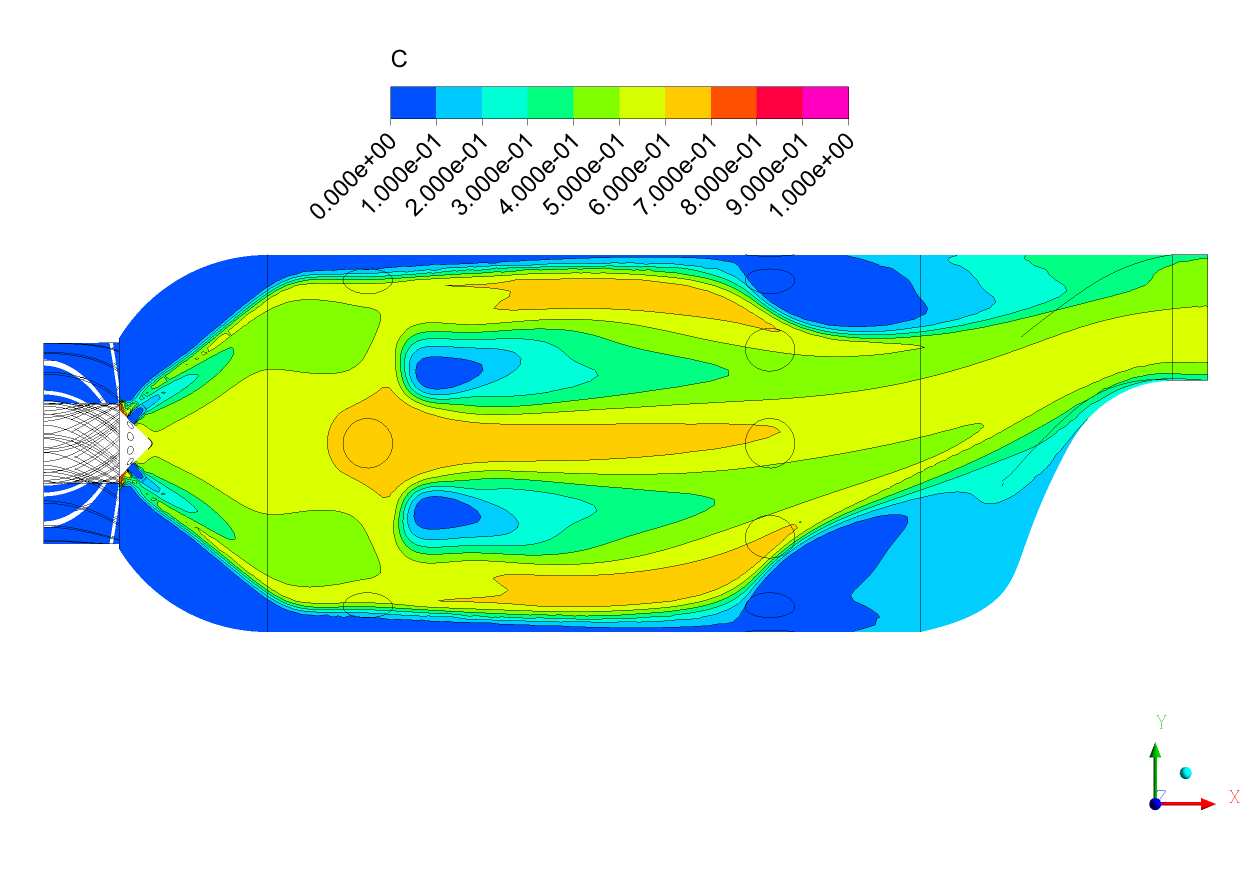}
		\caption{SST $k$–$\omega$}
		\label{fig:20c}
	\end{subfigure}
	\begin{subfigure}{0.48\linewidth}
		\includegraphics[width=\linewidth]{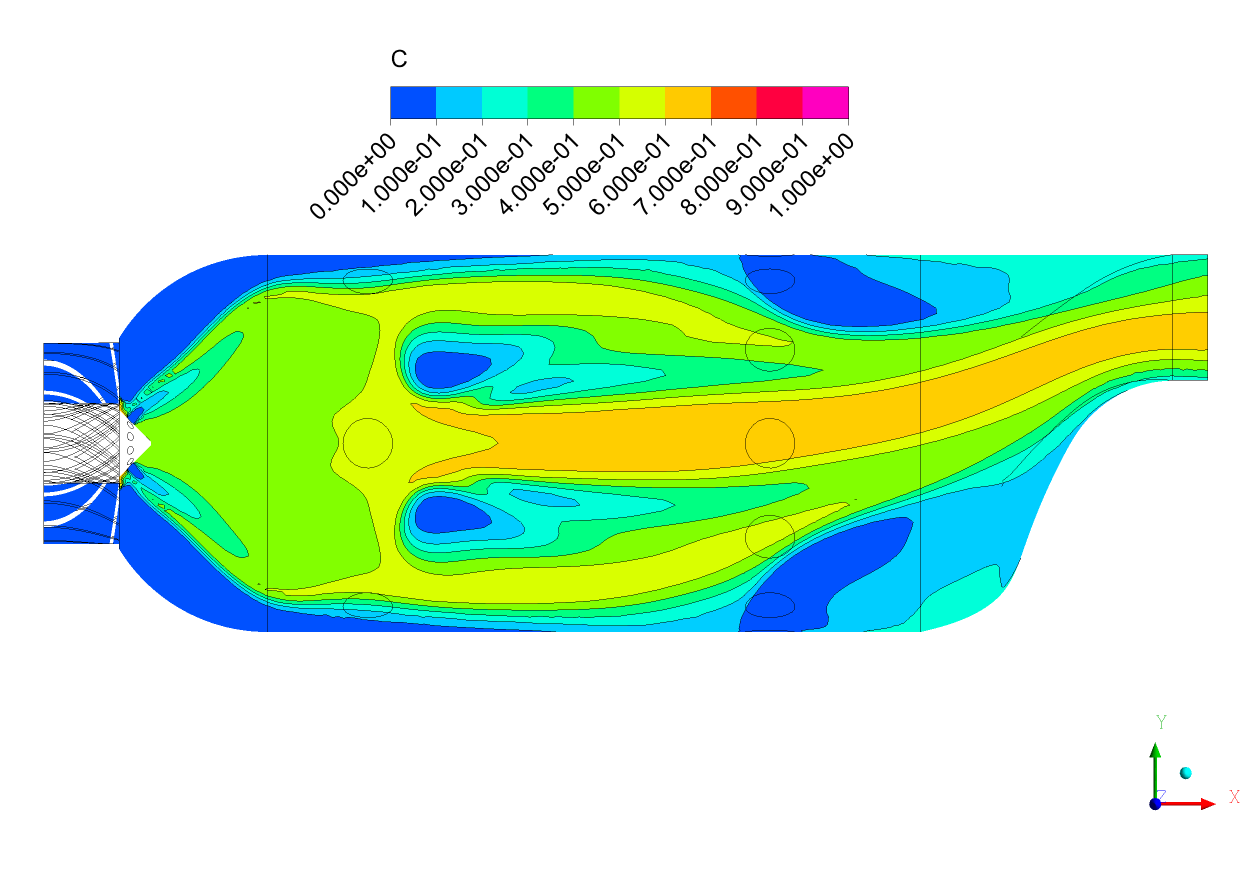}
		\caption{LPS–RSM}
		\label{fig:20d}
	\end{subfigure}
	\caption{Comparison of progress variables ($C$) predicted on the X-Y (front) plane using various turbulence models.}
	\label{fig:20}
\end{figure}

The temperature profiles on the primary holes (PH) plane predicted using the standard $k-\epsilon$, realizable $k-\epsilon$, SST $k-\omega$, and LPS-RSM models are shown in \fig\ref{fig:12a}. The SST $k-\omega$ model predicts the highest temperature, indicating the strongest mixing in the recirculation zone, consistent with its higher TKE prediction (\fig\ref{fig:5c}). This enhances flame stabilization and promotes more vigorous combustion. Consistently, the species distributions of \ce{C3H8}, \ce{CO2} and \ce{CO} in \figs\ref{fig:13a}-\ref{fig:15a} show that SST $k-\omega$ yields lower \ce{C3H8} (enhanced fuel consumption) and higher \ce{CO} (greater combustion completeness). It also predicts relatively higher \ce{CO}, likely due to intensified mixing leading to shorter residence time and limited oxidation of \ce{CO} to \ce{CO2}. This behavior can also be interpreted in terms of turbulent thermal diffusivity, defined as:
\begin{gather}
	\alpha_t = \frac{\mu_t}{\bar{\rho} \, Pr_t}
\end{gather}	
where $Pr_t$ is the turbulent Prandtl number, defined as the ratio of turbulent momentum diffusivity to turbulent thermal diffusivity, and is taken as 0.85 in the present study \cite{fluent2011ansys}. Turbulent thermal diffusivity quantifies the efficiency of turbulent heat transport \cite{Pope2000}. 
\newline
The turbulent thermal diffusivity ($\alpha_t$) profiles on the primary holes (PH) plane (\fig\ref{fig:16a}) show that the SST $k-\omega$ model predicts higher $\alpha_t$ in the CRZ ($r^\ast \lesssim \pm 0.5$), promoting stronger heat and species transport, a more intense reaction zone, and higher temperatures. Although the standard $k-\epsilon$ model also predicts relatively high $\alpha_t$, its larger eddy viscosity smooths velocity and scalar gradients, suppresses TKE (\fig\ref{fig:5c}), weakens air-fuel interaction, and thus yields lower temperatures than SST $k-\omega$. The realizable $k-\epsilon$ model predicts higher temperatures than the standard $k-\epsilon$ and LPS-RSM models, consistent with its higher $\alpha_t$ in the shear layer due to greater sensitivity to strain and higher TKE, which enhance mixing and combustion. However, its temperature, particularly near the CRZ core ($r^\ast \lesssim \pm 0.25$), remains lower than that predicted by the SST $k-\omega$ model.
\newline 
To further examine the influence of turbulence models on scalar fields, contours of mixture fraction (\figs\ref{fig:18a}-\ref{fig:18d}) and progress variable (\figs\ref{fig:20a}-\ref{fig:20d}) are presented for the standard $k-\epsilon$, realizable $k-\epsilon$, SST $k-\omega$, and LPS-RSM models. The mixture fraction (\eqn\ref{mixture-fraction}) indicates mixing quality and air-fuel availability, varying from $Z=0$ (pure oxidizer) to $Z=1$ (pure fuel). The SST $k-\omega$ model predicts lower mixture fraction values in the core region compared to the other models, indicating enhanced entrainment, dilution, and a more uniform mixture, which promotes more complete combustion and higher temperatures. In the present study, the progress variable is defined based on the combustion products of propane (\ce{C3H8})-air combustion as follows.
\begin{gather}
	C = \frac{Y}{Y_{\text{max}}},\qquad Y = (Y_{\ce{CO2}} + Y_{\ce{H2O}})
\end{gather}
This definition normalizes the progress variable such that $C=0$ corresponds to an unburned mixture and $C=1$ to a fully burned mixture. As seen in \figs\ref{fig:20a}–\ref{fig:20d}, the SST $k-\omega$ model predicts relatively higher $C$ in the primary holes (PH) region, indicating a greater extent of reaction and consistent with the higher temperature prediction.
%
\iffalse
\begin{figure}[!b]
	\centering
	\begin{subfigure}{0.48\linewidth}
		\includegraphics[width=\linewidth]{fig17a}
		\caption{Standard $k$–$\epsilon$}
		\label{fig:17a}
	\end{subfigure}
	\begin{subfigure}{0.48\linewidth}
		\includegraphics[width=\linewidth]{fig17b}
		\caption{Realizable $k$–$\epsilon$}
		\label{fig:17b}
	\end{subfigure}
	\begin{subfigure}{0.48\linewidth}
		\includegraphics[width=\linewidth]{fig17c}
		\caption{SST $k$–$\omega$}
		\label{fig:17c}
	\end{subfigure}
	\begin{subfigure}{0.48\linewidth}
		\includegraphics[width=\linewidth]{fig17d}
		\caption{LPS–RSM}
		\label{fig:17d}
	\end{subfigure}
	\caption{Comparison of mixture fraction ($Z$) predicted on the Z–Y (horizontal) plane using various turbulence models.The stoichiometric mixture fraction $Z =  0.06$ is depicted by the pink line.}
	\label{fig:17}
\end{figure}
\fi
%
\iffalse
\begin{figure}[!b]
	\centering
	\begin{subfigure}{0.48\linewidth}
		\includegraphics[width=\linewidth]{fig19a}
		\caption{Standard $k$–$\epsilon$}
		\label{fig:19a}
	\end{subfigure}
	\begin{subfigure}{0.48\linewidth}
		\includegraphics[width=\linewidth]{fig19b}
		\caption{Realizable $k$–$\epsilon$}
		\label{fig:19b}
	\end{subfigure}
	\begin{subfigure}{0.48\linewidth}
		\includegraphics[width=\linewidth]{fig19c}
		\caption{SST $k$–$\omega$}
		\label{fig:19c}
	\end{subfigure}
	\begin{subfigure}{0.48\linewidth}
		\includegraphics[width=\linewidth]{fig19d}
		\caption{LPS–RSM}
		\label{fig:19d}
	\end{subfigure}
	\caption{Comparison of progress variables ($C$) predicted on the Z-X (horizontal) plane using various turbulence models.}
	\label{fig:19}
\end{figure}
\fi
\newline
Nevertheless, the differences in predicted temperature, species concentrations, turbulent thermal diffusivity, mixture fraction, and progress variable across the models are primarily governed by their treatment of turbulent viscosity. The turbulent viscosity ($\mu_t$) directly controls mixing by dictating air entrainment, fuel-oxidizer interaction, and the structure and stability of the combustion zone. In the SST $k-\omega$ model, the strain-dependent term ($\tilde{S}_{ij}$) and blending function ($F_2$) regulate turbulence dissipation, yielding a more accurate distribution of $\mu_t$ in the swirl-dominated recirculation zone (CRZ) and shear layers. This sustains turbulence production, enhances TKE, and improves scalar mixing and flame stabilization, as reflected in lower mixture fraction and higher progress variable $C$ in the core, leading to higher temperatures compared to the standard and realizable $k-\epsilon$ and LPS-RSM models. In contrast, the LPS-RSM model predicts the lowest temperature (\fig\ref{fig:12a}) due to the dominance of slow and rapid pressure-strain terms, which primarily redistribute rather than generate Reynolds stresses. This weakens anisotropy-driven mixing, reduces TKE (globally; \figs\ref{fig:5c} and \ref{fig:6c}), limits fuel-air interaction, and results in lower heat release and temperature.
\newline
The temperature profiles indicate higher temperatures on the dilution holes (DH) plane (\fig\ref{fig:12b}) than on the primary holes (PH) plane (\fig\ref{fig:12a}), consistent with the overall progress variable distribution across all models (\figs\ref{fig:20a}-\ref{fig:20d}). This is because, in the primary zone, the flame is still stabilizing and mixing-controlled reactions are incomplete. As the flow moves downstream, increased residence time enables further oxidation, including conversion of \ce{CO} to \ce{CO2}, resulting in greater heat release. Additional air from the dilution holes further promotes combustion and elevates temperature. The realizable $k-\epsilon$ model predicts the highest temperature on the dilution plane. Temperature peaks occur in the shear layers ($r^\ast \approx \pm 0.75$), due to oxidation enhanced by dilution jets, and in the CVC core ($r^\ast \approx \pm 0.25$), due to sustained air entrainment and reaction. This is consistent with its strain-sensitive turbulent viscosity formulation, which yields higher $\alpha_t$ (\fig\ref{fig:16b}) and TKE (\fig\ref{fig:6c}), enhancing radial mixing and air-fuel interaction. Correspondingly, the realizable $k-\epsilon$ model predicts relatively low \ce{C3H8} (slightly higher than the standard $k-\epsilon$, but lower than SST $k-\omega$ and LPS-RSM; \fig\ref{fig:13b}), high \ce{CO2} (though lower than the standard $k-\epsilon$; \fig\ref{fig:14b}), and relatively low \ce{CO} (lower than both the standard $k-\epsilon$ and LPS-RSM; \fig\ref{fig:15b}), indicating enhanced but incomplete oxidation.
\newline
Following the realizable $k-\epsilon$ model, the standard $k-\epsilon$ model predicts the highest temperature on the dilution holes (DH) plane (\fig\ref{fig:12b}). This is attributed to its turbulent viscosity formulation, where $C_\mu$ is constant and thus less sensitive to strain compared to the realizable $k-\epsilon$ model. As a result, excessive turbulent diffusion in upstream shear layers is limited, allowing greater heat retention during downstream convection. Consequently, the standard $k-\epsilon$ model exhibits relatively uniform and high turbulent thermal diffusivity on both the primary and dilution holes (PH and DH) planes (\figs\ref{fig:16a} and \ref{fig:16b}). This sustains oxidation and heat release upstream of the dilution plane, maintaining higher temperatures. Although its predicted TKE is lower than that of the realizable $k-\epsilon$ model, it remains sufficient to support scalar transport and mixing. This behavior is further reflected in species distributions, as the standard $k-\epsilon$ model predicts the lowest \ce{C3H8} (\fig\ref{fig:13b}), highest \ce{CO2} (\fig\ref{fig:14b}), and elevated \ce{CO} levels, particularly near the CVC ($r^\ast \approx \pm 0.25$) (\fig\ref{fig:15b}). The higher \ce{CO} is attributed to rapid primary oxidation under strong mixing and high temperatures, coupled with local oxygen deficiency that limits the subsequent conversion of \ce{CO} to \ce{CO2}.
\newline 
The LPS-RSM model predicts the lowest temperature across the dilution plane (\fig\ref{fig:12b}). As noted earlier, its pressure-strain correlation primarily redistributes rather than produces Reynolds stresses, leading to reduced turbulent activity and lower TKE (\fig\ref{fig:6d}). This weakens turbulence and scalar transport, limits air-fuel interaction, and reduces temperature. Consistently, it predicts the highest \ce{C3H8} (\fig\ref{fig:13b}), lowest \ce{CO2} (\fig\ref{fig:14b}), and elevated \ce{CO}, indicating incomplete oxidation in the dilution region.
The SST $k-\omega$ model predicts intermediate temperatures, with peaks in the shear layers ($r^\ast \lesssim \pm 0.75$) and the CVC core ($r^\ast \lesssim \pm 0.25$) (\fig\ref{fig:12b}). It shows negligible \ce{C3H8} in the CVC core (\fig\ref{fig:13b}), indicating that temperature there is governed mainly by convection of combustion products rather than local reaction. Higher \ce{CO2} and lower \ce{CO} in the core suggest near-complete oxidation upstream, while finite \ce{C3H8} and elevated \ce{CO} in the shear layers ($r^\ast \approx \pm 0.75$; \figs\ref{fig:13b}, \ref{fig:14b}, and \ref{fig:15b}) indicate ongoing reactions.
Further, mixture fraction contours show that SST $k-\omega$ predicts a smaller stoichiometric region ($Z=0.06$) (\fig\ref{fig:18c}) compared to the other models, implying that most core reactions occur upstream, with remaining reactions confined to the shear layers near the dilution holes. This behavior arises from its shear-sensitive limiting and blending functions, which control turbulent viscosity and TKE, moderating turbulence in the core while preventing excessive amplification in the shear layers (\fig\ref{fig:6c}). Consequently, thermal diffusivity is better distributed, leading to moderate core temperatures ($r^\ast \approx \pm 0.25$) and peak temperatures in the shear layers ($r^\ast \approx \pm 0.75$).
\begin{table}[h!]
	\caption{Comparison of predicted mass fractions of \ce{C3H8} and \ce{CO} for different turbulence models at the combustor outlet.}	\label{outlet_species}
	\centering
	\renewcommand{\arraystretch}{1.5}
	\resizebox{0.8\textwidth}{!}{
	\begin{tabular}{|l|c|c|c|c|}
		\hline
		Mass fraction & Standard $k-\epsilon$ & Realizable $k-\epsilon$ & SST $k-\omega$ & LPS–RSM \\
		\hline
		$Y_{\ce{C3H8}}$ & $2.6395962 \times 10^{-6}$  &$1.7683729 \times 10^{-7}$ & $4.0727396 \times 10^{-8}$ & $1.1542677 \times 10^{-5}$
		 \\
		$Y_{\ce{CO}}$     & $5.2374177 \times 10^{-3}$ &$3.9994925 \times 10^{-3}$
		 & $2.2272459 \times 10^{-3}$ &  $4.3503798 \times 10^{-3}$ \\
		\hline
	\end{tabular}}
\end{table}
\newline
Overall, the SST $k-\omega$ model provides the most physically consistent prediction of the confined reacting swirling flow in the present gas turbine can combustor. Its viscosity limiters and blending function enable more accurate predictions of mean axial and transverse velocities, TKE, and shear stress. Although limitations of the isotropic eddy-viscosity assumption are evident in the dilution plane, where the flow becomes strongly three-dimensional and anisotropic, the model still offers the most realistic overall representation. It predicts higher temperatures on the primary holes plane, supported by \deleted{improved} \added{more accurately predicted} velocity, TKE, and shear stress fields, indicating a stronger and more stable CRZ, which is critical for air-fuel mixing and flame anchoring. The low \ce{C3H8} levels on the dilution plane confirm that most fuel oxidation occurs upstream. Mixture fraction contours further show a smaller stoichiometric region ($Z=0.06$), implying that reactions are largely completed within the CRZ and extended shear layers intersecting the dilution jets, consistent with experimental observations \citep{heitor1985experiments, heitor1986velocity}. This is also reflected in the lowest outlet levels of \ce{C3H8} and \ce{CO} (\tab \ref{outlet_species}). The progress variable contours (\fig\ref{fig:20c}) show high $C$ values in the CRZ and shear layers, confirming that combustion is largely completed upstream of the dilution holes and continues in the downstream shear layers, consistent with the stoichiometric mixture fraction distribution (\fig\ref{fig:18c}).
\newline
Examination of the mixture fraction contours for the standard $k-\epsilon$ and LPS-RSM models (\figs\ref{fig:18a} and \ref{fig:18d}) shows that the stoichiometric region ($Z=0.06$) extends to the combustor exit, indicating incomplete oxidation. This is consistent with the relatively higher outlet mass fractions of \ce{C3H8} and \ce{CO} (\tab \ref{outlet_species}). The standard $k-\epsilon$ model predicts higher \ce{CO} than LPS-RSM due to stronger turbulence, which promotes rapid initial oxidation but creates locally oxygen-deficient regions that limit complete conversion to \ce{CO2}. The progress variable contours for the standard $k-\epsilon$ model (\fig\ref{fig:20a}) show low values of $C$ in the CRZ and a broader reaction zone shifted downstream, indicating delayed, diffuse combustion and weaker flame anchoring, with reactions extending toward the outlet. The LPS-RSM model exhibits a similar trend, though with a slightly more confined reaction zone (\fig\ref{fig:20d}). For both models, the progress variable fields are consistent with their mixture fraction distributions, confirming slow oxidation and incomplete burnout. In contrast, the realizable $k-\epsilon$ model predicts a stoichiometric region that terminates upstream of the outlet (\fig\ref{fig:18b}), indicating near-complete oxidation within the combustor. The corresponding progress variable (\fig\ref{fig:20b}) shows high $C$ downstream of the CRZ due to enhanced air entrainment into the CVC, followed by gradual decay along the core as the flow moves downstream. Since $C$ represents the extent of reaction ($C=0$ unburned, $C=1$ fully burned), elevated values near the outlet indicate that most combustion is completed upstream. This is consistent with the mixture fraction field and the relatively low outlet levels of \ce{C3H8} and \ce{CO} (\tab \ref{outlet_species}).
\newline 
In summary, the comparative assessment of turbulence models demonstrates that model performance is strongly governed by the treatment of turbulent viscosity and its impact on mixing, reaction, and scalar transport. The SST $k-\omega$ model provides the most physically consistent representation of the confined reacting swirling flow, yielding improved predictions of velocity fields, TKE, shear stress, temperature, and species distributions. Its shear-sensitive formulation enables balanced turbulence production and dissipation, promoting effective air–fuel mixing, stable flame anchoring in the CRZ, and near-complete combustion within the combustor. The realizable $k-\epsilon$ model also performs reasonably well, capturing key mixing and reaction features, though with slightly less accuracy in core regions. In contrast, the standard $k-\epsilon$ and LPS-RSM models exhibit limitations, including excessive or insufficient turbulent mixing, leading to delayed or incomplete combustion, as evidenced by extended reaction zones and higher outlet levels of \ce{C3H8} and \ce{CO}. Overall, the results highlight the critical role of turbulence-chemistry interaction modeling in accurately predicting reacting swirling flows in gas turbine combustors.
\section{Concluding remarks}
This study presents a comprehensive computational fluid dynamics (CFD) analysis of combustion in a realistic can combustor, examining the influence of turbulence models on both velocity and scalar fields, including temperature and species concentrations. Non-premixed combustion is modeled using a presumed $\beta$-PDF approach coupled with a steady laminar flamelet formulation \deleted{based on} \added{with} the San Diego reaction mechanism, while turbulence is treated within the RANS framework. The performance of the standard $k-\epsilon$, realizable $k-\epsilon$, SST $k-\omega$, and LPS-RSM models is assessed in predicting mean axial and transverse velocities, TKE, and shear stress, along with their impact on temperature distribution, species evolution, and overall combustion characteristics within the combustor. Comparisons with experimental data show that the SST $k-\omega$ model predicts velocity and turbulence fields more accurately than the other models. Its viscosity limiter and models ($k-\epsilon$ and $k-\omega$) blending enable improved performance in confined swirling flows with adverse pressure gradients. Although limitations of the isotropic turbulence assumption persist at the dilution plane, particularly in TKE and shear stress predictions, the model captures key flow features more reliably. These include the CRZ, reflected in accurate mean axial velocity, and the CVC, indicated by improved transverse velocity prediction, represented through radial velocity contours ($\tilde{w}$).
Moreover, the SST $k-\omega$ model predicts higher temperatures in the primary zone (primary holes plane, near the CRZ), indicating that combustion is largely completed in this region. This is supported by lower \ce{C3H8} levels and higher TKE, both reflecting strong mixing and efficient combustion. Further downstream, on the dilution holes plane, the model predicts negligible \ce{C3H8}, reinforcing that combustion is completed upstream and indicating improved flame anchoring by the SST $k-\omega$ model. Furthermore, the SST $k-\omega$ model predicts the most compact stoichiometric mixture fraction region, primarily confined to the CRZ and shear layers intersecting the dilution holes, indicating that most air–fuel mixing and reaction completion occur there. The corresponding progress variable contours show high $C$ in this region, confirming near-complete combustion, with only a weak residual reaction zone downstream. This is consistent with negligible outlet levels of \ce{C3H8} and \ce{CO}, reinforcing effective combustion completion.
In contrast, the LPS-RSM model overpredicts mean axial and transverse velocities in the core region due to the action of pressure–strain correlations, which redistribute Reynolds stresses and suppress shear components. This reduces turbulence production, weakens shear layer development, and limits turbulent mixing. Consequently, outward momentum transport is restricted, leading to velocity overprediction, reduced TKE, diminished scalar transport, and weaker air–fuel interaction. As a result, lower temperatures are predicted, accompanied by higher outlet levels of \ce{C3H8} and \ce{CO}, indicating incomplete combustion.
The standard $k-\epsilon$ model underpredicts the negative axial velocity in the CRZ and misrepresents transverse velocity, indicating a weaker recirculation zone. This is due to its constant, isotropic turbulent viscosity, which limits its ability to capture strong gradients and curvature effects in swirling flows. Its over-dissipative nature also leads to lower TKE and shear stress, reducing scalar transport, weakening mixing, and slowing heat release, resulting in lower temperatures and higher outlet concentrations of \ce{C3H8} and \ce{CO}. The realizable $k-\epsilon$ model improves upon these predictions due to its strain-sensitive viscosity and modified dissipation equation, yielding better velocity, TKE, and shear stress fields. However, it still underpredicts the CRZ strength. Its relatively higher TKE enhances mixing and air–fuel interaction, leading to improved combustion, higher temperatures, and reduced outlet levels of \ce{C3H8} and \ce{CO} compared to the standard $k-\epsilon$ model.

%
%-------------- End of conclusion
\iffalse
\appendix
%\section*{Appendices}
\begin{figure}[!ht]
	\centering
	\includegraphics[width=1\linewidth]{FigA1}
	\caption{\added[id=AK]{Locations at combustor swirler exit where axial and transverse velocity are compared}}
	\label{Fig:A1}
\end{figure}

\begin{figure}[!tb]
	\centering
	\begin{subfigure}[t]{0.48\linewidth}
		\centering
		\includegraphics[width=\linewidth]{FigA2a}
		\caption{Left side}
		\label{fig:A2a}
	\end{subfigure}
	\hfill
	\begin{subfigure}[t]{0.48\linewidth}
		\centering
		\includegraphics[width=\linewidth]{FigA2b}
		\caption{Right side}
		\label{fig:A2b}
	\end{subfigure}
	\vspace{3mm}
	
	\caption{Predicted axial velocity on the left and right side of the swirler exit (see Figure \ref{Fig:A1}) under reacting conditions (refer \tab~\ref{tab:1}).}
	\label{fig:A2}
\end{figure}

\begin{figure}[!tb]
	\centering
	\begin{subfigure}[t]{0.48\linewidth}
		\centering
		\includegraphics[width=\linewidth]{FigA3a}
		\caption{Left side}
		\label{fig:A3a}
	\end{subfigure}
	\hfill
	\begin{subfigure}[t]{0.48\linewidth}
		\centering
		\includegraphics[width=\linewidth]{FigA3b}
		\caption{Right side}
		\label{fig:A3b}
	\end{subfigure}
	\vspace{3mm}
	
	\caption{Predicted transverse velocity on the left and right side of the swirler exit (see Figure \ref{Fig:A1}) under reacting conditions (refer \tab~\ref{tab:1}).}
	\label{fig:A3}
\end{figure}

\fi
%
\clearpage
%%%%%%%%%%%%%%%%%%%%%%%%%%%%%%%%%%%%%%%%%%
\section*{Declaration of Competing Interest}
%%%%%%%%%%%%%%%%%%%%%%%%%%%%%%%%%%%%%%%%%%
\noindent 
% All authors declare that they have no conflict of interest. 
The authors declare that they have no known competing financial interests or personal relationships that could have appeared to influence the work reported in this paper.
%
%The  authors  certify  that  they  have  NO  affiliations  with  or  involvement  in  any  organization or entity with any financial interest (such as honoraria; educational grants; participation in speakers’ bureaus; membership, employment, consultancies, stock ownership, or other equity interest; and expert testimony or patent-licensing arrangements), or non-financial interest (such as personal or professional relationships, affiliations, knowledge or beliefs) in the subject matter or materials discussed in this manuscript.
	%%%%%%%%%%%%%%%%%%%%%%%%%%%%%%%%%%%%%%%%%%%
%	\section*{Disclosure Statement}
	%\noindent 
%	This work is based, in part, on a preprint \citep{KumarXiv2026} authored by the same author(s). The present manuscript extends and substantially revises that earlier version.
	%%%%%%%%%%%%%%%%%%%%%%%%%%%%%%%%%%%%%%%%%%%
	\section*{Acknowledgements}
	%\noindent 
	AK acknowledges the receipt of the ``AICTE Post-Doctoral Fellowship'' from the All India Council for Technical Education (AICTE), Ministry of Education, Government of India.
	%
%%%%%%%%%%%%%%%%%%%%%%%%%%%%%%%%%%%%%%%%%%%
\section*{Authors Contributions Statement}
%%%%%%%%%%%%%%%%%%%%%%%%%%%%%%%%%%%%%%%%%%
%\noindent 
%
\vspace{-1em} 
\begin{table}[!h]
	%\begin{center}
	\renewcommand{\arraystretch}{1.5}
	\begin{tabular}{|l|p{0.75\linewidth}|}
		\hline
		Author  & Contribution(s) \\\hline
		1. Aishvarya Kumar	& 
		Conceptualization, Methodology, Software, Validation, Resources, Formal analysis, Investigation, Data Curation, Visualization, Writing - Original Draft\\\hline
		2. Ram Prakash Bharti	& 
		Supervision, Conceptualization, Methodology, Formal analysis, Writing - Review \& Editing\\\hline
%		3. Kamlesh Kumari	& Supervision, Funding, Writing - Review \& Editing\\\hline
	\end{tabular}
	%	\end{center}
	\end{table}
	\vspace{-1em} 
	%
	%--------------- Nomenclature
	%\clearpage
	\begin{spacing}{1.05}
%\input{Nomenclature.tex}
%\renewcommand{\nompreamble}{\vspace{1em}\fontsize{10}{8pt}\selectfont}
%{\printnomenclature[5em]}
\printnomenclature
%
%\clearpage
%\printglossaries 
%
%===========================
%\appendix
%===========================
%
%--------------- Bibliography
%
%\begin{thebibliography}{0000}
%\bibliographystyle{plainnat}
%\bibliographystyle{elsarticle/elsarticle-harv}\biboptions{authoryear}
\noindent \small
\bibliography{references}	
\end{spacing}
%===========================

% ---- Stop numbering ----
\nolinenumbers
\clearpage
% Start Supplementary pages
%
\appendix
%
%\renewcommand\thesection{Appendix~\Alph{section}}
%\renewcommand\thesubsection{\Alph{section}.\arabic{subsection}}
%\renewcommand{\thetable}{\Alph{section}.\arabic{table}} \setcounter{table}{0}
%\renewcommand{\thefigure}{\Alph{section}.\arabic{figure}} \setcounter{figure}{0} 
%
% Section numbering: S1, S2, ...
\renewcommand{\thesection}{S\arabic{section}} 
% Figure numbering: S1, S1, ...
\renewcommand{\thefigure}{S\arabic{figure}} \setcounter{figure}{0} 
% Figure numbering: S1.1, S1.2, ...
% \renewcommand{\thefigure}{\thesection.\arabic{figure}} \setcounter{figure}{0} 
% Table numbering: S1, S1, ...
\renewcommand{\thetable}{S\arabic{table}} \setcounter{table}{0} 
% Table numbering: S1.1, S1.2, ...
%\renewcommand{\thetable}{\thesection.\arabic{table}} \setcounter{table}{0} 
\renewcommand{\theequation}{S.\arabic{equation}} \setcounter{equation}{0} 
\renewcommand{\thepage}{S-\arabic{page}} \setcounter{page}{1}
\renewcommand{\thefootnote}{\fnsymbol{footnote}}
{\hfill\LARGE\bfseries Supporting Information}\\
\makebox[1\textwidth]{\hrulefill}\\
{\Large\bfseries Performance Evaluation of RANS-Based Turbulence Models in Predicting Turbulent Non-Premixed Swirling Combustion within a Realistic Can Combustor}\\[10pt]
{
Aishvarya Kumar$^\text{a}$, 
Ram Prakash Bharti,$^\text{b}$\footnote[1]{Corresponding author. E-mail address: rpbharti@iitr.ac.in (RP Bharti)} 
%Kamlesh Kumari$^\text{a}$
} \\
{\small $^a$ Department of Chemical Engineering, Sant Longowal Institute of Engineering and Technology (SLIET), Longowal 148016, Punjab, India;\\$^b$ Complex Fluid Dynamics and Microfluidics (CFDM) Lab, Department of Chemical Engineering, Indian Institute of Technology Roorkee, Roorkee - 247667, Uttarakhand, India} \\
\makebox[1\textwidth]{\hrulefill}
\section{Mathematical Modeling}\label{appendixA}
%--------------------------------------
%\section{Mathematical Modeling}
%--------------------------------------
%
%\setcounter{linenumber}{1} \linenumbers
\setlength{\abovedisplayskip}{6pt}
\setlength{\belowdisplayskip}{6pt}
\onehalfspacing
The mathematical model for the above-described physical problem under consideration is written as follows.
In turbulent flows involving combustion or significant heat transfer, substantial density fluctuations are induced due to the thermal heat release. For simulating these phenomena, Favre-averaging \citep{Liou1991} is the preferred approach due to its density-weighted formulation, which inherently accounts for the crucial aspects of turbulent combustion, i.e., compressibility effects and substantial density fluctuations. The simplification of non-linear terms in governing equations is a major advantage of Favre-averaging, leading to improved computational efficiency, enhanced numerical stability, and higher accuracy. Consequently, Favre averaging has become a crucial component of combustion simulation workflows.
\subsection{Favre-averaged governing equations}
%\newline
In turbulence modeling, Favre-averaged form of the governing equations allows for better handling of turbulence effects where the instantaneous quantities (${\psi}=\mathbf{u}, T, e, h, H$, etc) are split as $\psi=(\tilde{\psi} + {\psi}^{\prime\prime})$ into a Favre-averaged mean ($\tilde{\psi}$) and a fluctuating part (${\psi}^{\prime\prime}$), except for the density and pressure ($\phi=$ $\rho, p$) which are split as $\phi=(\overbar{\phi} + {\phi}^{\prime})$ into a Reynolds averaged (time, space or ensemble) mean ($\overbar{\phi}$) and a fluctuating part (${\phi}^{\prime}$). The Favre-averaging helps to eliminate the dependency of the equation on the fluctuating density as the Favre-averaged variables ($\tilde{\psi}$) denotes the density-weighted averaging of a quantity ($\psi$) defined \citep{Liou1991} as follows: $\tilde{\psi}=(\overbar{\rho\psi})/\overbar{\rho}$. More details about the relation between instantaneous quantities and averaged quantities can be found elsewhere \cite{Liou1991,Versteeg2007, Veynante2009,Blazek_2015,Lele2021}.
\newline
In a multicomponent gas mixture, the Favre-averaged density ($\overbar{\rho}$) is related to the species density and the mass fraction as follows.
\begin{gather}
	\overbar{\rho} =\left({\sum_{k} \frac{\tilde{Y}_k}{\rho_k}}\right)^{-1}
\end{gather}
where, subscript $k$ represents for $k$-th species,  $Y_k$ and $\rho_k$ represent the mass fraction and density of species. Assuming the ideal-gas behaviour, the density of each species can be expressed \cite{fluent2011ansys} as follows.
\begin{gather}
	\rho_k = \frac{\bar{p}_k M_k}{R\,\tilde{T}}
\end{gather}
where $p_k$ is the partial pressure of species (related to the total fraction through $p_k=X_k\bar{p}$, with $X_k$ being the mole fraction of the species), $M_k$ is the molecular weight of the species, $R$ is the universal gas constant ($=8.314$ kJ/kmol·K) and $\tilde{T}$ is the Favre-averaged  temperature.
\newline
The instantaneous local density ($\rho$) of the mixture depends on the local pressure, temperature, and species mass fractions of the reactants and products. It is calculated using the equation of state for an ideal gas, written in the Favre-averaged form as follows.
\begin{gather}
	\overbar{\rho} \approx
	\frac{\bar{p}}{R\,\tilde{T}\,\displaystyle\sum_{k}({\tilde{Y}_k}/{M_k})}
\end{gather}
where, $\rho_k$ is the density of species, $\overbar{p}$ is the average pressure, $\tilde{Y}_k$ is the Favre-averaged mass/mole fraction of species, and  ${M}_k$ is the molar mass of species. 
\subsubsection{Continuity equation}
The Favre-averaged form of the conservation of mass in a turbulent combustion is given as follows.
\begin{gather}
	\frac{\partial \overline{\rho}}{\partial t} 
	+ \frac{\partial (\overline{\rho} \tilde{u}_j)}{\partial x_j}
	= 0, 
	\quad \text{where} \quad 
	\tilde{u}_j = \frac{\overline{\rho u_j}}{\overline{\rho}}
\end{gather}
where $\overline{\rho}$ is the Reynolds-averaged (i.e., time-averaged or mean) density, $\tilde{u}_j$ is the $j$-th component of the Favre-averaged velocity vector ($\tilde{\mathbf{u}}$), $t$ is the time, and $u_j$ is the $j$-th component of the instantaneous velocity vector ($\mathbf{u}$).
\subsubsection{Momentum equation}
The Favre-averaged form of the conservation of momentum in turbulent combustion is given as follows.
\begin{gather} \label{Momentum}
	\frac{\partial (\overline{\rho}\,\tilde{u}_i)}{\partial t}
	+ \frac{\partial (\overline{\rho}\,\tilde{u}_i \tilde{u}_j)}{\partial x_j}
	= -\frac{\partial \overline{p}}{\partial x_i}
	+ \frac{\partial \big(\overline{\tau}_{ij} - \tilde{R}_{ij}\big)}{\partial x_j}
	%- \frac{\partial \big(\tilde{R}_{ij}\big)}{\partial x_j} 
	+ \overline{\rho} \overline{f_i}
\end{gather}
where, $\overline{p}$ is the mean pressure, $\overline{\tau}_{ij}$ is the Favre-averaged viscous stress tensor, {$\tilde{R}_{ij}$} is the mean Reynolds stress tensor. and $\overline{\rho} \overline{f_i}$ is the mean body force per unit volume.	
The Favre-averaged form of stress tensors ($\overline{\tau}_{ij}$, $\tilde{R}_{ij}$) are expressed as follows.
\begin{gather} \label{viscous-stress-tensor}
	\overline{\tau}_{ij} = \mu \left( \frac{\partial \tilde{u}_i}{\partial x_j} + \frac{\partial \tilde{u}_j}{\partial x_i} \right) - \frac{2}{3} \left(\mu \delta_{ij} \frac{\partial \tilde{u}_k}{\partial x_k}\right),
	\qquad
	\tilde{R}_{ij} = \overline{\rho u_i^{\prime\prime} u_j^{\prime\prime}}
\end{gather}
where, $\mu$ is the dynamic viscosity and $\delta$ is the Kronecker delta.	\eqn(\ref{Momentum}) includes an additional stress term (i.e., the Reynolds stress tensor, $\tilde{R}_{ij}$) arising due to turbulence, which often requires modeling using turbulence models ($k-\epsilon$, $k-\omega$, LES, RANS, etc) using turbulent viscosity to close the system of equations and solve for the turbulent flow field. In contrast, the Reynolds stress model (RSM) directly solves the Reynolds stresses ($\tilde{R}_{ij}$).
\subsubsection{Thermal energy equation}
The energy equation in the Favre-averaged enthalpy ($\tilde{H}$) form solved with the non-adiabatic non-premixed combustion model is written as follows.
\begin{gather} \label{enthalpy}
	\frac{\partial (\overline{\rho} \tilde{H})}{\partial t}  + \nabla \cdot (\overline{\rho} \tilde{\mathbf{u}} \tilde{H}) = \nabla \cdot \left(\frac{\overline{k}_t}{\overline{c}_p} \nabla \tilde{H}\right) + \overline{S}_h
\end{gather}
Assuming a unity Lewis number  ($Le=\alpha/\mathcal{D} \approx 1$), the conduction and species diffusion terms combine to give the right-hand side of \eqn(\ref{enthalpy}) and the contribution from viscous dissipation appears in the second term of the non-conservative form. The Favre-averaged  enthalpy ($\tilde{H}$) is defined as follows.
\begin{gather} \label{favre-averaged-enthalpy}
	\tilde{H} ={\sum_{k} \tilde{Y}_{k} \tilde{H}_{k}}
	\quad\text{where}\quad 
	{\tilde{H}_{k}} =  h_{k}^0 (T_{\text{ref},k}) + \int_{T_{\text{ref},k}}^{\tilde{T}} c_{p,k} dT 	
\end{gather}
where $\tilde{Y}_{k}$ and $\tilde{H}_{k}$ are the Favre-averaged mass fraction and  enthalpy of {$k$-th} species. The specific heat at constant pressure for each species ($c_{p,k}$), denoting the formation of enthalpy of each species ($h_k^0$) at the reference temperature ($T_{\text{ref},j}$) can be obtained from thermodynamic databases.
The turbulent thermal conductivity ($\overline{k}_t$) and the mean specific heat at constant pressure ($\overline{c}_p$) are expressed as follows.
\begin{gather}
	\overline{k}_t = \mu_t \frac{\overline{c}_p}{\text{Pr}_t}
	\qquad\text{and}\qquad
	\overline{c}_p = \sum_{k} \tilde{Y}_{k} \overline{c}_{p,k}
\end{gather}
where, $\text{Pr}_t$ is the turbulent Prandtl number, and $\overline{c}_{p,k}$ is the mean specific heat at constant pressure of $k$-th species.
\subsubsection{Mixture Fraction Theory}
The mixture fraction method provides a useful framework for modeling complex combustion processes (e.g., Non-premixed combustion, turbulent combustion, spray combustion, and engine combustion modeling). It is derived from a simple chemical reacting system (SCRS) considering the conservation equations for the reacting species and the assumption of a global combustion reaction focusing only on the final (and neglecting intermediate) species and reactions \cite{Versteeg2007, Veynante2009}.
In SCRS, the detailed kinetics is unimportant, and a global single-step, infinitely fast chemical reaction  with fuel-oxygen in stoichiometric proportion to form products. 
\begin{gather} \label{SCRS-equation}
	\ce{1 kg\ fuel + $s$ kg\ oxidant -> (1 + $s$) kg\ products} 
\end{gather}
where $s$ is the stoichiometric ratio of oxygen-fuel. \eqn\eqref{SCRS-equation} also shows that the rate of consumption ($\dot{\omega}$) of fuel (\ce{f})  and oxygen (\ce{o}), based on stoichiometry, are related as follows. 
\begin{gather} \label{one-step-reaction}
	{\dot{\omega}_\text{f} = ({1}/{s}) 	\dot{\omega}_\text{o}}
\end{gather} 
The transport equations for fuel (\ce{f})  and oxygen (\ce{o}) are written as follows.
\begin{gather} 
	\label{transport-fuel}%\label{transport-oxygen}
	\frac{\partial (\rho Y_{m})}{\partial t} + \frac{\partial (\rho u_j Y_{m} )}{\partial x_j} = \frac{\partial}{\partial x_j} \left(\Gamma_{m} \frac{\partial Y_{m}}{\partial x_j}   \right) +  \dot{\omega}_{m}\qquad\text{where}\qquad m = (\text{f, o})
\end{gather}
where, $\Gamma_{m}\ (= \rho \mathcal{D}_{m})$ and $Y_{m}$ are the diffusion coefficient and the mass fraction of $m$ (i.e., fuel or oxygen) in the mixture, i.e., $Y_\text{m}$ varies from 0 (no fuel or oxygen) to 1 (pure fuel or oxygen), and $\mathcal{D}_{m}$ is the diffusivity. 
Furthermore, oxidants typically contain inert species, such as \ce{N2}, which remain unaffected during combustion (except when \ce{NO_x} formation is considered), and thus the mass fraction of inert species ($Y_{in}$) remains constant before and after the reaction. 
Since the total mass fraction of reactants and products remains the same, the mass fraction of products can be obtained as $Y_{p} = 1 - (Y_{f} + Y_{o} +Y_{in})$ without solving any additional equation for $Y_{p}$.
\newline
Subsequently, assuming equal diffusivity ($\mathcal{D}_{f}=\mathcal{D}_{o}=\mathcal{D}_{i}$, i.e., $\Gamma_{f}=\Gamma_{o}=\Gamma_{i}$), introducing a new variable ($Z_{i} = sY_{f} - Y_{o}$), and considering one-step reaction (\eqn\ref{one-step-reaction}), the species transport equations (\eqn\ref{transport-fuel}) can be reduced into a single transport equation for $Z_i$, which is written as follows.
\begin{gather} \label{transport-z-updated}
	\frac{\partial (\rho Z_i)}{\partial t}  + \frac{\partial (\rho u_j Z_i )}{\partial x_j} = \frac{\partial }{\partial x_j} \left(\Gamma_i \frac{\partial Z_i}{\partial x_j}\right) 
\end{gather}
where, $Z_i$ is a passive scalar representing the elemental mass fraction of the element $i$ and obeys the scalar transport equation without source terms. Further, the conserved scalar mixture fraction ($f$) can uniquely characterize the instantaneous thermochemical states of the fluid, including temperature and species concentrations, which is defined as follows.
\begin{gather} \label{mixture-fraction}
	f = \left({\frac{Z_{i} - Z_{i, \text{o}}}{Z_{i, \text{f}}- Z_{i,  \text{o}}}}\right) \in [0,1]
\end{gather}
where, subscripts `\ce{o}' and `\ce{f}' denoting the inlet values of oxidizer and fuel stream, respectively.  
The mixture fraction ($f$) being linearly dependent on $Z_{i}$ (\eqn\ref{mixture-fraction}) is also a passive scalar and obeys the transport equation (\eqn\ref{transport-z-updated}) as follows.
\begin{gather} \label{transport-f}
	\frac{\partial (\rho f)}{\partial t}  + \frac{\partial (\rho u_j f )}{\partial x_j} = \frac{\partial }{\partial x_j} \left( \Gamma_f \frac{\partial f}{\partial x_j}\right)
	\qquad\text{where}\qquad \Gamma_f\  (=\mu/\sigma)
\end{gather}
Subsequently, the Favre-averaged form of the mixture fraction equation (\eqn\ref{transport-f}) in a conserved scalar transport model for combustion is written as follows.
\begin{gather} \label{transport-f-favre}
	\frac{\partial (\overbar{\rho} \tilde{f})}{\partial t}  + \frac{\partial (\overbar{\rho} \tilde{u_j} \tilde{f} )}{\partial x_j} = \frac{\partial }{\partial x_j} \left(\tilde{\Gamma}_{f} \frac{\partial \tilde{f}}{\partial x_j}\right);
	\qquad\text{where}\qquad
	\tilde{\Gamma}_{f} =\left(\frac{\mu}{\sigma}+ \frac{\mu_t}{\sigma_t}\right)
\end{gather}
where, $\Gamma_f$ is the diffusion coefficient, $\tilde{\Gamma}_{f}$ is the Favre-averaged effective dynamic diffusion coefficient, $\nu\ (=\mu/\rho)$ is the kinematic viscosity,   $\sigma$ is the {Schmidt number}, $\mu_t$ is the turbulent viscosity and {$\sigma_t$} is turbulent Schmidt number. 
%
%In turbulent flows, turbulent convection overwhelms molecular diffusion \cite{Veynante2009}, hence the equation \eqref{transport-f-favre} can be further simplified.
%
%\begin{gather} \label{transport-f-fluent}
%	\frac{\partial \overbar{\rho} \tilde{f}}{\partial t}  + \frac{\partial (\overbar{\rho} \tilde{u_j} \tilde{f} )}{\partial x_j} = \frac{\partial }{\partial x_j} \left( \frac{\mu + \mu_t}{\sigma_f} \frac{\partial \tilde{f}}{\partial x_j}\right)
%\end{gather}

%
\subsubsection*{(a) Relationship between mixture fraction ($f$) and equivalence ratio ($\phi$)} 
%
%In combustion processes, the equivalence ratio ($\phi$) is a key parameter that characterizes the fuel-air mixture and strongly influences pollutant formation (\ce{CO}, \ce{NO_x}, and unburned hydrocarbons (UHCs)), flame stability and structure, and overall combustion efficiency. Appropriate control of $\phi$ enables optimization of combustion performance, mitigation of emissions, and enhancement of system efficiency, thereby contributing to more sustainable and environmentally responsible operation. The equivalence ratio is defined as follows.
In combustion processes, the equivalence ratio ($\phi$) is a key parameter that characterizes the fuel-air mixture and is defined as follows.
\begin{gather}
	\phi = \frac{(F/A)_\text{actual}}{(F/A)_\text{stoichiometric}} 
\end{gather}
where, $F$ and $A$ represent the mass (or molar) flow rates of fuel and air, respectively. 
The equivalence ratio regimes are indicated as $\phi< 1$ for the lean mixture (excess air),  $\phi=1$  for the stoichiometric mixture (ideal air-fuel ratio), and $\phi > 1$ for the rich mixture (excess fuel), respectively. 
It strongly influences pollutant formation, such as \ce{CO}, \ce{NO_x}, and unburned hydrocarbons (UHCs), flame stability and structure, and overall combustion efficiency. Therefore, appropriate control of $\phi$ enables optimization of combustion performance, mitigation of emissions, and enhancement of system efficiency, thereby contributing to more sustainable and environmentally responsible operation. 
\newline	
The concept of mixture fraction ($f$) and equivalence ratio ($\phi$) can be correlated in reactive systems (\eqn\ref{SCRS-equation}) by considering a straightforward combustion system at stoichiometric conditions wherein \ce{F}, \ce{O} and \ce{P} symbolically representing fuel, oxidant and product streams as follows. 
\begin{gather} \label{general-reaction}
	\ce{$\phi$ F + $s$ O -> ($\phi$ + $s$) P}
\end{gather}
Examining \eqn\eqref{general-reaction}, the mixture fraction as a whole can be deduced to 
\begin{gather}
	f = \frac{\phi}{(\phi + s)}
\end{gather}
\subsubsection*{(b) Relationship between mixture fraction with species mass fraction, density, and temperature}
In non-adiabatic systems, where heat exchange may occur, the instantaneous values of various properties such as species mass fraction ($\phi_k$), density ($\rho$), and temperature ($T$), under the assumption of chemical equilibrium, depend on both the mixture fraction ($f$) and the instantaneous enthalpy ($H$), i.e., 
\begin{gather} \label{relationship}
	\phi_k = \phi_k (f, H);\qquad \tilde{\phi}_k = \tilde{\phi}_k(\tilde{f}, \tilde{H})
\end{gather}
%
%This relationship suggests that knowing the mixture fraction and the enthalpy allows us to determine the composition, density, and temperature of the system. The enthalpy $H$ accounts for energy changes due to chemical reactions and heat transfer, while the mixture fraction $f$ provides a measure of the fuel's presence in the mixture. By understanding these relationships, we can model and predict the behavior of reactive systems under various conditions, aiding in the design and control of combustion processes for efficiency and emission reduction.
%
%{and the Favre-averaging form,}
%
%\begin{gather}
%	\tilde{\phi}_k = \tilde{\phi}_k(\tilde{f}, \tilde{H})
%\end{gather}
where,  $\tilde{\phi}_k$ is the Favre-averaged mass fraction of $k$th species, $\tilde{f}$ is the Favre-averaged mixture fraction, and $\tilde{H}$ is the Favre-averaged enthalpy.
\subsubsection*{(c) Modeling of turbulence-chemistry interaction}
\eqn\eqref{relationship} describes the instantaneous relationship between mixture fraction and species fraction, density and temperature under the assumption of chemical equilibrium. In the simulation of turbulent flows, the main concern is predicting the averaged values of the fluctuating scalars. The Favre-averaged form of the mixture fraction (\eqn\ref{transport-f-favre}) is applied for turbulent combustion. The enthalpy equation in Favre-averaged form (\eqn\ref{enthalpy}) is solved in situations where the impact of radiation and other heat loss effects is significant. However, the mean species and temperature calculations using the field values of $\tilde{f}$ and $\tilde{H}$ are not as straightforward as for the laminar case. It is required to know the statistics of variable ($T, Y_k, \rho$) as a function of $f$ to compute the mean values of $\tilde{Y_k}$ and $\tilde{T}$, wherein an approach known as the presumed probability density function (PDF) comes into play in turbulent combustion calculation. 
%
%The connection between these averaged values and the instantaneous values relies on the turbulence-chemistry interaction model. Therefore, a presumed shape probability density function (PDF) approach is employed as the closure model for non-premixed combustion.
%
\subsubsection{Presumed Probability Density Function}
In modeling turbulent combustion, the probability density function (PDF), denoted as $p(f)$, represents the likelihood (or probability) of a fraction of time spent within a specified range ($\Delta f$) about any given value of the mixture fraction ($f$). Here, $f$ is the continuous random variable. The shape of the function $p(f)$ is influenced by the characteristics of turbulent fluctuations in $f$. In practice, the probability function, $p(f)$, is not directly known and is represented by a mathematical function designed to approximate the observed shapes of actual PDFs obtained through experimental observations.
\newline	
In non-premixed turbulent combustion modeling, the assumed PDF is often employed to efficiently account for the statistical impact of turbulence on scalar quantities like the mixture fraction. In lieu of directly solving full PDF of these quantities, which is computationally expensive, a predefined PDF shape is considered to approximate their statistical behavior. The commonly referred ``presumed beta PDF'' approach uses the ``$\beta$ distribution'' due to its flexibility in constraining the scalars between 0 and 1, which aligns well with variables like mixture fraction ($f$) in non-premixed combustion. The shape of the beta distribution is controlled by two positive shape parameters ($\alpha$ and $\beta$). The $\beta-$PDF is expressed as follows.
\begin{gather} \label{beta-pdf}
	p(f) = 	\frac{B_f(\alpha,\beta)}{B(\alpha, \beta)}, \\
	B_f(\alpha,\beta)  = \left[f^{\alpha-1} (1-f)^{\beta -1}\right], 
	\qquad B(\alpha, \beta) = \int_{0}^{1} B_f(\alpha,\beta) df \\
	\alpha = \mu_{f} X, \qquad \beta = (1-\mu_{f}) X,
	\qquad X = \left[ \frac{\mu_{f}(1-\mu_{f})}{\sigma_{f}}\right] -1
\end{gather}
where {$B(\alpha, \beta)$ is} the beta function, $\mu_f$ and $\sigma_f$ are the mean and variance of Favre-averaged mixture fraction ($\tilde{f}$, \eqn\ref{transport-f-favre}), respectively, which are defined \cite{Pope2000} as follows.
\begin{gather}
	\mu_f = \langle \tilde{f}\rangle = \left(\frac{\alpha}{\alpha + \beta}\right),
	%\end{gather}
	\qquad
	%\text{and}\quad 
	%\begin{gather}
	\sigma_f = \langle \tilde{f}^2\rangle -  \langle \tilde{f}\rangle^2  = \frac{\alpha \beta}{(\alpha + \beta)^2 (\alpha + \beta + 1)}
\end{gather}
Evidently, \eqn\eqref{beta-pdf} requires the mean and variance of $\tilde{f}$ (i.e., $\mu_f$ and $\sigma_f$). While the mean values ($\mu_f$) can be obtained using \eqn\eqref{transport-f-favre}, the variance is determined using the following transport equations for the variance ($\sigma_f$) of the Favre-averaged mixture fraction.
\begin{gather} \label{variance}
	\frac{\partial}{\partial t} (\overbar{\rho}{\sigma_f}) +  \frac{\partial (\overbar{\rho} \tilde{u_j} {\sigma_f} )}{\partial x_j} = \frac{\partial }{\partial x_j} \left( \Gamma_{{\sigma_f}} \frac{\partial {\sigma_f}}{\partial x_j}\right) + C_g \mu_t \cdot \left( \frac{\partial {\sigma_f}}{\partial x_j} \right) - C_d (\overbar{\rho} {\sigma_f}) \left(\frac{\epsilon}{k}\right)
\end{gather}
where, $C_g=0.286$, $C_d=2$, and $\Gamma_{{\sigma_f}}{=\Gamma_{f}}$ (\eqn\ref{transport-f}). 
Subsequently, after obtaining the mean and variance ($\mu_f$ and $\sigma_f$)  fields from \eqn(\ref{transport-f-favre}) and (\ref{variance}), the presumed PDF can be determined and used as a weighting function to determine the mean values of species mass fractions and density. 
In non-adiabatic systems, it is important to consider turbulent fluctuations using a joint PDF, represented as $p(f, H)$. However, computing joint PDF can be impractical for most applications. This challenge can be  simplified by assuming that enthalpy fluctuations are independent of the enthalpy level, i.e., $p(f, H) = p(f)\delta(H-\tilde{H})$. It means that heat exchange does not significantly impact turbulent fluctuations in enthalpy. Thus, the mean scalars {($\tilde{\phi}_i$ = $\tilde{T}$, $\tilde{Y}_{k}$)} are calculated using
\begin{gather} \label{non-adiabatic-pdf}
	\tilde{\phi}_i = \int_{0}^{1} \phi_i (f, \tilde{H}) p(f) \, df
\end{gather}
The Favre-averaged mean ethalphy ($\tilde{H}$, \eqn\ref{enthalpy}) is, however, essential to determine the scalar fields ($\tilde{\phi}_i$, \eqn\ref{non-adiabatic-pdf}). Furthermore, presumed PDF (\eqn\ref{beta-pdf}) serves as a key factor in calculating the mean time-averaged fluid density ($\overline{\rho}$) expressed as follows. 
\begin{gather}
	\frac{1}{\overline{\rho}} = \int_0^1 \frac{p(f)}{\rho(f)} \, df
\end{gather}
% (Main body of paper…)

\subsubsection{Chemistry Tabulation}
In the {CFD} solver employed for the current calculations \cite{fluent2011ansys},  look-up tables simplify and expedite simulations by pre-computing and storing key parameters{, avoiding the need to recalculating the entire} chemistry at every point in the flow. This process involves conducting numerous flamelet calculations for various possible mixture fractions, which represent local fuel-to-air (F/A) ratio.
\subsubsection{Look-up Tables}
Prior to performing the main simulations, appropriate combustion models and configurations are selected, including the fuel, oxidizer, reaction mechanism, and other relevant settings. A pre-processing step is then carried out to generate a look-up table through multiple flamelet calculations over a range of mixture fractions representing local  fuel-to-air (F/A) ratio.
For the present study, the look-up table is generated using \eqn{\eqref{non-adiabatic-pdf} } and the beta PDF equation (\eqn\ref{beta-pdf}). This pre-calculates the equation for various F/A ratio, storing the results in the table. The table stores pre-computed values of essential parameters like species mass fractions, density, and temperature. During the simulations, the solver calculates the local mixture fraction mean ($\mu_{f}$) using \eqn{\eqref{transport-f-favre}} and variance ($\sigma_{f}$) using \eqn{\eqref{variance}} and uses it to locate the corresponding values within the pre-computed table. In non-adiabatic systems, the values of each mass fraction, density, and temperature are determined from the calculated values of $\mu_{f}$, $\sigma_{f}$, and $\tilde{H}$. The readers may refer the source \cite{fluent2011ansys} for more details on the look-up tables.

\subsection{The Flamelet concept}
Turbulent flames represent a complex combustion regime, where turbulence and chemical reactions interact to produce intricate flame structures consisting of wrinkled, moving laminar sheets of reaction, characterized by localized heat release in specific regions. These regimes are narrow zones near the stoichiometric mixture fraction surface where combustion predominates (i.e., optimal fuel-oxidizer ratio), resulting in complex, dynamic flame behavior. The flamelet concept \cite{peters1984laminar} models a turbulent flame as an ensemble of laminar, locally one-dimensional reaction zones, referred to as flamelets, embedded within the turbulent flow field.
The validity of the flamelet concept is governed by the Damkohler number ($Da= {\tau_t}/{\tau_c}$), which compares the time scales of turbulent mixing ($\tau_t$) and chemical reactions ($\tau_c$). 	When $Da\gg 1$,  the flame is in the flamelet regime, where turbulence primarily wrinkles and stretches the flame without altering its internal structure. Under these conditions, the turbulent flame can be modeled as an ensemble of laminar flamelets convected by the flow. This concept is crucial in understanding turbulent combustion and has significant implications for engine design and optimization.
\newline
The flamelet model treats turbulent flames as locally laminar structures that are stretched and strained by turbulence. Although computationally efficient and chemically detailed, it assumes quasi-steady behavior, limiting its accuracy for transient or non-equilibrium phenomena such as ignition, extinction, and slow chemistry (e.g., \ce{NO_x}) \cite{fluent2011ansys}. Thermochemical properties (density, temperature, and species mass fractions) are precomputed as functions of mixture fraction ($f$) and scalar dissipation rate and stored in a flamelet library.
The flamelet library establishes relationships between scalar flow properties ($\phi$) and the mixture fraction ($f$), i.e., $\phi(f)$. However, turbulence-induced flame stretching modifies these relationships. To account for this effect, additional parameters such as strain rate or scalar dissipation rate ($\chi$) are incorporated into the library, improving modeling accuracy. The library includes detailed chemistry and reaction mechanisms, providing information on both major and minor species and enabling the evaluation of key combustion characteristics, including pollutant formation \cite{Versteeg2007}.
\newline
A commonly used laminar configuration for characterizing flamelets in turbulent flows is the counterflow diffusion flame, consisting of opposed, axisymmetric fuel and oxidizer jets. As the jet separation decreases or the jet velocity increases, the flame experiences increasing strain, departs from chemical equilibrium, and may ultimately extinguish. Owing to its self-similar structure, the governing equations reduce to a one-dimensional formulation along the jet axis, facilitating detailed evaluation of temperature and species mass fraction fields.
In a laminar counterflow flame, the mixture fraction ($f$) varies monotonically from unity (at the fuel stream) to zero (at the oxidizer stream). By transforming thermochemical quantities from physical space to mixture fraction space, species mass fractions and temperature can be parameterized in terms of the mixture fraction ($f$) and scalar dissipation rate ($\chi$). This formulation simplifies the chemical description and enables precomputation of flamelet solutions stored in look-up tables, thereby substantially reducing computational cost \citep{Versteeg2007, fluent2011ansys}. Detailed derivations and solution procedures for counterflow laminar diffusion flames are available in comprehensive reviews \citep{dixon1991structure,Bray1994}.
\subsubsection{Look-up table vs flamelet library}
In the present study, look-up table and flamelet library serve important distinct yet complementary functions. The flamelet library stores pre-computed relationships for laminar flamelets, encompassing comprehensive data on scalar flow properties ($\phi$), mixture fraction ($f$), strain rate, scalar dissipation rate ($\chi$), intricate chemical kinetics, and transport properties. In pre-computation, numerous flamelet calculations utilize the beta PDF to model mixture fraction distributions, representing local fuel-to-air ratio, and incorporating detailed chemical reaction mechanisms.
\newline	
Turbulent modeling variables ($k$ and $\varepsilon$), though not part of pre-computed data, influence flame characteristics, predicting turbulent flow fields that determine local conditions for applying pre-computed flamelet data. Thermodynamic equations for temperature, pressure, and density thermodynamic properties and heat release rates within the combustion zone. Additional equations for scalar dissipation rate ($\chi$), Favre-averaged mixture fraction ($\mu_{f}$) and its variance ($\sigma_{f}$), and distinguishing between adiabatic and non-adiabatic processes further refine temperature profiles and chemical reactions. These computations generate a look-up table for efficient retrieval during the simulation, {enhancing} computational efficiency by integrating detailed chemistry into combustion models \cite{Versteeg2007, fluent2011ansys}.

\subsubsection{Strain rate and Scalar Dissipation}
A characteristic strain rate ($a_{s}$) for a counterflow diffusion flamelet can be expressed as $a_s = (v/d)$, where $v$ denotes the relative speed between the fuel and oxidizer jets and $d$ denotes the distance between the jet nozzles. It quantifies the aerodynamic stretching of the flame, which signifies the flamelet behavior, including thickness and the interaction between chemical reactions and transport processes that directly influence the thickness, stability, and structure of the flame under varying flow conditions.
However, instead of using strain rate as a measure of departure from equilibrium, it is more convenient to use the scalar dissipation ($\chi$) defined as follows. 
\begin{gather}
	\chi = 2 {\mathcal{D}} |\nabla f|^2\qquad\text{where}\qquad 
	|\nabla f|^2 = \left[ \left(\frac{\partial f}{\partial x}\right)^2  + \left(\frac{\partial f}{\partial y}\right)^2  + \left(\frac{\partial f}{\partial z}\right)^2  \right]
	\label{scalar-dissipation-1}
\end{gather}
where, {$\mathcal{D}$} is the diffusion coefficient. 
The scalar dissipation ($\chi$) varies along the flamelet axis.  In the counterflow geometry, at the location where the stoichiometric mixture fraction ($f=f_{\rm st}$) is attained, the stoichiometric scalar dissipation  ($\chi = \chi_{\rm st}$) can be correlated \citep{peters1984laminar} to the strain rate ($a_s$) as follows.
\begin{gather} \label{scalar-dissipation}
	\chi_{\rm st} = (a_s/\pi){{\exp}\left(-2\zeta^2\right)},
	\qquad\text{where}\qquad \zeta = \texttt{erfc}^{-1}(2 f_{\rm st})
\end{gather}
where, $\texttt{erfc}$ is the complementary error function. As a flame experiences strain, its reaction zone narrows, intensifying the gradient  of $f$ at the stoichiometric point. The instantaneous stoichiometric scalar dissipation ($\chi_{st}$) captures non-equilibrium effects,  with $\chi\rightarrow 0$ indicating equilibrium, representing the reciprocal of a characteristic diffusion time. Conversely, increasing $\chi_{st}$ elevates the degree of non-equilibrium, surpassing critical threshold, potentially leading to local flamelet quenching.
\subsubsection{Embedding Diffusion Flamelets in Turbulent Flames}
A turbulent flame brush is modeled as a ensemble of discrete laminar flamelets. This approach simplifies turbulent flame interactions by leveraging pre-computed flamelet data and statistical distributions of key scalars parameters ($\phi=Y_{k}, T$), characterized by mixture fraction ($f$) and scalar dissipation rate ($\chi_{\rm st}$).
\newline
For adiabatic systems, the scalars ($\phi$) are expressed as follows.
\begin{gather}
	\phi = \phi(f, \chi_{\rm st})
\end{gather}
The Favre-averaged mean values of the scalars ($\tilde{\phi}$) in the turbulent flame are obtained using the  Favre-averaged joint PDF, $\tilde{p}(f,\chi_{\rm st})$, as follows.
\begin{gather} \label{joint-pdf-adiabatic}
	\tilde{\phi} = \frac{\overline{\rho \phi}}{\overline{\rho}} = \int \int \phi(f, \chi_{\rm st}) \tilde{p}(f,\chi_{\rm st}) \; df  d\chi_{\rm st}
\end{gather}
Since, $ f $ and $ \chi_{\rm st} $ are assumed to be statistically independent \cite{fluent2011ansys}, the joint PDF can be separated as follows.
\begin{gather}
	\tilde{p}(f,\chi_{\rm st}) = \tilde{p}_f(f) \tilde{p}_{\chi}(\chi_{\rm st})
\end{gather}
A $\beta$-probability density function ($\beta$-PDF) is assumed for the  Favre-averaged mixture fraction distribution, $\tilde{p}_f(f)$. The parameters of the $\beta$-PDF are determined from the transport 
equations for the Favre-averaged mixture fraction ($\tilde{f}$), and and its variance  ($\sigma_f$).
%where, a beta distribution is assumed for $\tilde{p}_f(f)$, which is specified by solving the transport equations for the Favre-averaged mixture fraction ($\tilde{f}$) and the Favre-averaged variance of the mixture fraction ($\sigma_{f}$).
%
Fluctuations in $\chi_{\rm st}$ are neglected, such that $\tilde{p}_{\chi}$ is approximated by a Dirac delta function centered at the Favre-averaged scalar dissipation rate ($\tilde{\chi}$), 
i.e.,
\begin{gather}
	\tilde{p}_{\chi} (\chi_{\rm st})= \delta (\chi - \tilde{\chi})
\end{gather}
For Reynolds-Averaged Navier-Stokes (RANS) simulations, the mean scalar dissipation rate ($\tilde{\chi}$) is modeled as follows.
\begin{gather} \label{mean-scalar-dissipation-rate}
	\tilde{\chi}_{\rm st} = \frac{C_{\chi} \epsilon {\sigma_{f}}}{\tilde{k}}
\end{gather}
where $ C_{\chi}=2$, $\epsilon$ is the turbulent dissipation rate, and $\tilde{k}$ is the Favre-averaged turbulent kinetic energy. 
\newline
To avoid runtime convolutions of PDFs, the integrations are pre-processed and stored in look-up tables. For adiabatic flows, these tables depend on three parameters ($\tilde{f}$, $\sigma_{f}$, $\tilde{\chi}_{\rm st}$). 
For non-adiabatic steady laminar flamelets, an additional parameter, enthalpy ($H$),  is required, which significantly increases the computational cost across a range of enthalpies. Therefore, adiabatic mass fractions are used \cite{binniger1998numerical, muller1994partially} assuming that heat exchange to the system has a negligible effect on species mass fractions. 
The temperature is computed using \eqn\eqref{favre-averaged-enthalpy} for the range of mean enthalpy ($\tilde{H}$) exchange.  Consequently, mean temperature and density PDF tables incorporate an additional dimension of mean enthalpy ($\tilde{H}$). In the special case where $\tilde{\chi}_{\rm st} = 0$, species mass fractions are computed as function of parameters ($\tilde{f}$, $\sigma_{f}$, $\tilde{H}$) representing the non-adiabatic equilibrium solution.
\subsubsection{Flamelet Generation}
The laminar counterflow diffusion flame equations can be reformulated  \citep{pitsch1998consistent} by transforming the independent variable from physical space ($x$) to mixture fraction space ($f$). 
The present solver facilitates efficient and accurate computation by using a simplified formulation in mixture fraction space \cite{pitsch1996three}, solving $N$ equations that govern the species mass fractions ($Y_k$), and a single equation for temperature ($T$) as follows.
\begin{gather}  \label{species-mass-fraction-mixture-fraction-space}
	\rho \frac{\partial {Y_k}}{\partial t} = \frac{1}{2} \rho \chi \frac{\partial ^2 {Y_k}}{\partial f^2} + {S_k}
\end{gather}
\begin{gather} \label{temperature mixture fraction space}
	\rho \frac{\partial T}{\partial t} = \frac{1}{2} \rho \chi \frac{\partial ^2 T}{\partial f^2} - \frac{1}{c_p} \sum_{k} {H_k S_k}  + \frac{1}{2 c_p} \rho \chi \left[\frac{\partial c_p}{\partial f } + \sum_{k} c_{p,k} \frac{\partial Y_k}{\partial f}\right] \frac{\partial T}{\partial f}
\end{gather} 
where, $c_{p,k}$ and $c_p$ denote the specific heat of the {$k$}-th species and the mixture-averaged specific heat, respectively,  {$S_{k}$} stands for the reaction rate of the {$k$}-th species which depends on the local concentration of reactants, temperature, and specific reaction mechanisms involved, and $H_{k}$ represents the specific enthalpy of the $k$-th species. 
\newline
The scalar dissipation ($\chi$) across the flamelet is modeled by extending \eqn\eqref{scalar-dissipation} to variable density ($\rho_r$) as follows  \citep{kim1997extinction}. 
\begin{gather} \label{scalar-dissipation-model}
	\chi(f) = \left(\frac{3a_s}{4 \pi}\right) \frac{\left(\rho_r + 1 \right)^2}{(2\rho_r + 1)} \exp\left(-2\zeta^2 \right), 
	\qquad \rho_r= \sqrt{\frac{\rho_{\infty}}{\rho}},\qquad 
	\zeta = \operatorname{erfc}^{-1}(2f)
\end{gather}
where, $\rho_{\infty}$ is the density of the oxidizer stream. 
%\newline
In the \textit{steady laminar flamelet method} (SLFM) approach, the time-dependent terms in \eqns \eqref{species-mass-fraction-mixture-fraction-space} and \eqref{temperature mixture fraction space} are omitted and justified as $Da \gg 1$ (i.e., $\tau_c \gg \tau_t$). This assumption makes SLFM particularly suitable for handling turbulence-induced non-equilibrium conditions primarily caused by aerodynamic strain (the deformation or stretching of the flame front due to velocity gradients in the turbulent flow) rather than by slow chemical kinetics.
\newline
A multiple flamelet files can be imported to convolve presumed PDFs (see \eqn\ref{joint-pdf-adiabatic}),  and construct look-up tables. The flamelet can be generated in the solver \citep{fluent2011ansys} or with separate stand-alone computer codes.  In the commercial solver used, multiple steady diffusion flamelets can be generated across a range of strain rates to account for variations in the strain field within multi-dimensional simulations. If the number of diffusion flamelets is specified to be greater than one, they are generated at scalar dissipation rates determined by \eqn\eqref{eq:conditional}.
\begin{gather}\label{eq:conditional}
	f(x) = 
	\begin{cases}
		{{10{\chi_{i-1}}} } & \text{for} \quad \chi_{i-1} < 1{~\rm s}^{-1} \\
		\chi_{i-1} + \Delta \chi  & \text{for} \quad \chi_{i-1}  \geq  1{~\rm s}^{-1}
	\end{cases}
\end{gather}
where, $i$ ranges from $1$ to the specified maximum number of diffusion flamelets, $\chi_0$ is the initial scalar dissipation rate, and $\Delta \chi$ is the dissipation rate step. Flamelets are generated until the maximum number is reached or they extinguish, with extinguished flamelets excluded from the library.

\subsubsection{Non-Adiabatic Steady Diffusion Flamelets}
For adiabatic steady diffusion flamelets, the methodology assumes that the species profiles of the flamelet remain unaffected by heat exchange \citep{binniger1998numerical, muller1994partially}, i.e., independent of thermal interactions. To model non-adiabatic effects, this limitation is relaxed by presenting the mean enthalpy as an additional parameter. By systematically varying the enthalpy field around the adiabatic reference state, the non-flamelet library is constructed, thus shifting the flame temperature and density without re-solving the detailed chemistry. This approach preserves the major species structure while incorporating the effects of thermal quenching and heat-transfer-induced variations in density and scalar dissipation.
Following the generation of diffusion flamelets, the flamelet profiles are convolved with assumed-shaped PDFs, as described by \eqn\eqref{joint-pdf-adiabatic}, and are then tabulated for look-up. The non-adiabatic PDF tables are characterized by the following functions.
\begin{gather}
	\widetilde{T} \left(\widetilde{f}, {\sigma_{f}}, \widetilde{H}, \widetilde{\chi}\right) \\
	\begin{cases}
		{\widetilde{Y_k}} \left(\widetilde{f}, {\sigma_{f}}, \widetilde{H} \right),  &
		\text{for } \chi = 0  \\
		{\widetilde{Y_k}} \left(\widetilde{f}, \sigma_{f}, \widetilde{\chi} \right),  &
		\text{for } \chi \neq 0 
	\end{cases}\\
	\overline{\rho} \left(\widetilde{f}, \sigma_{f}, \widetilde{H} , \widetilde{\chi} \right)
\end{gather}
During the solution process, the equations for the mean mixture fraction, mixture fraction variance, and mean enthalpy are solved. The scalar dissipation field is calculated from the turbulence field and the mixture fraction variance (\eqn\ref{mean-scalar-dissipation-rate}). The mean values of temperature, density, and species mass fraction are obtained from the PDF look-up table.

%--------------------------------------
\subsection{Turbulence Models}
%--------------------------------------
%
In a turbulent flow, the instantaneous velocity field ($u_i = \overline{u}_i + u_i'$) is decomposed into a time-averaged (mean) component ($\overline{u}_i$) and fluctuating component ($u_i^{\prime}$), representing deviations from the mean velocity. The non-linear interaction of fluctuating components can be interpreted as additional terms in the momentum equation, accounting for momentum transfer due to turbulence. 
These additional fluxes are captured by the Reynolds stresses ($R_{ij} = \overline{u_i^{\prime} u_j^{\prime}} \approx \overline{\rho u_i^{\prime\prime} u_j^{\prime\prime}}$), which represent the correlation between the velocity fluctuations in the $i$- and $j$-directions.  The Reynolds stresses are central to turbulence modeling and are typically linked to turbulence quantities such as  the turbulent kinetic energy ($k$) and dissipation rate  ($\epsilon$) through empirical or closure models.
\\
To close the system of Reynolds-Averaged Navier-Stokes (RANS) equations, turbulence models
%, such as $k-\epsilon$, $k-\omega$, and Reynolds Stress Models (RSM), etc., 
are typically used to estimate the Reynolds stresses ($R_{ij}$). These models aim to predict the turbulence effects without directly solving for the fluctuating velocity components. 
The turbulence models employed in this study, namely the standard $k$--$\epsilon$, realizable $k-\epsilon$ model, SST $k$--$\omega$, and Reynolds Stress Model (RSM), are formulated in their Favre-averaged forms. The conventional formulations of these models are detailed in our recent study \citep{kumar2024assessment}; therefore, only the Favre-averaged governing equations are presented here for completeness.
\subsubsection{Standard $k-\epsilon$ model}\label{sec:ske}
The Favre-averaged version of the $k-\epsilon$ model \citep{launder1972lectures}, which includes the equations for turbulent kinetic energy ($k$) and dissipation rate ($\epsilon$) is written as follows.
\begin{gather}\label{turbulent}
	\frac{\partial (\overbar{\rho} k)}{\partial t} + \frac{\partial (\overbar{\rho} \tilde{u}_j k)}{\partial x_j} = \frac{\partial}{\partial x_j} \left[ \left( \mu + \frac{\mu_t}{\sigma_k} \right) \frac{\partial k}{\partial x_j} \right] + P_k - \tilde{\rho} \epsilon
	%\end{gather}
	%
	\\[5pt]	
	%\begin{gather}
	\frac{\partial (\overbar{\rho} \epsilon)}{\partial t} + \frac{\partial (\overbar{\rho} \tilde{u}_j \epsilon)}{\partial x_j} = \frac{\partial}{\partial x_j} \left[ \left( \mu + \frac{\mu_t}{\sigma_\epsilon} \right) \frac{\partial \epsilon}{\partial x_j} \right] + C_{\epsilon 1} \frac{\epsilon}{k} P_k - C_{\epsilon 2} \frac{\epsilon^2}{k}\overbar{\rho} 
	\label{epsilon-equation}
\end{gather}
The turbulent production term ($P_k$), the turbulent viscosity ($\mu_t$), and turbulent dissipation rate ($\epsilon$) are expressed as follows.
\begin{gather}
	P_k = \mu_t \left( \frac{\partial \tilde{u}_i}{\partial x_j} + \frac{\partial \tilde{u}_j}{\partial x_i} \right) \frac{\partial \tilde{u}_i}{\partial x_j}
	%\end{gather}
	%
	\\[5pt]	
	%\begin{gather}
	\mu_t = \overbar{\rho} C_\mu \frac{k^2}{\epsilon} 
	\label{turbulent-viscosity}
	%\end{gather}
	%
	\\[5pt]	
	%\begin{gather}
	\label{turbulent-epsilon}
	{\epsilon} 
	= 2\,\nu\,\overline{S_{ij}^{\prime\prime} S_{ij}^{\prime\prime}}
	= \nu\,\overline{
		\left(
		\frac{\partial u_i^{\prime\prime}}{\partial x_j} 
		+ \frac{\partial u_j^{\prime\prime}}{\partial x_i}
		\right)
		\frac{\partial u_i^{\prime\prime}}{\partial x_j}}
\end{gather}
where, $\mu$, $\nu$ and $u_i^{\prime\prime}$ are the the molecular viscosity, kinematic viscosity and fluctuating velocity components, respectively.
In $k-\epsilon$ model (\eqns\ref{turbulent} -- \ref{turbulent-epsilon}), the values of the model constants ($C_{\mu} = 0.09$, $\sigma_{k} = 1.0$, $\sigma_{\epsilon} = 1.3$, $C_{\epsilon 1} = 1.44$, $C_{\epsilon 2} = 1.92$) have been experimentally determined \citep{launder1972lectures} for fundamental turbulent flows, including frequently encountered shear flows like boundary layers, mixing layers, and jets, as well as for decaying isotropic grid turbulence.
\subsubsection{Realizable $k-\epsilon$ model}
In the realizable $k-\epsilon$ model \cite{shih1994new},  the transport equation for the turbulent kinetic energy ($k$) retains the same form as in the standard $k-\epsilon$ model  (\eqn\ref{turbulent}). However, the transport equation for the turbulent dissipation rate ($\epsilon$) is reformulated and derived from the transport equation of the mean-square vorticity fluctuation, leading to a modified production term and a variable model coefficient, as follows.
\begin{gather}
	\frac{\partial (\overline{\rho}\,\epsilon)}{\partial t}
	+ \frac{\partial}{\partial x_j}\left(\overline{\rho}\,\tilde{u}_j\,\epsilon\right)
	= \frac{\partial}{\partial x_j}\left[\left(\mu + \frac{\mu_t}{\sigma_\epsilon}\right)
	\frac{\partial \epsilon}{\partial x_j}\right]
	+ \overline{\rho}\,C_1\,\tilde{S}\,\epsilon
	- \overline{\rho}\,C_2\,\frac{\epsilon^2}{k + \sqrt{\nu\,\epsilon}}
	\label{realizable-favre-epsilon}
	%\end{gather}
	%
	\\[5pt]	
	%\begin{gather}
	\text{where}\quad 
	C_1 = \max\left(0.43, \frac{\tilde{\eta}}{\tilde{\eta} + 5}\right),\quad
	\tilde{\eta} = \tilde{S}\left(\frac{k}{\epsilon}\right),\quad
	\tilde{S} = \sqrt{2\,\tilde{S}_{ij}\,\tilde{S}_{ij}},\\[4pt]
	C_2 = 1.9,\quad 
	C_{1\epsilon} = 1.44,\quad 
	\sigma_k = 1.0,\quad 
	\sigma_\epsilon = 1.2.
	\label{eq:favre-constants}
\end{gather}
where, $\mu_t$ and $\epsilon$ are defined using \eqns\eqref{turbulent-viscosity} and \eqref{turbulent-epsilon}, similar to standard $k-\epsilon$ model. 
\newline
For a model to be realizable, it must maintain non-negativity of the normal stress (\eqn\ref{non-negative-normal-stress}) and the ``Cauchy-Schwarz inequality” \cite{tennekes1972first} (\eqn\ref{schwarz-inequality}), expressed as follows.
\begin{gather}
	\widetilde{u_\alpha^{\prime\prime} u_\alpha^{\prime\prime}} \ge 0 \quad (\alpha=1,2,3), \label{non-negative-normal-stress}\\[4pt]
	\frac{
		\left(\widetilde{u_\alpha^{\prime\prime} u_\beta^{\prime\prime}}\right)^2
	}{
		\widetilde{u_\alpha^{\prime\prime} u_\alpha^{\prime\prime}}\;
		\widetilde{u_\beta^{\prime\prime} u_\beta^{\prime\prime}}
	}
	\le 1
	\qquad (\alpha = 1,2,3;\ \beta = 1,2,3). \label{schwarz-inequality}.
\end{gather}
Violating these conditions leads to non-physical Reynolds stress tensors, which the realizable 
$k-\epsilon$ model avoids by introducing a variable $C_\mu$.The Cauchy-Schwarz inequality constrains the magnitude of the Reynolds shear stresses relative to the normal stresses, thereby ensuring that the Reynolds stress tensor remains physically realizable. Although the eddy-viscosity hypothesis retains an isotropic form for the turbulent viscosity, realizability in the model is achieved by replacing the coefficient $C_\mu$ (constant in \eqn\ref{turbulent-viscosity}) with a variable formulation. The modified $C_\mu$ depends on local strain and rotation rates of the mean flow and incorporates additional terms and coefficients to improve turbulence prediction.
\begin{gather}
	C_\mu \;=\; \frac{1}{A_0 + A_s\!\left(\dfrac{k\,U^{*}}{\epsilon}\right)}
	\label{eq:Cmu_favre}
	%\end{gather}
	%
	\\[5pt]	
	%\begin{gather}
	U^{*} \equiv \sqrt{\;\tilde{S}_{ij}\,\tilde{S}_{ij} \;+\; \tilde{\Omega}_{ij}\,\tilde{\Omega}_{ij}\; }
	\label{eq:Ustar_favre}
	%\end{gather}
	%
	\\[5pt]	
	%\begin{gather}
	\tilde{S}_{ij} \;=\; \frac{1}{2}\left(\frac{\partial \tilde{u}_i}{\partial x_j} + \frac{\partial \tilde{u}_j}{\partial x_i}\right),
	\qquad
	\overline{\Omega}_{ij} \;=\; \frac{1}{2}\left(\frac{\partial \tilde{u}_i}{\partial x_j} - \frac{\partial \tilde{u}_j}{\partial x_i}\right),
	\qquad
	\widetilde{\Omega}_{ij} \;=\; \overline{\Omega}_{ij} - \epsilon_{ijk}\,\omega_k
	\label{eq:strain_vort_defs}
\end{gather}
where, $\epsilon_{ijk}$ is the Levi-Civita symbol, and $\omega_k$ is the system angular velocity. The model constant $A_0$ and $A_s$ are given as
\begin{gather}
	A_0 = 4.04, 
	\qquad
	A_s = \sqrt{6}\,\cos\phi,
	\qquad
	\phi = \frac{1}{3}\cos^{-1}(\sqrt{6}\,W),
	\label{eq:favre-phiW} \\[6pt]
	W = \frac{\widetilde{S}_{ij}\,\widetilde{S}_{jk}\,\widetilde{S}_{ki}}{\widetilde{S}^3}
	\qquad
	\widetilde{S} = \sqrt{\widetilde{S}_{ij}\,\widetilde{S}_{ij}},
	\label{eq:favre-Sij}
\end{gather}
%
%These terms aim to assist the model's turbulence-predicting ability.
%
\subsubsection{Shear stress transport (SST) $k-\omega$ model}
The SST $k-\omega$ model \citep{Menter1994,kumar2024assessment} provides accurate predictions for flow separation under adverse pressure gradients and ensures the correct asymptotic behavior in the near-wall region by effectively blending the $k-\omega$ model \citep{wilcox1988reassessment, Wilcox1998} in the near-wall region and the $k-\epsilon$ model \citep{launder1972lectures} in the free-stream region of the flow. 
The Favre-averaged form of the SST $k-\omega$ model, which includes the equations for turbulent kinetic energy ($k$) and specific dissipation rate ($\omega \equiv {\epsilon}/{k}$) is written as follows.
\begin{gather} \label{tke-sst}
	\frac{\partial (\overline{\rho} k)}{\partial t} + \frac{\partial}{\partial x_j} \left( \overline{\rho} k \tilde{u}_j \right) 
	= \frac{\partial}{\partial x_j} \left[ \left( \mu + \frac{\mu_t}{\sigma_k} \right) \frac{\partial k}{\partial x_j} \right] + {P_{k}} - {\epsilon_{k}}
	%\end{gather}
	\\[5pt]
	%\begin{gather} 
	\label{omega-transport}
	\frac{\partial (\overline{\rho} \omega)}{\partial t} + \frac{\partial}{\partial x_j} \left( \overline{\rho} \omega \tilde{u}_j \right) 
	= \frac{\partial}{\partial x_j} \left[ \left( \mu + \frac{\mu_t}{\sigma_\omega} \right) \frac{\partial \omega}{\partial x_j} \right] + {P_{\omega}} -  {\epsilon_{\omega}} + D_\omega
\end{gather}
where $\mu_t$ is the turbulent viscosity; $D_\omega$ is the cross-diffusion term; $P_k$ and $P_{\omega}$ denote the production of $k$ and $\omega$, respectively. %\hl{The model constant $\alpha$ corresponds to the dissipation in $\omega$, while $\beta^{\ast}$ and $\beta_{i}$ correspond to the production of $k$ and $\omega$, respectively.} 
The turbulent viscosity ($\mu_t$) is related with the mean strain rate tensor ($\tilde{S}_{ij}$) as follows.
\begin{gather} \label{turbulent-viscosity-SST-k-omega}
	\mu_t = \frac{\overline{\rho} k}{\omega} \left( \max \left[ \frac{1}{\alpha^*}, \frac{\tilde{S}_{ij} F_2}{\alpha_1 \omega} \right] \right)^{-1},
	%\end{gather}
	%
	\qquad
	%\begin{gather}
	\tilde{S}_{ij} = \frac{1}{2} \left( \frac{\partial \tilde{u}_i}{\partial x_j} + \frac{\partial \tilde{u}_j}{\partial x_i} \right)
\end{gather}
The coefficient $\alpha^{\ast}$ in \eqn\eqref{turbulent-viscosity-SST-k-omega} is unity ( $\alpha^{\ast}  = 1$) in the high Reynolds number ($Re$) form of the model; however, it reduces the turbulent viscosity{, introducing} a ``\textit{low $Re$ correction}'' as follows.
\begin{gather}
	\alpha^\ast = \alpha_\infty^\ast \left(\frac{\alpha_0^\ast + Re_k}{1 + Re_k}\right) \label{eq:26}  
	\qquad \text{where} \qquad  Re_k = \frac{Re_t}{R_k}, \qquad
	Re_t = \frac{\rho k}{\mu_t \omega}, \qquad \alpha_0^\ast = \frac{\beta_i}{3} %,  \nonumber
\end{gather}
The turbulent Prandtl numbers ($\sigma_k$ and $\sigma_{\omega}$) are expressed as follows.
\begin{gather}
	\sigma_k = \left( \frac{F_1}{\sigma_{k,1}} + \frac{1 - F_1}{\sigma_{k,2}} \right)^{-1}; \qquad
	\sigma_\omega = \left( \frac{F_1}{\sigma_{\omega,1}} + \frac{1 - F_1}{\sigma_{\omega,2}} \right)^{-1} 
\end{gather}
The blending functions ($F_i$) and auxiliary functions ($\phi_i$) are expressed as follows.
\begin{gather}
	F_1 = \tanh \left( \phi_1^4 \right) \label{F1}; \qquad
	F_2 = \tanh \left( \phi_2^2 \right)
	%\end{gather}
	\\[5pt]
	%\begin{gather}
	\phi_1 = \min \left[ \max \left( g_1, g_2 \right), \frac{4 \overline{\rho} k}{\sigma_{\omega,2} D_{\omega}^+ y^2} \right]; \qquad
	\phi_2 = \max \left( 2 g_1, g_2 \right) \\
	g_1 = \frac{\sqrt{k}}{0.09 \omega y}; \qquad
	g_2 = \frac{500 \mu}{\overline{\rho} y^2 \omega}; \qquad
	D_\omega^+ = \max \left[ 2 \overline{\rho} \frac{1}{\sigma_{\omega,2}} \frac{1}{\omega} \frac{\partial k}{\partial x_j} \frac{\partial \omega}{\partial x_j}, 10^{-10} \right]
\end{gather}
where $y$ is the distance to the nearest surface, and $D_\omega^+$ is the positive part of the cross-diffusion term ($D_\omega$).
The production terms, $P_{k}$ in \eqn\eqref{tke-sst} and $P_{\omega}$ in \eqn\eqref{omega-transport}, are expressed as follows.
\begin{gather} 
	P_{k} = 2 \mu_t (\tilde{S}_{ij} \tilde{S}_{ij}); \qquad P_{\omega} = \left( \frac{\alpha}{\nu_t} \right) 2 \mu_t (\tilde{S}_{ij} \tilde{S}_{ij})\\
	\text{where} \quad \alpha = \frac{\alpha_\infty}{\alpha^\ast} \left(\frac{\alpha_0 + Re_\omega}{1 + Re_\omega}\right), \qquad Re_\omega = \frac{\rho k}{\mu \omega}
	%\end{gather}
	\\[5pt]
	%\begin{gather*}
	\alpha_\infty=[F_1\alpha_{\infty,1}+\left(1-F_1\right)\alpha_{\infty,2\ }] 
	\label{eq:42}, \quad 
	\quad\text{where,}\quad 
	\alpha_{\infty,j}=\left(\frac{\beta_{i,j}\ }{\beta_\infty^\ast} -  \frac{\kappa^2}{\sigma_{\omega,j}\sqrt{\beta_\infty^\ast}}\right) \nonumber
\end{gather}
The local turbulence dissipation, $\epsilon_k$ (in \eqn\ref{tke-sst}) and $\epsilon_{\omega}$ (in \eqn\ref{omega-transport}), are accurately modeled as follows.
\begin{gather}
	\epsilon_k = \overline{\rho} \beta^\ast k \omega, \qquad \epsilon_{\omega}= \overline{\rho} \beta_i k \omega^2\\
	\text{where} \quad 
	\beta^\ast = \beta_i^\ast \left[1 + \zeta^\ast F\left(M_t\right)\right], \label{compressibility-correction} 
	\qquad \beta_i = F_1 \beta_{i,1} + (1 - F_1) \beta_{i,2} \\[5pt]
	\beta_i^\ast = \beta_\infty^\ast \left[\frac{(4/15) + \left(Re_{\beta}\right)^4}{1 + \left(Re_{\beta}\right)^4}\right], \quad
	Re_{\beta} = \frac{Re_t}{R_\beta}
\end{gather}
The compressibility correction function, $F(M_t)$ for $\beta^\ast$ (\eqn\ref{compressibility-correction}) is defined as follows:
\begin{gather}
	F\left(M_t\right) = 
	\begin{cases}
		0 & M_t < M_{t0} \\
		(M_t^2 - M_{t0}^2) & M_t > M_{t0}
	\end{cases}\\
	M_t^2 = ({2k}/{a^2}), \quad 
	a^2 = {\gamma R \tilde{T}}
	\label{speed-of-sound}
\end{gather}
where $a$ is the speed of sound, $\gamma$ is the adiabatic index, $R~(= {R_u}/{M})$ is the specific gas constant, $R_u$ is the universal gas constant, and $M$ is the molecular weight.
\\
The cross-diffusion term ($D_\omega$, \eqn\ref{omega-transport}) blends the standard $k-\omega$ and standard $k-\epsilon$ models as follows.
\begin{gather}	\label{Domega}
	D_\omega = 2 \left( 1 - F_1 \right) \overline{\rho} \frac{1}{\sigma_{\omega,2}} \frac{\partial k}{\partial x_j} \frac{\partial \omega}{\partial x_j}
\end{gather}
In SST $k-\omega$ model (\eqn\ref{tke-sst} - \ref{Domega}), the values of the model constants are $ \sigma_k = 2 $, $ \sigma_\omega = 2 $,   $\sigma_{k,1} = 1.176$, $\sigma_{\omega,1} = 2.0$, $\sigma_{k,2} = 1.0$, $\sigma_{\omega,2} = 1.168$, $\alpha = 0.31$, $\beta_{i,1} = 0.075$, $\beta_{i,2} = 0.0828$, $\beta_\infty^\ast = 0.09$, $\zeta^\ast = 1.5$, $R_k = 6$, $R_\beta = 8$, $\alpha_0^\ast = \beta_i/3$, $\beta_i = 0.072$, $\alpha_\infty^\ast = 1$, $\alpha_0 = 0.072$, $\alpha_\infty = 1$ and $M_{t0} = 0.25$. 

%--------------------------------------	
\subsubsection{Reynolds Stress Model (RSM)}
%--------------------------------------
%
The Reynolds stress Model (RSM) directly solves transport equations for the Reynolds stresses ($\tilde{R}_{ij} = \widetilde{u_i^{\prime\prime}u_j^{\prime\prime}} = {\overline{\rho\,u_i^{\prime\prime}u_j^{\prime\prime}}}/{\overline{\rho}}$) alongside the dissipation rate ($\epsilon$) transport equation (\eqn\ref{epsilon-equation} used in standard $k-\epsilon$ model). For three-dimensional flows, RSM approach requires solving seven additional equations, along with the mean flow (continuity and momentum) equations. 
\newline
The transport equation for the Favre-averaged Reynolds stresses ($\tilde{R}_{ij}$) is given as follows:
\begin{gather}
	\underbrace{\frac{\partial}{\partial t}\big(\overline{\rho}\,\tilde{R}_{ij}\big)}_{\text{Local Time Derivative}} 
	+
	\underbrace{\frac{\partial}{\partial x_k}\big(\overline{\rho}\,\tilde{u}_k\,\tilde{R}_{ij}\big)}_{\tilde{C}_{ij} \equiv \text{Convection Term}} 
	=
	- 
	\underbrace{\frac{\partial}{\partial x_k}\Big[\overline{\rho}\,\widetilde{u_i^{\prime\prime}u_j^{\prime\prime}u_k^{\prime\prime}}
		+ \overline{p' u_i^{\prime\prime}}\,\delta_{jk} + \overline{p' u_j^{\prime\prime}}\,\delta_{ik}\Big]}_{\tilde{D}_{T,ij} \equiv \text{Turbulent Diffusion}} 
	+ 
	\underbrace{\frac{\partial}{\partial x_k}\Big[\mu\,\frac{\partial \tilde{R}_{ij}}{\partial x_k}\Big]}_{\tilde{D}_{L,ij} \equiv \text{Molecular Diffusion}} 
	\nonumber \\[4pt]
	- 
	\underbrace{\overline{\rho}\Big(\tilde{R}_{ik}\frac{\partial \tilde{u}_j}{\partial x_k} + \tilde{R}_{jk}\frac{\partial \tilde{u}_i}{\partial x_k}\Big)}_{\tilde{P}_{ij} \equiv \text{Stress Production}} 
	+
	\underbrace{\overline{\,p'\!\left(\frac{\partial u_i^{\prime\prime}}{\partial x_j} + \frac{\partial u_j^{\prime\prime}}{\partial x_i}\right)\,}}_{\tilde{\phi}_{ij} \equiv \text{Pressure--Strain}} 
	- 
	\underbrace{2\mu\left(\overline{\frac{\partial u_i^{\prime\prime}}{\partial x_k}\frac{\partial u_j^{\prime\prime}}{\partial x_k}}\right)}_{\tilde{\epsilon}_{ij} \equiv \text{Dissipation}}
	\label{rsm-equation}
\end{gather}
\eqn\eqref{rsm-equation} requires special attention in modeling the three terms ($\tilde{D}_{T}$, $\tilde{\phi}$, $\tilde{\epsilon}$). The turbulent diffusion ($\tilde{D}_{T}$) and dissipation tensor ($\tilde{\epsilon}$)  are modelled as follows.
\begin{gather}
	\tilde{D}_{T,ij} = \frac{\partial}{\partial x_k}\left(\frac{\mu_t}{\sigma_k} \frac{\partial \widetilde{{u}_i^{\prime\prime} {u}_j^{\prime\prime}}}{\partial x_k}\right)
	\label{turbulent-diffusion-term}
	%\end{gather}
	%
	\\[5pt]
	%\begin{gather}
	\tilde{\epsilon}_{ij} = \overline{\rho} \epsilon \frac{2}{3} \delta_{ij} \left(1 + 2 \frac{k}{a^2}\right)
	\label{dissipation-tensor}
	%\end{gather}
	% 
	\\[5pt]
	%\begin{gather}
	k = \frac{1}{2} \mathrm{tr}(\tilde{R}_{ij}) = \frac{1}{2} \widetilde{{u}_i^{\prime\prime} {u}_i^{\prime\prime}}
	\label{eq:49}
\end{gather}
where $a$ is the speed of sound (\eqn\ref{speed-of-sound}), $\epsilon$ is the scalar dissipation rate (\eqn\ref{epsilon-equation}), and the turbulent kinetic energy ($k$) is obtained by taking the trace of the Reynolds stress tensor ($\tilde{R}_{ij}$).
%
%\subsubsection*{\deleted[id=AK]{(a)} Linear Pressure Strain - Reynolds Stress Model (LPS-RSM)}
%
\newline 
The pressure strain term ($\tilde{\phi}_{ij}$) of \eqn\eqref{rsm-equation} is modeled using Linear Pressure Strain - Reynolds Stress Model (LPS-RSM) model \citep{Gibson1978, Fu1987, Launder1989}. 
The LPS-RSM model linearly decomposes the pressure strain term ($\tilde{\phi}_{ij}$) into the slow pressure strain ($\tilde{\phi}_{ij,1}$), rapid pressure strain ($\tilde{\phi}_{ij,2}$), and wall reflection strain ($\tilde{\phi}_{ij,w}$) as follows.
\begin{gather}
	\tilde{\phi}_{ij} = \tilde{\phi}_{ij,1} + \tilde{\phi}_{ij,2} + \tilde{\phi}_{ij,w}
	\label{lps-pressure-strain} 
	%\end{gather}
	%
	\\[5pt]
	\text{where,} \qquad
	%
	% --- Slow pressure-strain term ---
	%\begin{gather}  
	\label{slow-pressure-strain} 
	\tilde{\phi}_{ij,1} = -2\,C_1\,\overline{\rho}\,\epsilon\,b_{ij}
	%\end{gather}
	\\[5pt]
	% --- Rapid pressure-strain term ---
	%\begin{gather} 
	\label{rapid-pressure-strain}
	\tilde{\phi}_{ij,2} = -C_2 \left[
	\left(\tilde{P}_{ij} - \tilde{C}_{ij}\right)
	- \frac{2}{3}\,\delta_{ij}
	\left(\tfrac{1}{2}\tilde{P}_{kk} - \tfrac{1}{2}\tilde{C}_{kk}\right)
	\right]
	%\end{gather}
	\\[5pt]
	% --- Wall reflection term ---
	%\begin{gather} 
	\label{wall-reflection-term}
	\begin{aligned}
		\tilde{\phi}_{ij,w}
		&\equiv C_1' \frac{\epsilon}{k}
		\left(
		\widetilde{u_k^{\prime\prime}u_m^{\prime\prime}} n_k n_m \delta_{ij}
		- \frac{3}{2}\widetilde{u_i^{\prime\prime}u_k^{\prime\prime}} n_j n_k
		- \frac{3}{2}\widetilde{u_j^{\prime\prime}u_k^{\prime\prime}} n_i n_k
		\right)
		\frac{C_l k^{3/2}}{\epsilon d} \\[4pt]
		&\quad + C_2'
		\left(
		\tilde{\phi}_{km,2} n_k n_m \delta_{ij}
		- \frac{3}{2}\tilde{\phi}_{ik,2} n_j n_k
		- \frac{3}{2}\tilde{\phi}_{jk,2} n_i n_k
		\right)
		\frac{C_l k^{3/2}}{\epsilon d}
	\end{aligned}
\end{gather}
where, $C_1 = 1.8$, $C_2 = 0.60$, $C_\mu = 0.09$,  $C_1^\prime = 0.5$, $C_2^\prime = 0.3$, $\tilde{P}_{ij}$ and $\tilde{C}_{ij}$ are defined in \eqn\eqref{rsm-equation},  $C_l = \left({C_\mu^{3/4}}/{\kappa}\right)$, $n_k$ is the $x_k$ component of the unit normal to the wall, $d$ is the normal distance to the wall, $\kappa = 0.4187$ is the von Karman constant. The Reynolds stress anisotropy tensor ($b_{ij}$) which quantifies deviations from isotropic turbulence in terms of turbulent kinetic energy is expressed as follows.
\begin{gather} 
	\label{anisotropy-tensor}
	\tilde{b}_{ij} = - \left( \frac{-\overline{\rho}\widetilde{ u^{\prime\prime}_i u^{\prime\prime}_j} + \frac{2}{3} \overline{\rho} k \delta_{ij}}{2 \overline{\rho} k} \right)
\end{gather}
%
%===========================
\end{document}